\documentclass[acmtog,screen,nonacm]{acmart}
\usepackage{subcaption}
\usepackage{graphicx}
\usepackage{algorithm}
\usepackage{algpseudocode}
\usepackage{multirow}
\usepackage{wrapfig}
\usepackage{pdfpages}
\usepackage{graphicx}
\algrenewcommand\textproc{}
\newcommand{\modify}[1]{\textcolor{black}{#1}}

\newcommand{\mbx}{\ensuremath{\mathbf{x}}}
\newcommand{\mbq}{\ensuremath{\mathbf{q}}}
\newcommand{\realNum}{\ensuremath{\rm I\!R}}
\DeclareMathOperator*{\argmin}{arg\,min}
\AtBeginDocument{%
	\providecommand\BibTeX{{%
			\normalfont B\kern-0.5em{\scshape i\kern-0.25em b}\kern-0.8em\TeX}}}

\begin{document}
	
	% \title{Fast Unified GPU-IPC Framework for Affine-Deformable Coupling}
	%Advancing GPU IPC for Stiff Elastodynamics}
	\title{StiffGIPC: Advancing GPU IPC for Stiff Affine-Deformable Simulation}
	
	\author{Kemeng Huang}
        %\authornotemark[2]
        %\authornotemark[2]
        %\authornote{project lead}
	\orcid{0000-0001-9147-2289}
	\email{kmhuang@connect.hku.hk}
	\email{kmhuang819@gmail.com}
        \affiliation{%
		\institution{Carnegie Mellon University}
		\country{USA}
	}
	\affiliation{%
		\institution{The University of Hong Kong,  TransGP}
		\country{Hong Kong SAR}
	}
	
	\author{Xinyu Lu}
        %\authornote{equal coding contribution}
	\email{lxy819469559@gmail.com}
	\affiliation{%
            \institution{TransGP}
		%\institution{TransGP}
		\country{Hong Kong SAR}
	}
	
	\author{Huancheng Lin}
	\orcid{0000-0003-4446-1442}
	\email{lamws@connect.hku.hk}
        \affiliation{%
		\institution{Carnegie Mellon University}
		\country{USA}
	}
	\affiliation{%
		\institution{The University of Hong Kong,  TransGP}
		\country{Hong Kong SAR}
	}
	
	\author{Taku Komura}
	\orcid{0000-0002-2729-5860}
	\email{taku@cs.hku.hk}
	\affiliation{%
		\institution{The University of Hong Kong, TransGP}
		\country{Hong Kong SAR}
	}
	
	\author{Minchen Li}
	\orcid{0000-0001-9868-7311}
	\email{minchernl@gmail.com}
	\affiliation{%
		\institution{Carnegie Mellon University}
		\country{USA}
	}
	
	% \affiliation{%
	
	% \affiliation{%
	
	% \affiliation{%
	
	% \affiliation{%
	
	% \affiliation{%
	
	% \affiliation{%
	
	% \affiliation{%
	
	\renewcommand{\shortauthors}{Huang et al.}
	
	\newcommand{\mc}[1]{\textcolor{blue}{Minchen: #1}}
	\newcommand{\todo}[1]{\textcolor{red}{\textbf{Todo: #1}}}
	
	\begin{abstract}

		Incremental Potential Contact (IPC) is a widely used, robust, and accurate method for simulating complex frictional contact behaviors. However, achieving high efficiency remains a major challenge, particularly as material stiffness increases, which leads to slower Preconditioned Conjugate Gradient (PCG) convergence, even with the state-of-the-art preconditioners. In this paper, we propose a fully GPU-optimized IPC simulation framework capable of handling materials across a wide range of stiffnesses, delivering consistent high performance and scalability with up to 10$\times$ speedup over state-of-the-art GPU IPC methods.
		Our framework introduces three key innovations: 1) A novel connectivity-enhanced Multilevel Additive Schwarz (MAS) preconditioner on the GPU, designed to efficiently capture both stiff and soft elastodynamics and improve PCG convergence at a reduced preconditioning cost. 2) A $C^2$-continuous cubic energy with an analytic eigensystem for \modify{inexact} strain limiting, enabling more parallel-friendly simulations of stiff membranes, such as cloth, without membrane locking. 3) For extremely stiff behaviors where elastic waves are barely visible, we employ affine body dynamics (ABD) with a hash-based \modify{two-level} reduction strategy for fast Hessian assembly and efficient affine-deformable coupling.
		We conduct extensive performance analyses and benchmark studies to compare our framework against state-of-the-art methods and alternative design choices. Our system consistently delivers the fastest performance across soft, stiff, and hybrid simulation scenarios, even in cases with high resolution, large deformations, and high-speed impacts. %Our framework will be fully open-sourced upon acceptance.
		
	\end{abstract}
	
	\begin{CCSXML}
		<ccs2012>
		<concept>
		<concept_id>10010147.10010371.10010352.10010379</concept_id>
		<concept_desc>Computing methodologies~Physical simulation</concept_desc>
		<concept_significance>500</concept_significance>
		</concept>
		<concept>
		<concept_id>10010147.10010169.10010170</concept_id>
		<concept_desc>Computing methodologies~Parallel algorithms</concept_desc>
		<concept_significance>300</concept_significance>
		</concept>
		</ccs2012>
	\end{CCSXML}
	
	\ccsdesc[500]{Computing methodologies~Physical simulation}
	\ccsdesc[300]{Computing methodologies~Parallel algorithms}

	\keywords{GPU Programming, Incremental Potential Contact, Elastodynamics, Finite Element Method, Affine Body Dynamics, Preconditioning, Cloth Simulation}
	
	\begin{teaserfigure}
		\includegraphics[width=\textwidth, , trim=100 0 100 0, clip]{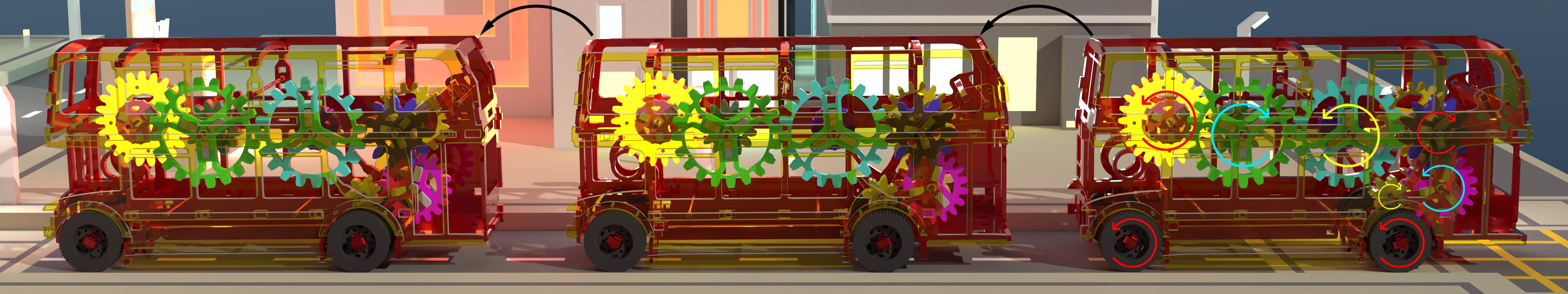}
		\caption{\textbf{London bus.} We model a $0.7m \times 1.2m \times 2.3m$ bus, including its frame, gears, and wheels, based on the puzzle instructions from Wooden City$^\circledR$. Despite the \textit{eye-norm} modeling accuracy, our simulation robustly captures the realistic non-smooth motion caused by potential interference in the low-precision gear system and the self-adjustment of motor speed (averaging $3m/s$) relative to resistance (see our supplemental video). The simulation includes 265K surface triangles and an average of 118K rapidly changing contact pairs, achieving efficient performance at $1.56s$ per time step ($\Delta t = 5ms$).}
		\label{fig:teaser}
	\end{teaserfigure}
	
	
	\maketitle
	
	\section{Introduction}
	Incremental Potential Contact (IPC)~\cite{ipc} is a cutting-edge elastodynamic contact simulation method widely used in computer graphics, computational mechanics, robotics, etc. Despite its robustness, accuracy, and differentiability in simulating complex frictional contact behaviors, IPC's efficiency remains a significant bottleneck, limiting its full potential. Several variants have been proposed to address IPC's efficiency issues, often at the expense of accuracy~\cite{medial_ipc, 2ndDec_ipc, li2023subspace,pd_ipc}. 
	
	To accelerate IPC without sacrificing accuracy, \citet{gipc} introduced GIPC with a GPU-friendly redesign of the numerical algorithms. This included replacing direct factorization with a Preconditioned Conjugate Gradient (PCG) solver and proposing a Gauss-Newton approximation of the barrier Hessian matrices with analytic eigensystems. While GIPC is effective, its efficiency deteriorates significantly as object stiffness increases. This is mainly due to the growing condition number of the global linear system, which requires more PCG iterations to solve. 
   To improve the PCG convergence of stiff material, Wu et al.~\shortcite{mas} proposed the multilevel additive Schwarz (MAS) preconditioner \cite{mas}.
   Their approach involves sorting the nodes based on Morton codes and building a hierarchy by grouping the nodes at each level.    
   Despite its effectiveness, the lack of consideration  
   for mesh connectivity during reordering leads to suboptimal domain hierarchy construction. This results in high construction costs, additional overhead for deformable simulations
   and challenges for GPU optimization.

	When simulating stiff elastic thin shells like cloth, another challenge arises, which is membrane locking. With linear triangle elements, the stiff membrane energy (Young's modulus around $10MPa$ for cloth \cite{penava2014determination}) will often result in nonnegligible extra artificial bending resistance, forming sharp creases and plastic appearances in the simulation results (\autoref{fig:memLocking} top). Simulating cloth with smaller stiffness can result in more realistic wrinkles, but it will suffer from over-elongation issues. To tackle this challenge, \citet{cipc} propose to augment soft membrane energy with a barrier-based strain-limiting term to prevent cloth from over-stretching while avoiding membrane locking. This strategy enables realistic cloth simulation within the IPC framework, but the required exact strain limit satisfaction necessitates a backtracking-based line search \textit{filtering} scheme, as the updated strain has a complicated relation to the step size, making analytic expressions unavailable. Additionally, numerical eigendecomposition is needed for computing a positive semi-definite approximation of the strain-limiting term's Hessian matrix, which further complicates GPU optimization.
	
	For even stiffer problems where elastic waves are barely visible, objects can be treated as rigid \cite{ferguson2021intersection} or stiff affine \cite{abd} bodies in the IPC framework for a reduced number of degrees of freedom (DOF). But to accurately simulate contact behaviors, surface elements from the original input geometry are used, which also makes simulating rigid-deformable coupled scenarios convenient.
	\citet{ubarrier} introduced a unified Newton barrier method for stiff affine-deformable simulation, possibly with articulation constraints. However, although some components are GPU-accelerated, the primary simulation processes still execute on the CPU, leading to suboptimal performance.
	ZeMa~\cite{zema} is another GPU IPC framework for stiff affine-deformable simulation, with most processes parallelized on the GPU, except for the linear system, which is solved on the CPU using a direct solver. However, ZeMa lacks a well-optimized contact Hessian assembly algorithm, as it accumulates the $12\times12$ dense contact Hessian matrices to the affine body DOFs atomically, where conflicting operations can significantly impede the performance, especially when there are a large number of contacts. Moreover, direct solvers often fall short in large-scale simulations.
	
	In summary, \modify{there is still plenty of room} for optimizing linear solver preconditioners, strain limiting, global Hessian matrix assembly, etc., for realizing a highly GPU-optimized IPC framework that can efficiently simulate large-scale affine-deformable coupled scenarios.
	In this paper, we propose such a framework, achieving up to $10\times$ speedup compared to GIPC via the following 3 major innovations:
	
	\begin{itemize}
		\item A novel connectivity-enhanced MAS preconditioner on the GPU that achieves improved PCG convergence at a lower precomputation and per-iteration cost (\autoref{sec:preconditioner}). Our preconditioner consistently performs effective and well-structured aggregations, which supports smaller blocksizes and further GPU optimizations based on warp reduction.
		\item A $C^2$-continuous cubic \modify{inexact} strain-limiting energy with an analytic eigensystem, enabling realistic cloth simulation without membrane locking (\autoref{sec:membrane_energy}). As numerical eigendecomposition and line search \textit{filtering} for the feasibility of the strain limits are not needed, our model supports highly GPU-parallelized computations.
		\item 
		% We propose a heuristic based on Young's modulus and time step size to decide whether to reduce an object to the affine space and 
		A hash-based \modify{two-level} reduction strategy for fast Hessian matrix assembly \modify{in affined-deformable coupled systems} (\autoref{sec:global_hessian}). Our strategy significantly reduces the number of numerical operations, and it enables the development of a memory-efficient symmetric blockwise Sparse Matrix-Vector Multiplication (SpMV) method to further boost PCG performance.
        % \modify{\item 
        % This is the first full-GPU implementation of unified IPC framework designed for stiff affine-deformable simulation. In this framework, deformable objects are simulated using FEM, while rigid objects are simulated using ABD. This design enables efficient handling of rigid objects and their seamless coupling with deformable objects.}

	\end{itemize}
	In \autoref{sec:exp}, we perform extensive and rigorous performance analyses and benchmark studies to validate our framework and compare it to state-of-the-art GPU IPC systems and alternative design choices that may seem reasonable but suffer from suboptimal performance in practice. Our framework exhibits the fastest performance in soft, stiff, and hybrid simulation scenarios, even with high resolution, extreme deformation, and high-speed impacts. %Our system will be fully open-sourced upon acceptance.
	
	\section{Related Works}
	
	\paragraph{Contact Simulation}
	Simulating frictional contact for (nearly) rigid and deformable solids has been an extensively studied topic in both computer graphics and computational mechanics. Starting from a few decades ago, various methods have been developed, ranging from impulse-based methods \cite{bridson2002robust,harmon2009asynchronous}, impact zone methods \cite{harmon2008robust,tang2018cloth,li2020p}, and more constraint-based methods \cite{kaufman2008staggered,otaduy2009implicit,macklin2019non,verschoor2019efficient,allard2010volume} to fictitious domain methods \cite{jiang2017simplicial,muller2015air,misztal2012topology,pagano2008self}, etc. These methods are effective and efficient in many situations, but they lack guarantees on algorithmic convergence and generating penetration-free results, especially when simulating challenging examples where extensive parameter tuning is often also required when varying set-ups. We refer to \citet{SiggraphContact22} for a comprehensive review.
	
	More recently, \citet{ipc} proposed Incremental Potential Contact (IPC), which simulates the nonsmooth frictional contact behaviors of deformable solids by approximating them using smooth constitutive models with bounded error. Equipped with a filter line search scheme and the projected Newton method, penetration-free results and algorithmic convergence are guaranteed within an optimization time integrator. IPC has shown effectiveness in various application scenarios, including cloth reconstruction \cite{zheng2024physavatar}, material modeling of interlocked rigid components \cite{tang2023beyond}, multi-material coupled simulations \cite{xie2023contact,jiang2022hybrid,li2024dynamic}, etc. Despite being robust and accurate, improving the efficiency of IPC while maintaining its reliability is still a major challenge.
	
	\paragraph{Accelerating IPC Performance}
	Over the past few years, several methods have been proposed to enhance IPC's efficiency. \citet{medial_ipc} reduced the DOFs of deformable bodies to their medial axis and derived contact and friction forces on the associated slab primitives. With significantly fewer DOFs, the efficiency improved substantially at the expense of accuracy, and penetration-free results were only guaranteed on their medial representation. \citet{pd_ipc} incorporated IPC into the projective dynamics framework on the GPU, delaying contact constraint set updates and continuous collision detection (CCD) to once per set of inner iterations to efficiently generate penetration-free results. However, their spring-based approximation to the barrier-based contact model limited their performance in challenging scenarios. \citet{2ndDec_ipc} applied a stencil-wise coordinate descent method to solve the IPC system. With a penetration-free warm start and graph coloring-based GPU parallelization, their method achieved fast performance. However, as a block coordinate descent method, its performance drops quickly with stiff materials or a large number of contacts. \citet{wang2023fast} and \citet{li2023subspace} both applied time-splitting techniques to decouple contact and elasticity simulation, achieving faster performance but requiring small time steps for stiff problems due to the lack of unconditional stability.
	These methods trade accuracy for efficiency using simplified models or specialized strategies, often introducing extra parameters for tuning. 
	
	More recently, \citet{gipc} introduced GIPC, a fully GPU-optimized IPC method with an inexact Gauss-Newton solver using the MAS preconditioner \cite{mas}. GIPC also derived approximated contact energy Hessians with analytic eigensystems, avoiding GPU-unfriendly numerical eigendecomposition. 
    Concurrently, \citet{zema} proposed ZeMa, another GPU IPC framework supporting the coupling of deformable bodies with stiff affine bodies. Both GIPC and ZeMa achieved significant speed-ups without sacrificing accuracy. However, as discussed earlier, their high performance could unavoidably degrade when there are stiff materials or a large number of contacts. We thus follow this path to keep advancing GPU IPC for stiff elastodynamics without sacrificing accuracy and robustness.
	
	\paragraph{Multiresolution Methods}
	One of the most critical design choices on the GPU is to apply iterative solvers instead of direct solvers for linear systems.
	% , since direct solvers based on matrix factorization, which often result in much denser matrices, are memory-bound and challenging to parallelize. 
	Multigrid is a popular GPU-friendly preconditioner often used in iterative linear solvers, such as the PCG method, for efficiently solving ill-conditioned systems.
	% However, iterative linear solvers, such as the PCG method, need effective preconditioners for efficiently solving ill-conditioned systems. Multigrid is a popular GPU-friendly preconditioner often used in physics-based animations.
	The major challenge in multigrid methods is the development of effective restriction and prolongation operators, especially for unstructured meshes. Researchers in computer graphics have explored various geometric multigrid methods (GMG) \cite{wang2020hierarchical,wang2018parallel,xian2019scalable,zhu2010efficient} by constructing mesh hierarchies, where numerous implementation issues are often encountered. Compared to GMG, another type of \modify{multigrid}, the algebraic multigrid (AMG) methods \cite{naumov2015amgx,demidov2019amgcl,tamstorf2015smoothed}, do not require explicitly constructing coarser geometries. However, they often require a significant amount of precomputation to analyze the structure of the matrix, making them more expensive when handling problems with varying matrix structures. This inspires \citet{mas} to propose a multilevel additive schwarz (MAS) \cite{dryja1990multilevel} preconditioner, incorporating the idea of domain decomposition to achieve fast convergence with low per-iteration costs when applied to PCG. But its node reordering based on Morton code is not aware of the mesh connectivity, which may lead to suboptimal aggregation. We thus propose a connectivity-enhanced MAS based on METIS \cite{METIS} to tackle this issue and achieve even faster performance.

    \modify{\paragraph{Sparse Matrix-Vector Multiplication} 
    Optimizing sparse matrix-vector multiplication (SpMV), the primary bottleneck in iterative linear solvers, is as crucial as accelerating the convergence. \citet{csx-sym} introduced a symmetric variant of the Compressed Sparse eXtended (CSX) \cite{kourtis2011csx} format, leveraging nonzero indexing compression to reduce memory traffic in symmetric SpMV. \citet{bcsr} proposed optimizations using the Block Compressed Row Storage (BCSR) format, while GIPC \cite{gipc} developed a matrix-free SpMV method tailored to their PCG solver. A recent survey on Sparse Matrix-Matrix Multiplication (SpMM) \cite{SpMM} further discusses relevant optimization strategies. Currently, cuSPARSE\footnote{\href{https://developer.nvidia.com/cusparse}{https://developer.nvidia.com/cusparse}} provides the fastest SpMV implementation via the Block Sparse Row (BSR) format, utilizing GPU-optimized instructions and warp reduction. However, taking full advantage of BSR SpMV requires an efficient Hessian assembly strategy, such as \citet{tang2016cama}. Furthermore, since SpMV is memory-bound and the Hessian is symmetric, leveraging symmetry can enhance performance, but cuSPARSE does not support this optimization. To address these limitations, we introduce a fast Hessian assembly method using a sorted block coordinate format, enabling efficient warp reduction and symmetry exploitation, surpassing cuSPARSE BSR SpMV in performance.}
	
	\paragraph{Strain Limiting and Eigenanalysis}
	To realistically simulate cloth using linear elements, strain limiting is often applied to avoid membrane locking. Most existing methods \cite{narain2012adaptive,wang2010multi,thomaszewski2009continuum,english2008animating,goldenthal2007efficient} enforce a bound constraint on the strain measure per element using augmented Lagrangian approaches or post-projection techniques. While these methods do not guarantee strict satisfaction of strain limits, they can effectively simulate membrane locking-free behaviors. However, when coupled with frictional contact, these methods often introduce artifacts due to the need for smaller time steps or difficulties in achieving solver convergence.
	More recently, \citet{cipc} applied a barrier method to handle strain-limiting constraints, guaranteeing exact constraint satisfaction in a monolithic manner. However, it requires expensive backtracking-based line search \textit{filtering} and numerical eigendecomposition, which are both GPU-unfriendly. \modify{Here, two backtracking line searches are needed in each Newton iteration: the first, often called line search \textit{filtering}, determines the maximum step size that maintains a feasible configuration, while the second starts from this step size and further refines it to find a configuration with a lower objective function.}
	With the development of analytical eigenanalysis on distortion energies, such as the isotropic \cite{smith2019analytic,lin2022isotropic} and anisotropic \cite{anisotropic_eigen} elasticity, membrane \cite{panetta2020analytic,kim_baraf_witkin}, bending \cite{AeigenBending_wang,eigen_bend}, and contact \cite{shi2023unified,gipc} energies, they have become popular components of modern GPU simulation methods.
	We thus propose a cubic \modify{inexact} strain-limiting energy with analytic eigensystems so that realistic cloth dynamics can be more efficiently simulated without the need of numerical eigendecomposition and backtracking-based line search \textit{filtering}. \modify{Concurrently, \citet{cubicBarrier} proposed a cubic barrier energy to strictly enforce penetration-free and strain-limiting constraints, with the aim of avoiding vanishing search steps when the constraint gaps are tiny.}
	
	% \paragraph{GPU Optimization}
	% The GPU optimization of physical simulation
	% is sought for accelerating high-quality and computationally demanding animations \citet{li2020p,muller2003particle,tang2016cama,tang2018cloth,wang2021gpu}. For deformable objects, researchers have considered accelerating collision response
	% based on impact zones [Harmon et al. 2008] on the GPU (see, e.g.,
	% Tang et al. [2018b] and Li et al. [2020]), while others (e.g., Chitalu
	% et al. [2020], Lauterbach et al. [2010]) accelerate proximity queries.
	% Parallel techniques like Wang [2021] also simulate sub-millimeter
	% level cloth deformation with regular grids fitted to the garment.
	
	\section{Background and preliminaries}
	\label{sec:background}

	\begin{figure*}[htbp]
		\centering
		\begin{subfigure}[b]{0.8\textwidth}
			\centering
			\includegraphics[width=\columnwidth, trim=2 0 2 0, clip]{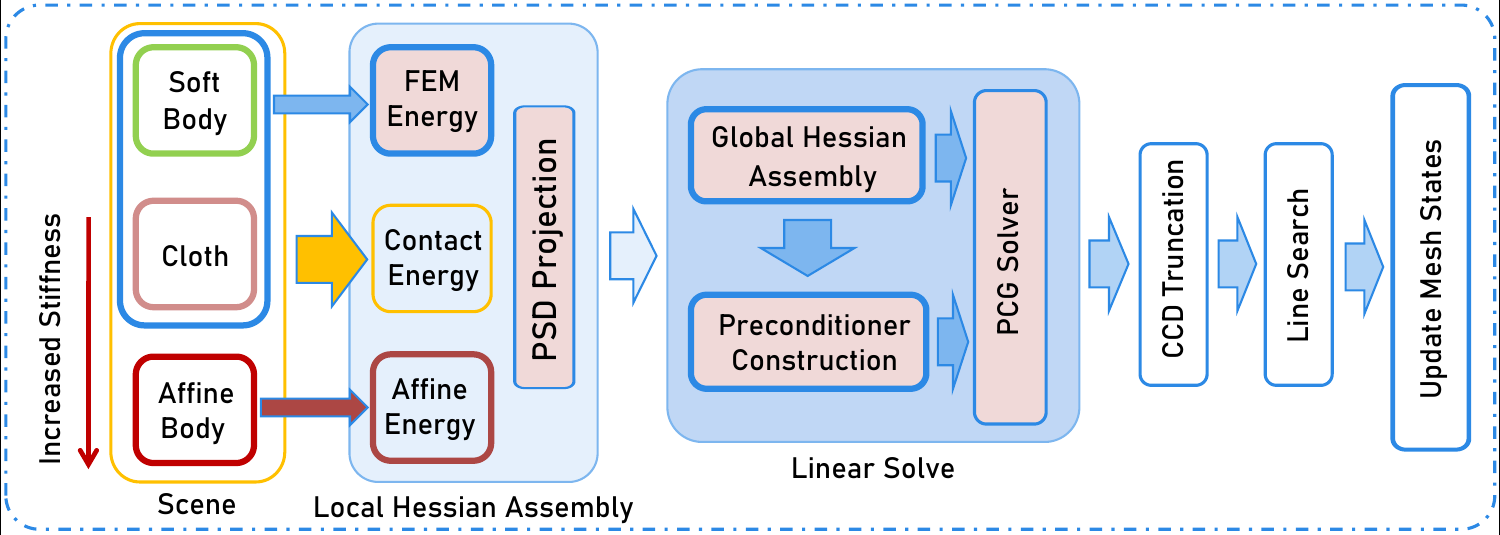}
		\end{subfigure}
		\caption{\label{fig:framework}\textbf{An overview of 1 Newton iteration in our unified GPU IPC framework.} Our simulation focuses on cloth, soft bodies with varying stiffness, and stiff affine bodies \cite{abd}. Cloth and soft bodies are modeled using FEM. 
			% When a soft body's stiffness exceeds the limit of observable deformation, it is simulated as a stiff affine body \cite{abd}. 
			Contact between all objects is handled using IPC \cite{ipc}. We solve the system with a Newton-PCG solver and use Continuous Collision Detection (CCD) to filter the search direction and prevent penetration, applying backtracking line search to ensure energy decrease. Our contributions, highlighted in light pink blocks, include a novel inexact strain limiting energy with an analytic eigensystem for efficient PSD projection, an optimized linear solver with a connectivity-enhanced MAS preconditioner, and a highly-parallelized global Hessian assembly strategy for affine-deformable coupling.}
	\end{figure*}
	
	\subsection{Unified Incremental Potential}
	IPC~\cite{ipc} formulates implicit time integration of elastodynamic contact as a minimization problem:
	\begin{equation}\label{eq:energy-min-deformable}
	\mathbf{x}^{t+\Delta{t}}\approxeq\argmin_{\mathbf{x} \in \realNum^{3n}} E_s(\mathbf{x}),
	\end{equation}
	where $\mathbf{x}$ is the world-space positions of the $n$ nodes, followed by a velocity update $\mathbf{v}^{t+\Delta{t}} = (\mathbf{x}^{t+\Delta{t}}-\mathbf{x}^t)/\Delta t$, {taking implicit Euler as an example}. The objective function
	\begin{equation}\label{eq:E_s}
	E_s(\mbx) = \frac{1}{2}(\mathbf{x} -\mathbf{\hat{\mathbf{x}}})^T\mathbf{M}_s(\mathbf{x} -\mathbf{\hat{\mathbf{x}}})+\Delta t^2\Psi_s(\mathbf{x})+B(\mathbf{x})+D(\mathbf{x})
	\end{equation}
	is the Incremental Potential (IP), where $\hat{\mathbf{x}} = \mathbf{x}^t+\Delta t \mathbf{v}^t+\Delta t^2 \mathbf{M}_{s}^{-1} \mathbf{f}_e$, $\mathbf{M}_s$ is the mass matrix, $\mathbf{f}_e$ is the external force, $B$ is the contact barrier potential, and $D$ is the approximated friction potential. 
	{The total elasticity energy $\Psi_s(\mathbf{x}) = \Psi_{vol}(\mbx) + \Psi_{memb}(\mbx)+\Psi_{bend}(\mbx)+\Psi_{strain}(\mbx)$ contains volumetric strain energies ($\Psi_{vol}(\mbx)$) as well as membrane ($\Psi_{memb}(\mbx)$), bending ($\Psi_{bend}(\mbx)$), and strain-limiting ($\Psi_{strain}(\mbx)$) energies for thin shells.}
	% be defined as
	% \begin{equation}\label{eq:psi_s}
	% \Psi_s(\mbx) = \Psi_{vol}(\mbx)+\Psi_{shell}(\mbx),
	% \end{equation}
	% { where $\Psi_{vol}(\mbx)$ is the total elasticity energy for volumetric bodies} and $\Psi_{shell}(\mbx)$ for thin shells includes membrane, bending, and strain-limiting terms~\cite{cipc}:
	% \begin{equation}\label{eq:psi_shell}
	% \Psi_{shell}(\mbx) = \Psi_{memb}(\mbx)+\Psi_{bend}(\mbx)+\Psi_{strain}(\mbx).
	% \end{equation}
	
	The IP for simulating affine bodies~\cite{abd} is defined as
	\begin{equation}\label{eq:E_r}
	E_r(\mbq) = \frac{1}{2}(\mathbf{q} -\mathbf{\hat{\mathbf{q}}})^T\mathbf{M}_r(\mathbf{q} -\mathbf{\hat{\mathbf{q}}})+\Delta t^2\Psi_r(\mathbf{q})+B(\mathbf{x}(\mathbf{q}))+D(\mathbf{x}(\mathbf{q})),
	\end{equation}
	where $\mbq \in \realNum^{12\alpha}$ is the reduced space coordinates of the $\alpha$ affine bodies. For affine body $j$, $\mbq_j = [\mathbf{p}_j^T, \mathbf{A}_{j1}^T, \mathbf{A}_{j2}^T, \mathbf{A}_{j3}^T]^T$, with $\mathbf{p}_j \in \realNum^3$ the translation vector and $\mathbf{A}_j \in \realNum^{3\times 3}$ the affine deformation matrix. Here, $\mathbf{x}_i(\mathbf{q}_j) = \mathbf{A}_j\bar{\mathbf{x}}_i+\mathbf{p}_j$ is the current position of full-space node $i$ on affine body $j$ for measuring contact energies, with $\bar{\mathbf{x}}_i$ the rest-state full space coordinates. The Jacobian matrix $\mathbf{J}_{ij} = \frac{\partial \mathbf{x}_i}{\partial \mathbf{q}_j} \in \realNum^{3\times12}$ maps information between the reduced and full spaces. For example, the mass matrix $\mathbf{M}_r$ of affine body $j$ is calculated as $\mathbf{M}_r = \sum_i m_i\mathbf{J}_{ij}^T\mathbf{J}_{ij},$ where $m_i$ is \modify{the lumped mass corresponding to} each node $i$. Similarly, the contact gradient of affine body $j$ is calculated as $\nabla_{\mathbf{q}_j} B = \sum_i \mathbf{J}_{ij}^T\nabla_{\mathbf{x}_i} B.$ From this, we can see that computing contact forces and Hessian matrices for the reduced DOFs could require accumulating values from a significant amount of contact pairs, which can be expensive. 
	
	Combining \autoref{eq:E_s} and \autoref{eq:E_r}, denoting nodal positions of unreduced deformable solids as $\mathbf{x}_s$, the unified {affine-deformable} coupled IP~\cite{ubarrier} is defined as
	\begin{align}\label{eq:E_all}
		E(\mbq, \mbx_s) & =\frac{1}{2}(\mathbf{x}_s -\mathbf{\hat{\mathbf{x}}}_s)^T\mathbf{M}_s(\mathbf{x}_s -\mathbf{\hat{\mathbf{x}}}_s)+\Delta t^2\Psi_s(\mathbf{x}_s) 
		\nonumber\\ &\, \,
		+\frac{1}{2}(\mathbf{q} -\mathbf{\hat{\mathbf{q}}})^T\mathbf{M}_r(\mathbf{q} -\mathbf{\hat{\mathbf{q}}})+\Delta t^2\Psi_r(\mathbf{q}) \nonumber\\ &\, \,
		+B([\mathbf{x}_s^T,\mathbf{x}(\mathbf{q})^T]^T)+D([\mathbf{x}_s^T,\mathbf{x}(\mathbf{q})^T]^T),
	\end{align}
	and we minimize this energy w.r.t. $\{\mathbf{q}, \mathbf{x}_s\}$ per time step to simulate the coupled system:
	\begin{equation}\label{eq:energy-min}
	(\mathbf{q}, \mathbf{x}_s)^{t+\Delta{t}}\approxeq\argmin_{\mathbf{x}_s \in \realNum^{3n}, \mathbf{q} \in \realNum^{12\alpha}} E(\mathbf{q}, \mathbf{x}_s).
	\end{equation}
	
	To accelerate local Hessian computation on the GPU, we utilize potential energies with analytic eigensystems (\autoref{sec:ip_min_linsol}). Specifically, for collision energy $B$ and friction energy $D$, we employ the eigenanalysis from GIPC \cite{gipc}. The Stable Neo-Hookean model~\cite{snk} is applied for $\Psi_{vol}$, while the bending energy from \cite{eigen_bend} is used for $\Psi_{bend}$. 
	To efficiently assemble $\Psi_{memb}$ and $\Psi_{strain}$, we use \citet{kim_baraf_witkin}'s membrane energy and propose a cubic \modify{inexact} strain-limiting energy with analytic eigensystem, detailed in \autoref{sec:membrane_energy}.
	
	\subsection{IP Minimization and Linear Solves}\label{sec:ip_min_linsol}
	To minimize $E(\mathbf{u})$, where $\mathbf{u} = \{\mbq, \mbx\}$, line search methods are often used, which \modify{compute} a descent direction by minimizing a local quadratic proxy of the IP in iteration $i$:
	\begin{equation}\label{eq:tailor}
	E_i(\mathbf{u}) = E(\mathbf{u}_i) + \left(\mathbf{u} - \mathbf{u}_i\right)^T\nabla E(\mathbf{u}_i) + \frac{1}{2}\left(\mathbf{u} - \mathbf{u}_i\right)^T \mathbf{P}(\mathbf{u}_i) \left(\mathbf{u} - \mathbf{u}_i\right),
	\end{equation}
    \modify{and then search along this direction to find a new iterate.}
	The descent direction is obtained by solving the linear system $\mathbf{P}(\mathbf{u}_i)  \mathbf{d} = -\nabla E(\mathbf{u}_i)$, where $\mathbf{d} = \mathbf{u}_i^* - \mathbf{u}_i$ and $\mathbf{u}_i^* = \argmin_{\mathbf{u}} E_i(\mathbf{u})$. Here, $\mathbf{P}$ is a symmetric positive definite (SPD) proxy matrix approximating $\nabla^2 E$ for fast convergence, with projected Newton~\cite{ipc} or Gauss-Newton~\cite{gipc} approximations shown effective. The new iterate is computed as $\mathbf{u}_{i+1} = \mathbf{u}_i + \alpha\mathbf{d}$, where $\alpha$ is the step size calculated via backtracking line search, with filtering to ensure feasibility imposed by contact and strain limits \cite{cipc}. The minimization terminates when the iterate is sufficiently close to the local minimum, e.g. when $\|\mathbf{d}\| \le \varepsilon_d$, where $\varepsilon_d$ is the Newton tolerance.

	To solve the linear system $\mathbf{P}(\mathbf{u}_i) \mathbf{d} = -\nabla E(\mathbf{u}_i)$, Cholesky factorization, using e.g. CHOLMOD~\cite{chen2008algorithm}, is a popular solution. These direct solvers perform well on ill-conditioned systems but their performance decreases with increasing DOFs, especially for large-scale simulations. Additionally, the high memory cost of direct solvers results in worse performance on the GPU compared to the CPU~\cite{medial_ipc}. Conversely, iterative solvers, such as the preconditioned conjugate gradient (PCG) method, iteratively approach the solution via matrix-vector multiplications (SpMV), making them more practical for large-scale simulations.

	Although PCG has superior convergence rate compared to many non-Krylov iterative solvers, such as Jacobi and Gauss-Seidel, it still requires many iterations to reach low error when the linear system is ill-conditioned. An effective preconditioner that approximates the inverse of the proxy matrix can make the system better conditioned and easier to solve. MAS~\cite{mas} is currently the best option considering both convergence speed and the overhead of computing and applying the preconditioner. Additionally, the high computational cost of matrix assembly and SpMV in each iteration often makes PCG the major bottleneck in simulation. Thus, designing GPU-parallel algorithms for PCG is also critical. See \autoref{fig:framework} for an overview of our simulation framework.

	% This paper makes contributions in all three of these aspects, encompassing the unified framework design, resulting in a super-efficient simulator that excels in all types of stiff simulations.

	\section{Connectivity-Enhanced MAS Preconditioner with GPU Optimization}
	\label{sec:preconditioner}
	The multilevel additive Schwarz (MAS) preconditioner \cite{mas} enables efficient linear solver convergence with excellent scalability w.r.t. problem size and stiffness in elastodynamics problems. However, its node reordering has some limitations, which makes it less effective in cases with complex geometry or softer materials.
	We start by a comprehensive analysis of these limitations (\autoref{sec:mas_limit}), and then propose a connectivity-enhanced MAS construction (\autoref{sec:geometry_mas}) together with further GPU optimizations (\autoref{sec:gpu_geometry_mas}) to achieve a $2.2\times$ faster performance in average compared to \citet{mas}.
	% Although the convergence speed of MAS may not be the fastest, it outperforms other popular preconditioners by having a relatively low computational cost and excellent scalability w.r.t. problem size and stiffness.
	% Recent studies on GPU MAS~ demonstrate its significant superiority in these aspects.

\begin{figure}[htbp]
		\centering
		\begin{subfigure}[b]{0.3\textwidth}
			\centering
			\includegraphics[width=\columnwidth, trim=0 0 0 0, clip]{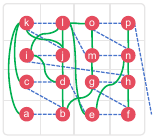}
		\end{subfigure}
		\caption{\label{fig:mas_morton}\textbf{A node sorting example using Morton code.} Red disks represent nodes, green lines show topology connections, and dotted blue lines indicate Morton code order. After sorting, nodes will follow the character sequence shown. Note that this is a contrived example for clarity.}
	\end{figure}

	\subsection{Discussions of MAS}\label{sec:mas_limit}
	
	MAS constructs the domain hierarchy by first sorting all mesh vertices using Morton codes, and then grouping these vertices into subdomains of size $N$. Next, connected vertices within each subdomain are merged to create the next-level simulation domain. These merged vertices, called super nodes, are then grouped again to form the domains at subsequent levels until a maximum level is reached, no further merges could be performed, or only a single subdomain is present. During preconditioning, the inverse of each subdomain's Hessian matrix is multiplied with the mapped input vector, and the resulting vectors are mapped and summed to produce the output \modify(see our supplemental document for more details). With a fixed number of levels, maximizing node connectivity within each subdomain is critical to making dimensionality reduction more effective, which allows information to propagate farther at coarser levels and ensures faster convergence. See our supplemental document for more details.
	
	We identify two major areas for potential improvement in MAS:
	1) The convergence of MAS heavily relies on node reordering using Morton codes. While Morton codes capture spatial information, they do not consider mesh connectivity, which can result in slower convergence.
	2) To obtain denser node connectivity within each subdomain and improve aggregation efficiency, a subdomain size of 32 vertices is typically used in MAS. However, this leads to significant overhead in terms of precomputation and preconditioning, especially when using double-precision floating-point numbers. Reducing the subdomain size to 16 vertices can decrease computational costs, but it may also slow down convergence, as fewer nodes are merged within each subdomain, potentially resulting in worse overall performance \cite{mas}.

	\begin{figure}[htbp]
		\centering
		\begin{subfigure}[b]{0.48\textwidth}
			\centering
			\includegraphics[width=\columnwidth, trim=0 0 0 0, clip]{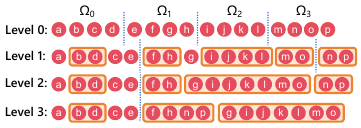}
		\end{subfigure}
		\caption{\label{fig:mas_level_old}\textbf{MAS aggregation of \autoref{fig:mas_morton} using Morton code.} Within each subdomain (size 4 here, separated by dashed blue lines), connected nodes are merged into a super node (orange block) at the next coarser level. The order of the super node is determined by the lowest order of the merged nodes. For example, in level 1, nodes $b$ and $d$ are merged and positioned before $c$ because $b$ has a lower order. The subdomains are then reconstructed from sets of 4 consecutive super nodes, and here this process continues until no further merging is possible at a certain level.
		}
	\end{figure}

	%\subsubsection{Ideal MAS construction}
	
	\paragraph{Case Study}
	Let's examine the first issue more closely with an example (\autoref{fig:mas_morton}). In this case, we assume a domain size of 4 for MAS, meaning the nodes are grouped into subdomains $\Omega = \Omega_0 \cup \Omega_1 \cup \Omega_2 \cup \Omega_3$, each with 4 nodes, at level 0, as shown in \autoref{fig:mas_level_old}. In $\Omega_0 = \{a, b, c, d\}$, only nodes $\{b, d\}$ are connected. Consequently, at the coarser level 1, these two nodes are merged into a super node. Similarly, nodes $\{f, h\}$, $\{i, j, k, l\}$, $\{m, o\}$, and $\{n, p\}$ are merged at level 1. However, starting from level 1, there is no connectivity between any super nodes in the first subdomain, and only a few can be merged in the other subdomains. This results in 3 levels for these 16 nodes, with inefficient aggregation, leading to poor performance of the MAS preconditioner. In fact, the performance can even be worse than that of a simple block Jacobi preconditioner in practice.
	
	\begin{figure}[htbp]
		\centering
		\begin{subfigure}[b]{0.48\textwidth}
			\centering
			\includegraphics[width=\columnwidth, trim=0 0 0 0, clip]{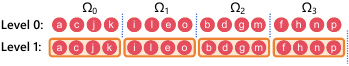}
		\end{subfigure}
		\caption{\label{fig:mas_level_new}\textbf{Our connectivity-enhanced MAS aggregation of \autoref{fig:mas_morton}.} The orange blocks represent super nodes at the coarser levels. At level 0, our method groups nodes based on mesh connectivity, and so each group becomes 1 super node at level 1, completing the aggregation.}
	\end{figure}
	
	Since the aggregation is based on node connectivity, a natural improvement is to group the nodes directly by their connectivity at level 0, ensuring that the 4 nodes in each subdomain are fully connected and can be merged into a single super node, as shown in \autoref{fig:mas_level_new}. In this way, at level 1, there will be only 4 super nodes that cover all the original nodes in the mesh, allowing us to terminate the aggregation with just one subdomain. As a result, this method constructs an efficient two-level MAS that can be used to effectively approximate the global Hessian. %In contrast, using Morton codes for grouping at level 0 produces suboptimal results, leading to slower convergence and degraded overall performance.

	\subsection{Connectivity-Enhanced MAS Construction}\label{sec:geometry_mas}
	The example above demonstrates the importance of considering mesh connectivity during node reordering for optimal node aggregation. However, in practice, achieving perfect aggregation is both expensive and challenging. Therefore, we propose a practical connectivity-enhanced strategy that efficiently approximates optimal aggregation. Our core idea is to use METIS~\cite{METIS} to partition the mesh nodes before the simulation and then reorder them with padding at level 0, optimizing node connectivity within each subdomain.
	
\begin{figure}[htbp]
		\centering
		\begin{subfigure}[b]{0.20\textwidth}
			\centering
			\includegraphics[width=\columnwidth, trim=0 0 0 0, clip]{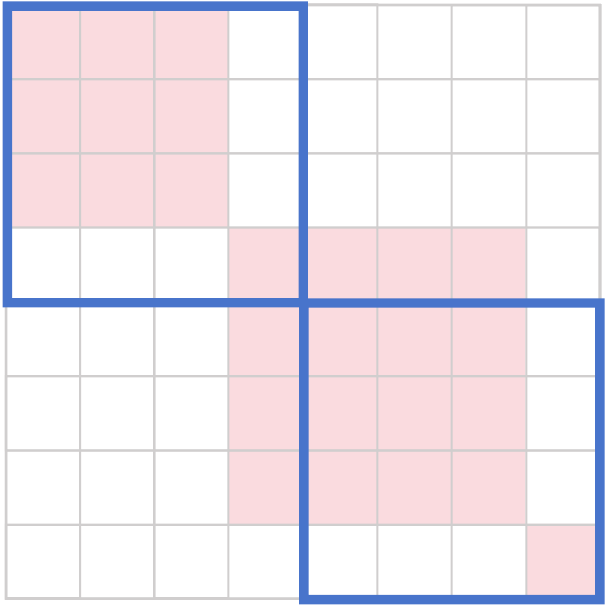}
			\subcaption{\label{fig:metis_old}}
		\end{subfigure}
		\begin{subfigure}[b]{0.20\textwidth}
			\centering
			\includegraphics[width=\columnwidth, trim=0 0 0 0, clip]{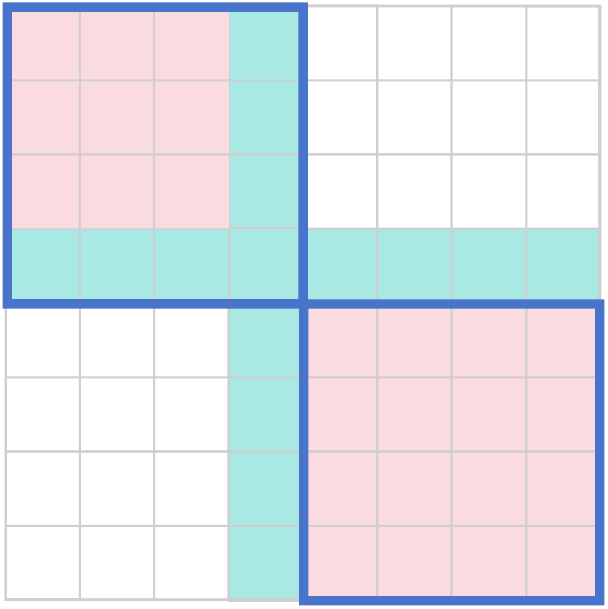}
			\subcaption{\label{fig:metis_new}}
		\end{subfigure}
		\caption{\textbf{Aligning METIS partitions and MAS subdomains.} The red blocks represent subdomains from the METIS partition, the deep blue rectangle shows the MAS subdomain, and the green-blue blocks represent inserted zero entries as padding. (a) An example where the METIS partition sizes are 3 and 4, while the MAS subdomain size is 4, leading to isolated nodes within subdomains and potentially suboptimal aggregation (\autoref{sec:mas_limit}). (b) Our solution inserts zero entries to prevent isolated nodes caused by misalignment between the METIS partition and MAS subdomains.}
		\label{fig:align_metis_mas}
	\end{figure}

	METIS is a fast graph partitioning method that can divide mesh nodes into a specific number of partitions, minimizing inter-partition connectivity and often leading to dense connectivity within each partition. However, METIS does not guarantee that each partition will have exactly the same number of nodes, as achieving perfectly balanced partitions is expensive and sometimes impossible. Handling blocks of varying sizes on the GPU can degrade performance and complicate implementation.
	Thus, we can treat METIS partitioning as a reordering step, and then distribute all nodes evenly into each MAS subdomain. However, this can still result in suboptimal aggregation since the nodes in the same subdomain may come from different METIS partitions as shown in \autoref{fig:metis_old}.
 
% due to isolated nodes within the subdomains as shown in \autoref{fig:metis_old}.
	
	To address this issue and ensure effective aggregation, we propose an index mapping scheme to assign each partition to a separate MAS subdomain and pad the empty entries as illustrated in \autoref{fig:metis_new}.
% We first conduct METIS partition such that the number of nodes in each partition matches the size of the MAS subdomain and simply pad in case below the size (see \autoref{fig:metis_new}).
 %pad each partition to reach $N$ nodes and propose an index mapping scheme to assign each partition to a separate MAS subdomain, as illustrated in \autoref{fig:metis_new}.
	We first set the number of METIS partitions $M$ as
	\begin{align}\label{eq:num_subdomains}
		M &= \left\lceil \frac{V}{N - N_o} \right\rceil,
	\end{align}
	where $N$ is the number of nodes in each MAS subdomain, $V$ is the total number of mesh nodes, $\left\lceil \cdot \right\rceil$ is the ceiling operator, and $N_o$ is a slack variable that ensures the node count in each METIS partition does not exceed $N$. During the precomputation stage, we search for the smallest possible $N_o$, starting from $0$. If any partition exceeds $N$ nodes, we increment $N_o$ and repartition. In almost all cases, $N_o = 1$ is sufficient.

\begin{figure}[htbp]
		\centering
		\begin{subfigure}[b]{0.465\textwidth}
			\centering
			\includegraphics[width=\columnwidth, trim=0 0 0 0, clip]{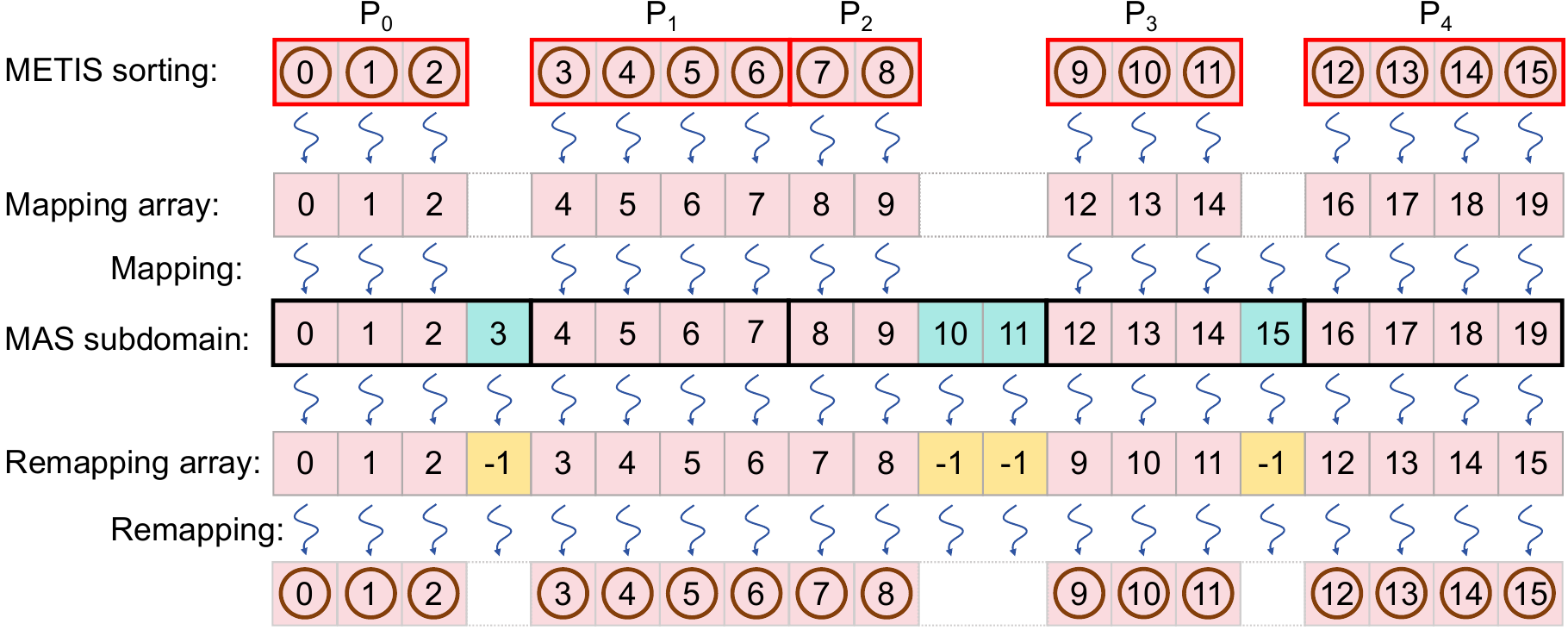}
		\end{subfigure}
		\caption{\label{fig:mapping} \textbf{Our mapping and remapping scheme.} Nodes in each METIS partition are mapped to a separate MAS subdomain using a mapping array (stored in contiguous memory, aligned here for clarity) with padding shown in blue. Our remapping array maps MAS subdomain nodes to the corresponding mesh nodes for easy access to geometric data.}
	\end{figure}

	To implement this approach, we first construct a mapping array of size $N \cdot M$, which maps each node to a slot in its corresponding MAS subdomain, as shown in \autoref{fig:mapping}. This mapping array is then used to build the MAS preconditioner. To ensure correct data access during construction, we also create a remapping array, which helps GPU threads access the correct node data for each entry in the mapping array. Padded entries, where no nodes are mapped, are marked with the value $-1$ (see \autoref{fig:mapping}). 
	The remaining MAS levels are constructed using the original method based on node connectivity. \modify{Considering the connectivity introduced by the current contact pairs in each Newton iteration}, our approach reduces connectivity issues and enables more effective aggregation, improving the convergence speed of the MAS-preconditioned linear solver by up to 2.4$\times$. \modify{Note that the METIS partitioning is not updated during the simulation, as we observed that it does not provide further speedup.}

	Another significant advantage of our method is the ability to use smaller subdomain sizes, resulting in substantial performance gains when computing the inverse of diagonal blocks and applying the preconditioner. In our implementation, we use a subdomain size of 16, compared to the size 32 used 
 in \cite{mas}. Despite the smaller size, our method remains effective as node connectivity within each subdomain is optimized. This enables cost-effective preconditioning while maintaining a high convergence speed, achieving up to 3$\times$ speedup in the PCG solve.

	\subsection{GPU Warp Reduction for Connectivity-Enhanced MAS}\label{sec:gpu_geometry_mas}
	Our connectivity-enhanced MAS construction ensures that the nodes within each subdomain at level 0 are densely connected. This property opens up new opportunities for additional GPU optimizations in efficiently computing the proxy Hessian matrix at coarser MAS levels.

\begin{algorithm}
		\caption{Compute MAS preconditioner}
		\begin{algorithmic}[1]
			\Statex \textbf{Input:} Global Hessian matrix $\mathbf{AG}$ \Comment{{\color[rgb]{0,0.61,0.33} an array of tuples: [block row index, block column index, $3\times3$ block]}}
			
			\Statex \textbf{Output:} MAS preconditioner $\mathbf{M}_{MAS}^{-1}$
			%\Statex $\mathbf{A}_M(l) \leftarrow \mathbf{0}$
			
			%\Function{Construct level 0 sub-matrix}{$\mathbf{A}$}
			\vspace{0.1cm}
			\Statex{\textbf{Method:}}
			\For{ $GlobalThreadId$ $=$ $0$, $1$, ..., $len(\mathbf{AG})-1$} \textbf{in parallel}
			% \Statex {\color[rgb]{0,0.61,0.33}// load each $3\times3$ sub-matrix with indices information}
			\State $[r,c, \mathbf{M3}] = \mathbf{AG}(GlobalThreadId)$ \Comment{{\color[rgb]{0,0.61,0.33} get $3\times 3$ block}}
			% \Statex {\color[rgb]{0,0.61,0.33} // map the node indices to MAS subdomain}
                \State $[domain\_Idx] = \mathbf{Map}(r)$
			\State $[domain\_Idy] = \mathbf{Map}(c)$ \Comment{{\color[rgb]{0,0.61,0.33} align METIS and MAS subdomain with padding (\autoref{fig:align_metis_mas})}}
			\State $m = domain\_Idy/N$ \Comment{{\color[rgb]{0,0.61,0.33}get MAS subdomain id }}
			\State $l = 0$ \Comment{{\color[rgb]{0,0.61,0.33}MAS level, starting from 0}}
			% \Statex {\color[rgb]{0,0.61,0.33} // check if the mapped indices belong to same MAS subdomain}
			\If{$domain\_Idx/N == domain\_Idy/N$} \Comment{{\color[rgb]{0,0.61,0.33} if nodes $r$ and $c$ belong to the same level-0 MAS subdomain, write $\mathbf{M3}$ to the level-0 MAS matrix}}
			%\State $i_m = i\%N$ \quad $j_m = j\%N$
			%\State $j_m = j\%N$
			% \Statex {\color[rgb]{0,0.61,0.33}// write $\mathbf{M3}$ to corresponding MAS block matrix in level 0}
			\State $\mathbf{M}_{0, MAS}(m)[domain\_Idx\%N, domain\_Idy\%N] = \mathbf{M3}$
			\Else \Comment{{\color[rgb]{0,0.61,0.33} add $\mathbf{M3}$ to higher level MAS matrix if needed}}
			%\Statex // map the node indices to MAS subdomain shown in \autoref{fig:mas_level_new}
			%\For{(l++) < levels - 1}
			%\State $i = l==0?go\_next(I):go\_next(i)$
			%\State $j = l==0?go\_next(J):go\_next(j)$
			%\State $atomicAdd(\mathbf{M}_l(j/N)[i\%N, j\%N], \mathbf{M3})$
			%\EndFor
			\For{$l = 1, 2, ..., levels - 1$}
			
			%\State $I = go\_next(I)$ \quad $J = go\_next(J)$
			%\State 
			%\State $m_{IJ} = J/N$
			%\State $I_m = I\%N$
			%\State $J_m = J\%N$
			\If{$r$ and $c$ in the same subdomain at level $l$}
			\State AtomicAdd $\mathbf{M3}$ to $\mathbf{M}_{l, MAS}$
			\EndIf
			\EndFor
			%\State atomicAdd $\mathbf{M3}$ to corresponding $\mathbf{BM}$ in higher levels $l$
			\EndIf
			\EndFor
			%\EndFunction
			
			%\Function{propogate level 0 sub-matrix}{$\mathbf{M}$}
			\For{$BId$ $=$ $0$, $1$, ..., $len(\mathbf{M}_{0, MAS})-1$} \textbf{in parallel} \Comment{{\color[rgb]{0,0.61,0.33} accumulate all $3 \times 3$ blocks in $\mathbf{M}_{0, MAS}$ by warp reduction and add it to all higher-level MAS matrices}}
			\State $\mathbf{M3} = {warpReduction}(\mathbf{M}_{0, MAS}(BId))$ \Comment{{\color[rgb]{0,0.61,0.33} the warpReduction here utilizes the method of \citet{gpumpm}}}

			\For{$l=1,2,...,levels-1$}
			\State AtomicAdd $\mathbf{M3}$ to $\mathbf{M}_{l, MAS}$
			\EndFor
			\EndFor
			%\EndFunction
			\State return $\mathbf{M}_{MAS}^{-1}$ \Comment{{\color[rgb]{0,0.61,0.33} invert unpadded parts}}
		\end{algorithmic}\label{alg:Matrix_construction}
	\end{algorithm}

	In GPU MAS~\cite{mas}, since nodes in each subdomain can be mapped to an arbitrary number of super nodes at the next coarser level, it is nontrivial to design an efficient reduction scheme for accumulating the matrix entries to assemble the proxy matrix at the next level.
	Instead, they construct a mapping array to establish connections between mesh nodes and super nodes at different levels, and then apply atomic operations to compute the proxy matrices at higher levels. Take \autoref{fig:mas_level_old} as an example, node $\{f\}$ in level 0 maps to $\{f, h\}$ in level 1 and 2, and then to $\{f, h, n, p\}$ in level 3.
	The data associated with $\{f\}$ will be added to the super nodes at coarser levels via atomic operations. However, as we see, as the structure becomes coarser, more atomic additions of a larger number of matrix entries are needed, which will result in more writing conflicts and increase computational cost, especially when using double-precision floating-point numbers.
	
	In our case, the nodes within the same level-0 subdomain map to a small number (often the same) super nodes in coarser levels. This allows for the application of a warp reduction method to accumulate the level-0 matrix blocks. The accumulated matrix is then used to construct the sub-matrix of the next coarser levels using atomic operations. This approach effectively reduces the cost of MAS matrix precomputation, since significantly less atomic operations are needed as the level-1 matrix can be efficiently computed using warp reduction. This also allows for more levels in the MAS preconditioner with lower computational cost. See Algorithm \ref{alg:Matrix_construction} and our supplemental document for more details. \modify{For meshes with fewer than 16 nodes, our method may group nodes from different meshes to minimize the number of subdomains. In such special scenarios, we revert to the original MAS assembly method in \citet{mas}.}

	\section{Cubic {Inexact} Strain-Limiting energy with Analytic eigensystem}
	\label{sec:membrane_energy}
	We develop our cubic \modify{inexact} strain-limiting energy based on the FEM Baraff-Witkin (FBW) constitutive model \cite{kim_baraf_witkin}. Inspired by \citet{cipc} which augments soft membrane energy with a barrier-based strain-limiting term to avoid membrane locking, we aim to develop a barrier-free \modify{inexact} strain-limiting energy with analytic eigensystems so that expensive backtracking-based line search \textit{filtering} for the strain limit and numerical eigendecomposition are both avoided. We will first introduce FBW and our model (\autoref{sec:memb_form}), and then derive the analytic eigensystem of our Hessian matrix (\autoref{sec:memb_eigen}).
	
	\subsection{Formulation}\label{sec:memb_form}
	In the FBW method \cite{kim_baraf_witkin}, the membrane energy is defined as the sum of a stretching term and a shearing term:
	\begin{equation}\label{eq:psi_shell}
	\Psi_{memb} = \Psi_{stretch}+\Psi_{shear}.
	\end{equation}
	The stretching term $\Psi_{stretch}$ is defined using two orthonormal basis, $\mathbf{n}_u = [1, 0]^T$ and $\mathbf{n}_v = [0, 1]^T$, and based on $I_5(\mathbf{F}, \mathbf{n}) = \mathbf{n}^T\mathbf{F}^T\mathbf{F}\mathbf{n}$ \cite{anisotropic_eigen}, where $\mathbf{F} = [\mathbf{F}_0|\mathbf{F}_1] \in \realNum^{3x2}$ is the deformation gradient:
	\begin{equation}\label{eq:FBW}
	\Psi_{stretch} = \lambda a_t(\sqrt{I_5(\mathbf{F}, \mathbf{n}_u)}-1)^2+\lambda a_t(\sqrt{I_5(\mathbf{F}, \mathbf{n}_v)}-1)^2.
	\end{equation}
	Here, $a_t$ is the volume weight, i.e. the product of triangle area and the thickness, and $\lambda$ is the stretching stiffness. In this paper, we simply use the same $\lambda$ for both $\mathbf{n}_u$ and $\mathbf{n}_v$ directions. But note that different stiffnesses can be applied to model more anisotropic behaviors, which will not affect the computation and eigenanalysis.

	We follow \citet{kim_baraf_witkin} and use $\sqrt{I_5(\mathbf{F}, \mathbf{n})}$ to define our \modify{inexact} strain-limiting energy, as it effectively represents the stretch factor in the direction $\mathbf{n}$. Directly incorporating $I_5$ into the barrier-based strain-limiting model of \citet{cipc} enables reusing the eigenanalysis from \citet{kim_baraf_witkin}. 
    \modify{However, backtracking-based line search \textit{filtering} is still required since no analytic bound for feasible step sizes is available for the strain. This method is inefficient for parallelization because its iterative halving process, applied to each triangle, requires the global line search iterations to continue until all triangles satisfy the strain limit.} Additionally, because the barrier function diverges rapidly near the strain limit, solver performance becomes highly sensitive to the barrier stiffness, complicating parameter setting.
	To address this, we apply a cubic penalty function to \modify{inexactly} handle the inequality constraint imposed by the strain limit. This approach allows for slight constraint violations while providing stretching resistance on the same scale as practical settings.

	To simplify the explanation, let's consider the $\mathbf{n}_u$ direction in \autoref{eq:FBW} as an example. We construct a $C^2$-continuous clamped cubic \modify{inexact} strain-limiting energy using $I_5$:
	\begin{equation}\label{eq:memb_u}
	% \Psi_{SL}^u = \lambda' a_t S(I_5(\mathbf{F}, \mathbf{n}_u))(\sqrt{I_5(\mathbf{F}, \mathbf{n}_u)}-1)^3,
    \Psi_{SL}^u = \lambda' a_t \max(\sqrt{I_5(\mathbf{F}, \mathbf{n}_u)}-1, 0)^3,
	\end{equation}
	where $\lambda'$ is the stiffness,
 %    and $S$ is a piecewise function defined as	\begin{equation}\label{eq:S_u}
	% S(s) =
	% \begin{cases}  
	% 1, &s > 1,  \\  
	% 0, &0 \le s \le 1,
	% \end{cases}
	% \end{equation}
	and the $\max$ operator ensures that the \modify{inexact} strain-limiting force is only applied when the cloth is stretched. \modify{This idea aligns with the tension field theory \cite{Tension-field, inflatable,chen2023complex}, which deactivates stiffness under compression to obtain a convex energy. Additionally, \citet{kikuuwe2009edge} activates a cubic term under compression to mitigate the element inversion issue.} 
    
    % \modify{Equivalently, \autoref{eq:memb_u} can also be written as $$\Psi_{SL}^u = \lambda' a_t \max(\sqrt{I_5(\mathbf{F}, \mathbf{n}_u)}-1, 0)^3.$$} 
    By similarly defining the cubic term in the $\mathbf{n}_v$ direction, we obtain our complete cubic \modify{inexact} strain-limiting energy 
	$$\Psi_{SL} = \Psi_{SL}^u + \Psi_{SL}^v.$$ 
	This energy is well-defined even if the strain limit is exceeded, eliminating the need for line search \textit{filtering}. Given that the Young's modulus of real fabrics ranges from $1MPa$ to $10MPa$ \cite{penava2014determination} 
    when the cloth is stretched, we set our $\lambda' = 5MPa$ \modify{by default}, as it provides stretching resistance at the same scale when the cloth approaches the strain limit, e.g., $5\%$. Thus, our model prevents over-elongation in practical cloth manipulation scenarios. 
    \begin{wrapfigure}{r}{0.48\linewidth}
		\centering
		\includegraphics[width=0.48\columnwidth, trim=0 0 0 0, clip]{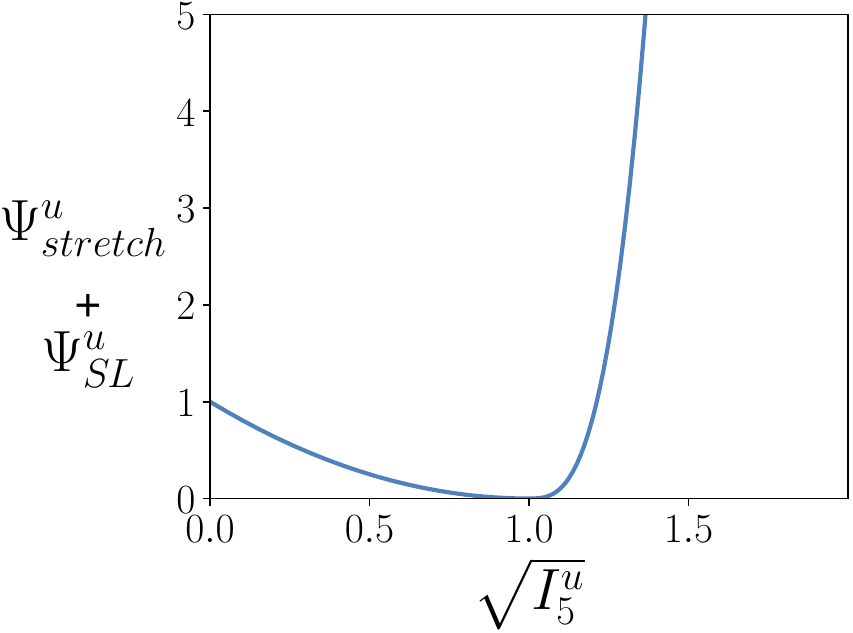}
		\caption{\label{fig:mem_func}\textbf{Our composed strain energy in the $\mathbf{n}_u$ direction.} Combined with the membrane energy from \citet{kim_baraf_witkin}, our cubic \modify{inexact} strain-limiting energy enables using soft compression resistance and stiff stretching resistance to avoid membrane locking and over-elongation issues. Here, $\lambda a_t = 1$ and $\lambda^{'}=100\lambda$ are used for visualization.}
	\end{wrapfigure} 
    For $\Psi_{stretch}$, we can now safely use a small stiffness ($\lambda = 0.05MPa$ in all our examples) to avoid membrane locking. Refer to \autoref{fig:mem_func} for a plot of our composed membrane energy in the $\mathbf{n}_u$ direction. Since the stiffness for the shearing term $\Psi_{shear}$ primarily depends on fabric type and, based on our observations, has minimal impact on the over-elongation or membrane locking issues, we simply set it to 30\% of the stretching stiffness and apply the original eigenanalysis from \citet{kim_baraf_witkin} to compute its PSD Hessian.

	% Since the shearing term $\Psi_{shear}$ mainly resist elongation in the diagonal direction, we directly use FBW with real stiffness parameters ($5MPa$\todo{confirm} in all our examples), and the original eigenanalysis from \citet{kim_baraf_witkin} remains applicable.

	\subsection{Eigenanalysis}\label{sec:memb_eigen}
	Following \citet{anisotropic_eigen, kim_baraf_witkin}, we can analytically derive the eigensystem of our $I_5$-based \modify{inexact} strain-limiting energy's Hessian matrix with respect to the deformation gradient $\mathbf{F}$. For simplicity, we will omit $\lambda'$ and $a_t$ in the following discussion.
	
	The eigenvalues of $I_5$-based energies are:
	\begin{align}\label{eq:lambda}
		e_1(I_5) &= 4I_5\frac{\partial^2 \Psi}{\partial {I_5}^2} + 2\frac{\partial \Psi}{\partial {I_5}}, \nonumber\\ 
		e_{2,3}(I_5) &= 2\frac{\partial \Psi}{\partial {I_5}}.
	\end{align} 
	Since $\mathbf{n}_u$ and $\mathbf{n}_v$ are orthogonal, we can substitute $I_5(\mathbf{F}, \mathbf{n}_u)$ and $I_5(\mathbf{F}, \mathbf{n}_v)$ into \autoref{eq:lambda} separately to obtain all six eigenvalues for our $\Psi_{SL}$ when the cloth is stretched ($I_5 > 1$):
	\begin{align}\label{eq:e1}
		e_1(I_5(\mathbf{F}, \mathbf{n}_u)) = 6(\sqrt{I_5(\mathbf{F}, \mathbf{n}_u)} - 1),
	\end{align}
	\begin{align}\label{eq:e23_0}
		e_{2,3}(I_5(\mathbf{F}, \mathbf{n}_u)) &= 3\left(\frac{1}{\sqrt{I_5(\mathbf{F}, \mathbf{n}_u)}} + \sqrt{I_5(\mathbf{F}, \mathbf{n}_u)} - 2\right),
	\end{align}
	\begin{align}\label{eq:e4}
		e_4(I_5(\mathbf{F}, \mathbf{n}_v)) = 6(\sqrt{I_5(\mathbf{F}, \mathbf{n}_v)} - 1),
	\end{align}
	\begin{align}\label{eq:e56_0}
		e_{5,6}(I_5(\mathbf{F}, \mathbf{n}_v)) &= 3\left(\frac{1}{\sqrt{I_5(\mathbf{F}, \mathbf{n}_v)}} + \sqrt{I_5(\mathbf{F}, \mathbf{n}_v)} - 2\right).
	\end{align}
	From these equations, we see that all eigenvalues are positive when $I_5 > 1$, indicating that our cubic \modify{inexact} strain-limiting energy $\Psi_{SL}$ is convex with respect to $\mathbf{F}$ (recall that $\Psi_{SL} = 0$ when $I_5 \leq 1$). Since $\mathbf{F}$ is a linear function of the degrees of freedom $x$, $\Psi_{SL}$ is also convex with respect to $x$.
	
	Thus, when the the cloth is stretched, we can directly compute the PSD Hessian matrix of $\Psi_{SL}$ in the $\mathbf{n}_u$ and $\mathbf{n}_v$ directions separately, applying the chain rule with $I_5$ as the intermediate variable:
	\begin{align}\label{eq:Stretch_Hessian}
		\mathbf{H}_{SL,*} &= 3 \left(1 - \frac{1}{\sqrt{I_5(\mathbf{F}, \mathbf{n}_*)}}\right)(\sqrt{I_5(\mathbf{F}, \mathbf{n}_*)} - 1)\mathbf{H}_* 
		\nonumber\\ &\, \, 
		+ \frac{3(I_5(\mathbf{F}, \mathbf{n}_*) - 1)}{I_5(\mathbf{F}, \mathbf{n}_*)^{3/2}}(\mathbf{f}_*\mathbf{f}_*^T).
	\end{align}
	Here, $\mathbf{f}_* = \text{vec}(\mathbf{F}\mathbf{L}_*)$, $\mathbf{L}_* = \mathbf{n}_*(\mathbf{n}_*)^T$, $\mathbf{H}_* = \mathbf{L}_* \otimes \mathbf{1}_{3\times3}$, $\otimes$ denotes the tensor product, $\mathbf{I}_{3\times3}$ is the $3\times 3$ identity matrix, and $\text{vec}(\cdot)$ is the vectorization operator~\cite{defomDynamic_kim}, with $*$ representing either $u$ or $v$.
	All of these computations can be efficiently parallelized on the GPU, unlike numerical eigendecomposition.

	\section{Fast global hessian operations on the GPU}
	\label{sec:global_hessian}
	
	In this section, we introduce our \modify{two-level} reduction strategy for fast Hessian matrix assembly in {affine-deformable} coupling simulations (\autoref{sec:multi_layer_reduction}). \modify{Specifically,} using a custom hash-based parallel reduction algorithm (\autoref{sec:hash_based_reduction}), our strategy significantly reduces the number of numerical operations and write conflicts, and it enables the development of a memory-efficient symmetric sparse blockwise matrix-vector multiplication method (\autoref{sec:spmv}) to further boost PCG performance.

	\subsection{Two-Level Reduction for ABD Hessian Assembly}\label{sec:multi_layer_reduction}

	As previously mentioned, in affine body simulation, the main bottleneck lies in transforming and accumulating the contact Hessian matrices (collision and friction) from mesh node DOFs to the affine body DOFs within the global Hessian matrix. For contact pairs involving nodes belonging to affine bodies, consider the example of contact between two affine bodies. In this case, each $3\times3$ block $\mathbf{C}(r,c)$ in the contact Hessian, where $r$ and $c$ are node indices, must be mapped to a $12\times12$ matrix for the affine body pair $\{AF(r), AF(c)\}$ via multiplication with the Jacobian matrices $\{\mathbf{J}_r, \mathbf{J}_c\}$. All resulting $12\times12$ Hessian matrices for each affine body pair are then accumulated \cite{abd,ubarrier}. 

    \begin{wrapfigure}{r}{0.46\linewidth}
		\centering
		\includegraphics[width=\linewidth, trim=0 0 0 0, clip]{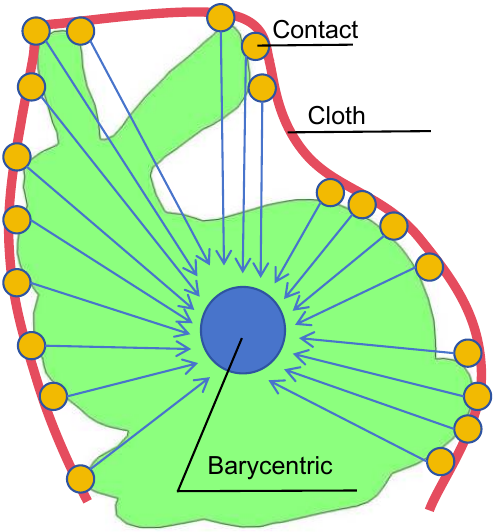}
		\caption{\label{fig:rigidBunny}Example of accumulating contact information for reduced affine bodies.}
	\end{wrapfigure}
	However, the number of mapped Hessians typically far exceeds the number of affine body pairs, \modify{which becomes even more pronounced in the affine-deformable coupling cases} (see \autoref{fig:rigidBunny}), leading to significant accumulation overhead, 
    let alone the frequent write conflicts.
	To address this issue, we first accumulate all $3\times3$ contact Hessian blocks with the same node index pair $\{r,c\}$ through reduction, then apply a single product with $\{\mathbf{J}_r, \mathbf{J}_c\}$ to transform it into a $12\times12$ matrix. This process yields a sequence of $12\times12$ matrices for the blocks of the global Hessian corresponding to the affine body pair $\{AF(r), AF(c)\}$. A second layer of reduction is then used to accumulate these $12\times12$ Hessians in parallel, significantly reducing write conflicts and the number of values to be accumulated, resulting in faster performance.

	\subsection{Hash-based Parallel Reduction}\label{sec:hash_based_reduction}
	
	To further enhance accumulation efficiency, we developed a parallel reduction technique using a hashing method based on either the node index pair $\{r,c\}$ or the affine body index pair $\{AF(r), AF(c)\}$. Each matrix block is encoded with a 64-bit hash value, where the first index forms the higher 32 bits and the second forms the lower 32 bits. Sorting the hash values enables consecutive memory placement of matrix blocks sharing the same hash \cite{hkm_sharedMemory, huang_hash}, facilitating efficient parallel accumulation. The key insight here is to build a custom reduction scheme that is based on registers rather than shared memory and avoid global synchronization by taking advantage of warp-level operations \cite{gwarpRed, gpumpm, gipc}.
	
	Our algorithm is constructed using basic parallel primitives:
	\begin{enumerate}
		\item Device run-length encoding
		\item Device exclusive sum
		\item Warp segmented reduction
	\end{enumerate}
	The first two primitives are standard in parallel computing. For warp-level segmented reduction, we use the \modify{CUB\footnote{\href{https://nvidia.github.io/cccl/cub/}{https://nvidia.github.io/cccl/cub/}}} intrinsic function \emph{\textbf{HeadSegmentedReduce(values, tags)}}. 
	Consider a simple example with a warp size of 8: if $\emph{\textbf{values}} = [1, 1, 1, 1, 1, 1, 1, 1]$ and $\emph{\textbf{tags}} = [1, 0, 0, 1, 0, 0, 1, 0]$, the output will be $\emph{\textbf{outPut}} = [3, \bullet, \bullet, 3, \bullet, \bullet, 2, \bullet]$, where $ \bullet $ indicates invalid entries. Here, the \emph{\textbf{tags}} array specifies start and end positions for reduction in \emph{\textbf{values}}, with 1 marking the start. Reduction results (valid entries in \emph{\textbf{outPut}}) are then written to the result array.
	The overall procedure, \textit{FastHashReduction}, is detailed in Algorithm~\ref{alg:fast_hash_reduction}. First, the map array $O$ is constructed in parallel, mapping duplicate keys to their unique key indices to ensure reduction results are written to the correct locations (lines 2–6). The \textit{FastSegmentReduction} algorithm (Algorithm~\ref{alg:fast_segment_reduction}) then efficiently computes the matrix.

	%\begin{algorithm}
	%\caption{Warp Segmented Reduction Definition}\label{alg:warp_segmented_reduction_definition}
	%\begin{algorithmic}[1]
	%\Statex \textbf{Input:} 
	%\Statex $F$ \Comment{A bit field, a set bit tells the start of a new segment}
	%\Statex $V$ \Comment{The value to be reduced}
	%\Statex \textbf{Output:} 
	%\Statex $I$  \Comment{Reduction result}
	%\State $I \gets$ \Call{\textbf{WarpSegmentedReduction}}{$F$,$V$}
	%\end{algorithmic}
	%\end{algorithm}
	
	%In Algorithm \autoref{alg:fast_hash_reduction}, two arrays are used as input: the first array contains the hash keys, and the second array holds the corresponding values. The key-value pairs are already sorted by the hash keys. The algorithm consist of two parts: the first part distributes the unique hash key offsets to every key in the input array, indicating the start of a segment. The second part is a segmented reduction using the distributed offsets and the values.
	
	\begin{algorithm}
		\caption{FastHashReduction}\label{alg:fast_hash_reduction}
		\begin{algorithmic}[1]
			\Statex \textbf{Input:}
			\Statex \hspace{0.4cm} $K$ \Comment{{\color[rgb]{0,0.61,0.33}an array of sorted hash keys}}
			\Statex \hspace{0.4cm} $V$ \Comment{{\color[rgb]{0,0.61,0.33}an array of values, sorted by the hash keys}}
			\Statex \textbf{Output:}
			\Statex \hspace{0.4cm} $\mathbf{AG}$ \Comment{{\color[rgb]{0,0.61,0.33}global Hessian matrix}}
			%\Statex
			% \Statex {\color[rgb]{0,0.61,0.33}// $UK$ is unique key, $NK$ is the count of keys with the same value}
			
			\vspace{0.1cm}
			\Statex{\textbf{Method:}}
			\State $UK \gets$ \Call{RunLengthEncode}{$K$} \Comment{{\color[rgb]{0,0.61,0.33} get unique keys ($UK$), which are used to recover the row/column indices in the global Hessian}}
			% \Statex {\color[rgb]{0,0.61,0.33}// get the length of unique key array}
			\State $Num \gets$ \Call{len}{K}
			% \Statex {\color[rgb]{0,0.61,0.33}// get prefix sum , start from 0}
			%\State $ESum \gets$ \Call{ExclusiveSum}{$NK$} \Comment{{\color[rgb]{0,0.61,0.33} get prefix sum, starting from 0}}
			% \Statex {\color[rgb]{0,0.61,0.33}// init $P$ array with 0}
			\Statex {\color[rgb]{0,0.61,0.33}// assign the unique key index to $O$ for each key in $K$:}
			%\State $P \gets $ \Call{Zeros}{len(K)}
			\For {$tid = 0, 1, ..., Num-2$} \textbf{in parallel} 
			%\State $P[ESum[tid] + NK[tid] - 1] \gets 1$ 
                \State $P[i] \gets K[i]\neq K[i+1]$ 
			\EndFor
			
			\State $O \gets$ \Call{ExclusiveSum}{$P$} \Comment{{\color[rgb]{0,0.61,0.33}map each $K$ to unique key}}
			
			\State $R \gets$ \Call{FastSegmentReduction}{$O$,$V$}
                \State $\mathbf{AG} \gets$ \Call{ConstructGlobalHessian}{$UK$, $R$}
		\end{algorithmic}
	\end{algorithm}
	
	\paragraph{FastHashReduction}
	Our FastHashReduction algorithm is designed to handle various value types. For ABD Hessian matrices, each value is a $12 \times 12$ matrix, while for FEM Hessian matrices, each value is a $3 \times 3$ matrix. In a hybrid FEM-ABD model, there are four types of Hessian matrix blocks: ABD-ABD ($12 \times 12$), FEM-FEM ($3 \times 3$), ABD-FEM ($12 \times 3$), and FEM-ABD ($3 \times 12$).
	To unify these four block matrix types, we decompose them into multiple $3 \times 3$ sub-blocks, resulting in:
	\begin{enumerate}
		\item 16 hash keys and 16 sub-blocks of $3 \times 3$ matrices for ABD-ABD Hessian matrix blocks.
		\item 4 hash keys and 4 sub-blocks of $3 \times 3$ matrices for ABD-FEM and FEM-ABD Hessian matrix blocks.
	\end{enumerate}
	The FEM-FEM Hessian matrix blocks remain unchanged. This approach allows us to launch a single, unified CUDA kernel to reduce all types of local Hessians, maximizing parallelism and minimizing branching. 
	% Consequently, this method facilitates efficient coupling of stiff affine bodies (ABD) with deformable objects (FEM) in a unified GPU framework.

	\begin{figure}[htbp]
		\centering
		\begin{subfigure}[b]{0.45\textwidth}
			\centering
			\includegraphics[width=\columnwidth, trim=0 0 0 0, clip]{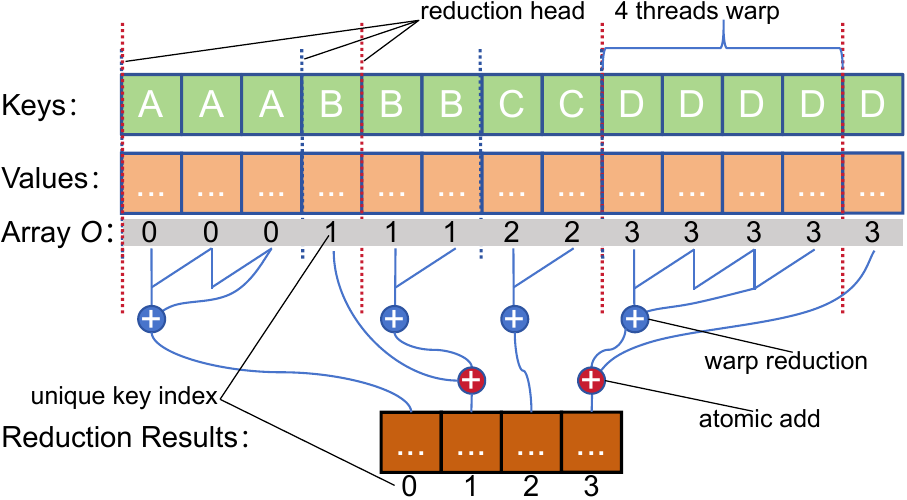}
		\end{subfigure}
		\caption{\label{fig:fsreduction} \textbf{Illustration of FastSegmentReduction.} The warp-level reduction is performed within segments, using Array $O$ to identify segment boundaries. For segments spanning multiple warps (B and D), atomic operations are applied to avoid data conflicts, while direct writing is used otherwise.
		}
	\end{figure}

	\paragraph{FastSegmentReduction}
	In Algorithm \autoref{alg:fast_hash_reduction}, we create a mapping array $O$ to associate each key-value pair with its target memory location, as illustrated in \autoref{fig:fsreduction}. Array $O$ is also used to generate tag information for the CUB intrinsic function \emph{\textbf{HeadSegmentedReduce(values, tags)}}. The \emph{\textbf{tags}} are determined by comparing adjacent entries in $O$, as described in Algorithm \autoref{alg:fast_segment_reduction}.
	{Here, $b^{-}$ and $b$ represent the previous and current unique key indices, respectively. A new segment begins when $b^{-} \neq b$ or when the corresponding thread is the head of its warp. This is indicated using the \emph{\textbf{IsHead}} flag. This flag is set per thread, but only the thread with the flag equal to $1$ is allowed to write to the output array.}

	This hash key reduction method efficiently constructs the global Hessian matrix, critical for the subsequent linear system solve and preconditioner computation. Additionally, our Hessian assembly method ensures that \(3 \times 3\) sub-blocks in each row are sorted in the memory by column index, enabling an efficient SpMV algorithm with warp reduction to enhance PCG solver performance.

	\begin{algorithm}
		\caption{FastSegmentReduction}\label{alg:fast_segment_reduction}
		\begin{algorithmic}[1]
			\Statex \textbf{Input:}
			\Statex \hspace{0.4cm} $O$ \Comment{{\color[rgb]{0,0.61,0.33}The result in line 6 of \autoref{alg:fast_hash_reduction} }}
			\Statex \hspace{0.4cm} $V$ \Comment{{\color[rgb]{0,0.61,0.33}An array of values, sorted by the hash keys}}
			\Statex \textbf{Output:}
			\Statex \hspace{0.4cm} $R$ \Comment{{\color[rgb]{0,0.61,0.33}An array of reduced values, one value for each hash key}}
			%\State $M \gets$ \Call{Len}{V}
			
			\vspace{0.1cm}
			\Statex{\textbf{Method:}}
			\State \textbf{define} BD = BLOCK\_DIM
			%\State \textbf{define} WD = WARP\_DIM
			\State BlockCount $\gets$ (\Call{len}{V} + BD - 1) / BD
			\For {GlobalThreadId $=$ $0$, $1$, ..., BlockCount$\times$BD-1 } \textbf{in parallel}
			
			%\State LocalThreadId $\gets$ GlobalThreadId \% BLOCK\_DIM
			%\State WarpId $\gets$ LocalThreadId / WARP\_DIM
			\Statex \hspace{0.4cm} {\color[rgb]{0,0.61,0.33}// get tags for \emph{\textbf{cub::WarpReduce(values, tags)}}:} 
                \State $b^{-},b \gets -1$ \Comment{{\color[rgb]{0,0.61,0.33}$-1$ for invalid value}}
			\State LaneId $\gets$ GlobalThreadId \% WARP\_DIM
                %\State $b^{-},b,b^{+} \gets -1$
			
			%\State IsValid, IsHead $\gets$ 0
                \State IsHead $\gets$ 0
			\State Value $\gets$ $\mathbf{0}$
			%\If {$0 <$ GlobalThreadId $<$ \Call{len}{O}}
			%\State $b^{-} \gets O[$GlobalThreadId $-1]$
			%\EndIf
			%\If {GlobalThreadId $<$ \Call{len}{V} - 1}
			%\State $b^{+} \gets O[$GlobalThreadId $+1]$
			%\EndIf
			\If {GlobalThreadId $<$ \Call{len}{O}}
			\State $b \gets O[$GlobalThreadId$]$
			\State Value $\gets V[$GlobalThreadId$]$
                    \If {GlobalThreadId $>$ 0}
                        \State $b^{-} \gets O[$GlobalThreadId $-1]$
                    \EndIf
			%\State IsValid $\gets 1$
			\EndIf
			\If {LaneId $== 0$ \textbf{Or} $b^{-} \neq b$}
			%\State IsHead $\gets 1$, IsCrossWarp $\gets b^{-} == b$
                \State IsHead $\gets 1$
			%\Else
			%\State IsHead $\gets b^{-} \neq b$
%			\If {LaneId $=$ WARP\_DIM $- 1$}
			%\State IsCrossWarp $\gets b^{+} == b$
			%\EndIf
			\EndIf
			%\State Tuple $\gets$ (IsValid, IsCrossWarp)
			%\State \modify{IsValid $\gets$ {\emph{\textbf{cub::WarpReduce}}}(IsValid, IsHead)}
			% \Statex {\color[rgb]{0,0.61,0.33}// conduct reduction on the values}
			\State Value $\gets$ {\emph{\textbf{HeadSegmentedReduce}}}(Value, IsHead)
			\Statex \hspace{0.4cm} {\color[rgb]{0,0.61,0.33}// write the result to the corresponding memory location:}
			\If {IsHead \textbf{and} GlobalThreadId $<$ \Call{len}{O}}
			%\If {IsCrossWarp}
			\State \Call{AtomicAdd}{$R[b]$, Value}
			%\Else
			%\State $R[b] \gets$ Value
			%\EndIf
			\EndIf
			\EndFor
		\end{algorithmic}
	\end{algorithm}

	\subsection{Symmetric Reduce-By-Key (SRBK) SpMV}\label{sec:spmv}
	
	SpMV is a frequently called subroutine in scientific computing, where a conjugate gradient solver, for example, may invoke hundreds or thousands of SpMV operations for each linear solve. Thus, speeding up SpMV is critical for improving the overall performance. The NVIDIA cuSparse library's SpMV on Block Sparse Row (BSR) format is the fastest for general sparse matrices, but it lacks support for symmetric matrices. Since the SpMV algorithm is memory-bound, we aim to leverage symmetry to reduce memory access and achieve significant speedup.
	
	We propose a symmetric Reduce-By-Key (SRBK) SpMV method (Algorithm~\autoref{alg:srbk_spmv}) that is approximately $2\times$ faster than cuSparse BSR. Our algorithm works on a format called Sorted Symmetric Block Coordinates, storing only the diagonal and upper triangular blocks of the matrix. All block row and column indices are sorted, a free lunch from our FastHashReduction algorithm for assembling the matrix.
	
	\begin{algorithm}
		\caption{Symmetric Reduce-By-Key (SRBK) SpMV}\label{alg:srbk_spmv}
		\begin{algorithmic}[1]
			\Statex \textbf{Input:}
			% \Statex {\color[rgb]{0,0.61,0.33}// upper triangle block matrix/row index/column index array}
			\Statex \hspace{0.4cm}  $\mathbf{Rid}$, $\mathbf{Cid}$, $\mathbf{AU}$
			\Comment{{\color[rgb]{0,0.61,0.33} block row index, block column index, and value of upper-triangular and diagonal blocks}}
			\Statex \hspace{0.4cm} $\mathbf{V}_{input}$ \Comment{{\color[rgb]{0,0.61,0.33}SpMV input vector}}
			\Statex \textbf{Output:}
			\Statex \hspace{0.4cm} $\mathbf{V}_{output}$ \Comment{{\color[rgb]{0,0.61,0.33}SpMV output vector}}
			
			\vspace{0.1cm}
			\Statex{\textbf{Method:}}
			\State \textbf{define} BD = BLOCK\_DIM
			\State BlockCount $\gets$ (\Call{len}{$\mathbf{AU}$} + BD - 1) / BD
			\For {GlobalThreadId $=$ $0$, $1$, ..., BlockCount$\times$BD$-1$ } \textbf{in parallel}

            \State $b^{-},b \gets -1$
            \State LaneId $\gets$ GlobalThreadId \% WARP\_DIM
                \State IsHead $\gets$ 0
			\State Value $\gets$ $\mathbf{0}$
			\If {GlobalThreadId $<$ \Call{len}{$\mathbf{AU}$}}
			\State $b \gets \mathbf{Rid}[$GlobalThreadId$]$
			%\State Value $\gets V[$GlobalThreadId$]$

                \State j $\gets \mathbf{Cid}[$GlobalThreadId$]$  
			\State $\mathbf{H} \gets \mathbf{AU}[$GlobalThreadId$]$
			% \Statex {\color[rgb]{0,0.61,0.33}// Handle upper triangle}
			\State Value $\gets \mathbf{H} \cdot \mathbf{V}_{input}[j]$ \Comment{{\color[rgb]{0,0.61,0.33} multiply with upper-triangular and diagonal blocks}}
			% \Statex {\color[rgb]{0,0.61,0.33}// Handle lower triangle}
			\If {$b \neq j$} \Comment{{\color[rgb]{0,0.61,0.33}If not diagonal}}
			\State \Call{AtomicAdd}{$\mathbf{V}_{output}[j]$, $\mathbf{H}^T \cdot \mathbf{V}_{input}[b]$} \Comment{{\color[rgb]{0,0.61,0.33} multiply with lower-triangular blocks}}
			\EndIf
			%\EndIf

                    \If {GlobalThreadId $>$ 0}
                        \State $b^{-} \gets \mathbf{Rid}[$GlobalThreadId$ -1]$
                    \EndIf
			\EndIf
			\If {LaneId $== 0$ \textbf{Or} $b^{-} \neq b$}
                \State IsHead $\gets 1$

			\EndIf
			%\If {GlonalThreadId<len(AU)} \Comment{{\color[rgb]{0,0.61,0.33}GlobalThreadId is within valid data range}}
			
			%\Statex
			%\State IsValid, IsCrossWarp, Value $\gets$ \Comment{Apply reduction as \autoref{alg:fast_segment_reduction} do.}
			%\State Value $\gets$ \Call{WarpSegmentedReduction}{Value, IsHead}
			%\If {IsHead \textbf{and} IsValid}
			%    \If {IsCrossWarp}
			%        \State \Call{AtomicAdd}{$\mathbf{Y}[i]$, Value}
			%    \Else
			%        \State $\mathbf{Y}[i] \gets$ Value
			%    \EndIf
			%\EndIf
			\State Value $\gets$ {\emph{\textbf{HeadSegmentedReduce}}}(Value, IsHead)
			\Statex \hspace{0.4cm} {\color[rgb]{0,0.61,0.33}// write the result to the corresponding memory location:}
			\If {IsHead \textbf{and} GlobalThreadId $<$ \Call{len}{$\mathbf{AU}$}}
			%\If {IsCrossWarp}
			\State \Call{AtomicAdd}{$\mathbf{V}_{output}[b]$, Value}
			%\Else
			%\State $R[b] \gets$ Value
			%\EndIf
			\EndIf 
			\EndFor
		\end{algorithmic}
	\end{algorithm}
	
	Our SRBK SpMV and FastSegmentReduction algorithms have a similar structure, both using a "Reduction By Key" parallel primitive. Sorted row indices serve as hash keys, while the product of the block matrix and the input vector represents the value to be reduced (lines 4-23). The main difference is in how we access the lower triangular part of the matrix, where we atomically add the multiplication results to the output vector (line 14). The data from the lower triangular part is directly transposed from the upper part, reducing global memory access by nearly half, and thus significantly enhancing performance.

	%This algorithm might be slow if segments cover too many warps, but our multi-layer reduction strategy typically limits segment size to less than one warp, making conflicting operations relatively rare.
	
	%\section{Convergence Analysis of Barrier Energy Designs}
	%\label{sec:barrier}
	
	%\todo{computation-wise, will $1/x$ better than $\log^2$?}
	
	% if time allows:
	% wood fracture with anisotropic elasticity?
	% Titanic sinking?
	% nerf and 3d gaussian splatting?
	% robot cutting?
	
	\section{Experiments and Analysis}\label{sec:exp}
	We present our evaluation and results in this section. The experiments were conducted with a 24-core Intel Core i9 13900KF CPU
	with 64GB of RAM, and an NVIDIA RTX 4090 GPU with 24GB of RAM.
	For the simulations, we use modified PCG, with a relative error tolerance of $10^{-4}$, which has shown to be sufficient for IPC \cite{gipc}.
	
	Our experiments include ablation and comparative studies on preconditioners, strain-limiting energies, SpMV, and GPU IPC methods (\autoref{sec:comparisons}), followed by stress tests (\autoref{sec:stress_tests}). The simulation statistics are presented in \autoref{tab:MAS_Preconditioner}, \autoref{tab:memlocking}, and \autoref{tab:Overall_comparison}. 
	Regarding alternative GPU IPC methods, we compared our system against GIPC \cite{gipc}, using the same linear solver and termination criteria as ours, but with hybrid block-diagonal or GPU MAS \cite{mas} preconditioner. Additionally, we compared our system to the GPU-based ABD method~\cite{abd}, where only CHOLMOD compiled with Intel MKL LAPACK and BLAS is running on the CPU. Finally, we conducted a comparison with the state-of-the-art {affine-deformable} coupling IPC framework, ZeMa~\cite{zema}, which also uses the CPU-based CHOLMOD solver, while leveraging GPU acceleration for other components. All simulations were performed using double-precision floating-point arithmetic.
	
	% this is less very relevant here, both in experiments or technical parts
	% To suggest whether a volumetric object should be treated as a stiff affine body or an FEM deformable body, we employ a heuristic based on Young's modulus $Y$, density $\rho$, and time step size $\Delta t$. As we know, the speed of elastic waves equals to the sound speed in a solid, which can be estimated as $v = \sqrt{Y/\rho}$. Now, if the elastic wave travels across the entire object we are simulating in just a single time step, we will barely see the elastic behaviors. This suggests an intuitive heuristic per object $i$ that, if $L_i/\sqrt{Y_i/\rho_i} < dt$, where $L_i$ is the length of the maximum bounding box side of object $i$, then it should be simulated as an affine body. Otherwise, it should be simulated as a deformable body to capture the elastic behaviors.
	% To provide an example, when simulating a 1m-wide cube with $\rho=1000kg/m^3$ and $\Delta t = 0.01s$, it should be treated as an affine body when $Y > 10MPa$.
	% Note that we only apply our heuristic to volumetric objects, and interestingly, it still allows capturing large-scale or high-frequency elastic behaviors when simulating stiff objects with large dimensions and/or tiny time step sizes.
	
	\begin{figure}[htbp]
		\centering
		\begin{subfigure}[b]{0.23\textwidth}
			\centering
			\includegraphics[width=\columnwidth, trim=500 280 400 600, clip]{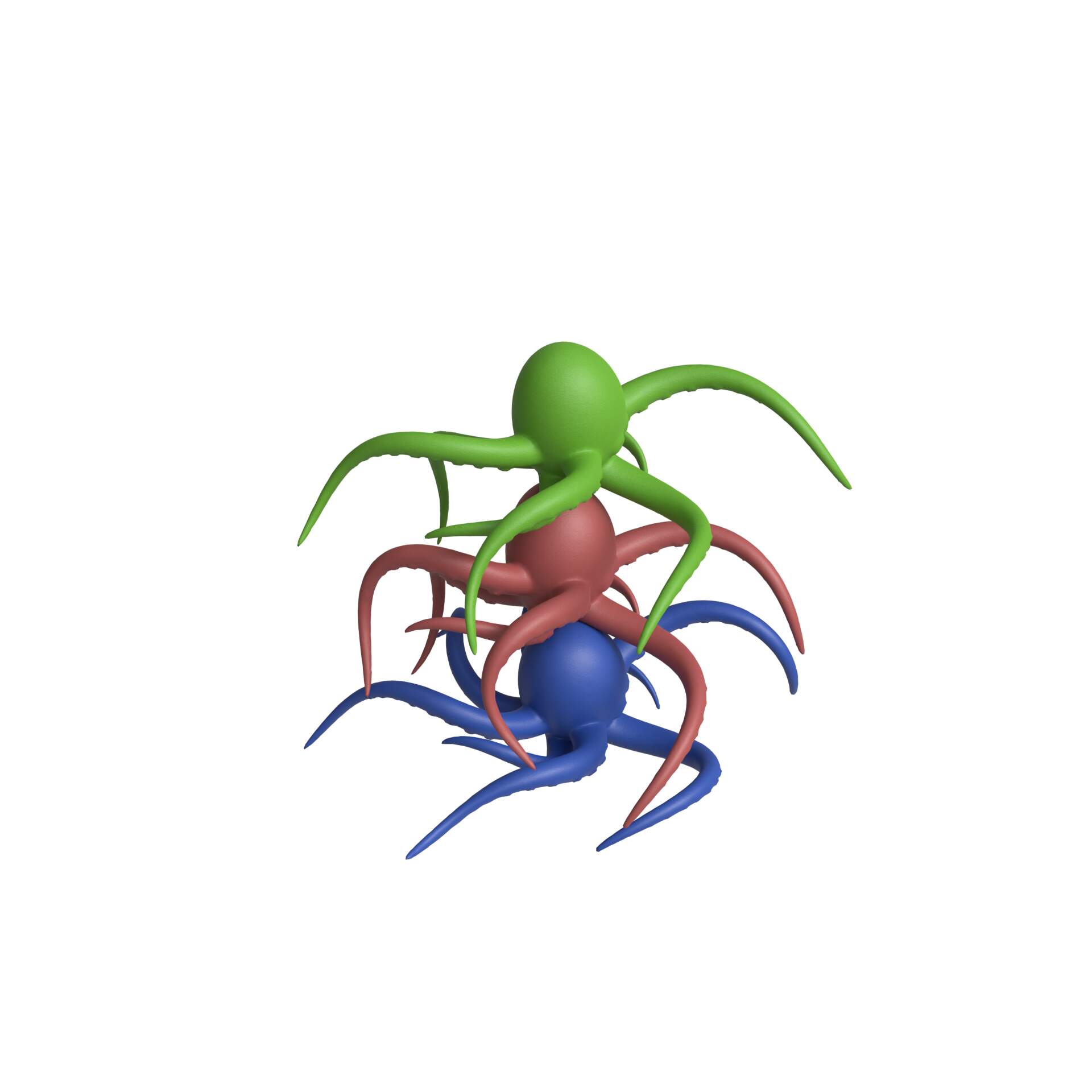}
			%\subcaption{\label{fig:memLocking_5e6}}
		\end{subfigure}
		\begin{subfigure}[b]{0.23\textwidth}
			\centering
			\includegraphics[width=\columnwidth, trim=500 80 400 600, clip]{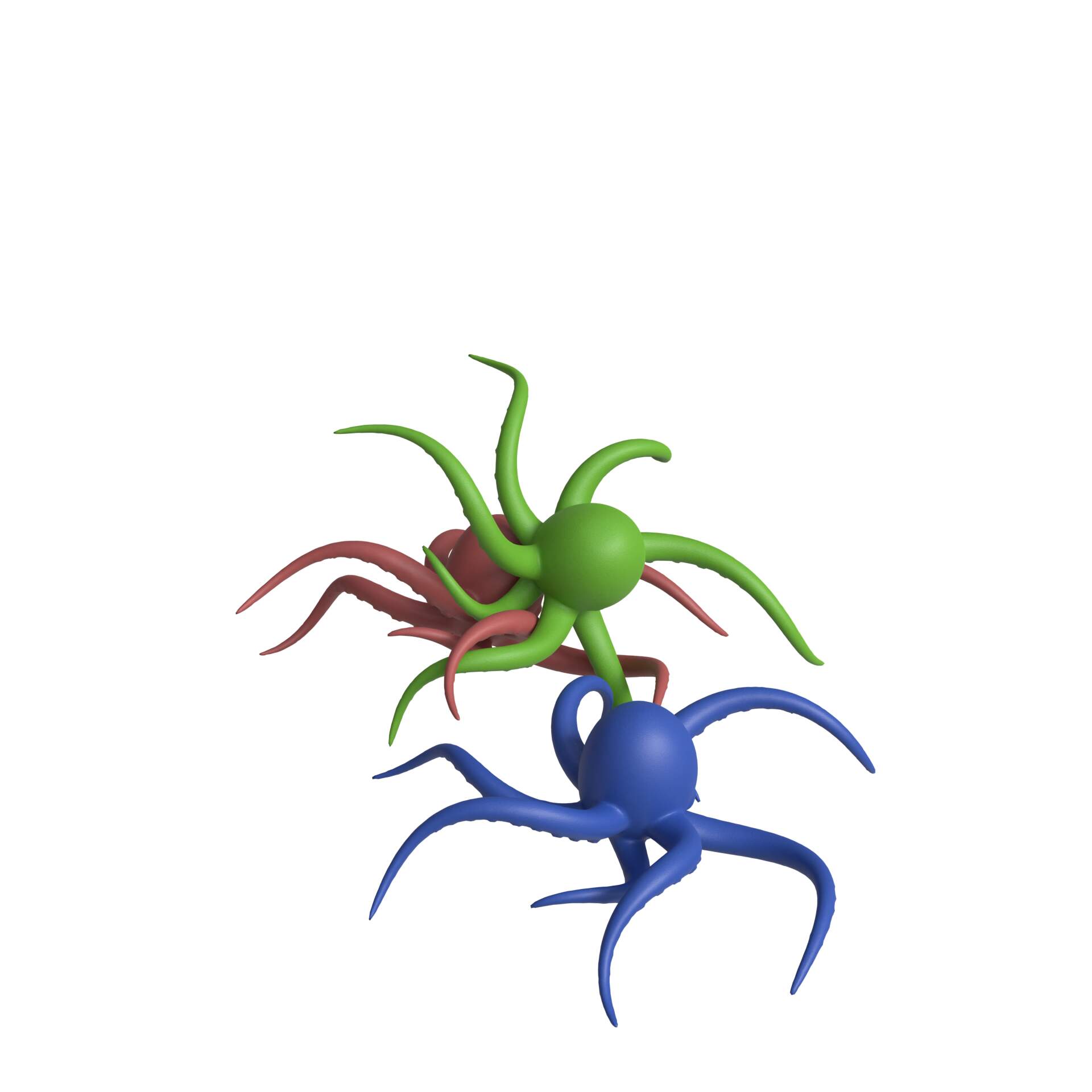}
			%\subcaption{\label{fig:memLocking_5e5}}
		\end{subfigure}
		\caption{\label{fig:octopus}\textbf{Octopus stack.} The simulation uses $\hat{d} = 10^{-3}l$, $\rho=1000 \, \text{kg}/\text{m}^3$, $\Delta t = 5\times10^{-3} \, \text{s}$, and a Newton tolerance of $10^{-2}l\Delta t$. This test case validates the performance improvement of our connectivity-enhanced MAS preconditioner, which achieves a $1.8\times$ speedup in linear solve.}
	\end{figure}
	
	\subsection{Ablation and Comparative Studies}\label{sec:comparisons}

	\paragraph{Preconditioner.} 
	To evaluate the performance of our connectivity-enhanced MAS preconditioner, we construct several scenes with different geometric structures (see \autoref{fig:octopus}, \autoref{fig:bunny2}, \autoref{fig:sqball}, and \autoref{fig:cardhouse}). The timing breakdown and CG iteration counts are reported in \autoref{tab:MAS_Preconditioner}. For a fair comparison, we implemented alternative preconditioners into our simulation framework, ensuring that all other components remain the same. The block-diagonal preconditioner was also tested as a reference. 
	Our results show that, with more effective aggregation, our method using a subdomain size of 32 provides the most significant improvement in PCG convergence compared to MAS \cite{mas} (see avg. \#cg column in \autoref{tab:MAS_Preconditioner}). Even with smaller subdomain sizes (16), our method converges faster than MAS and performs the best overall, achieving up to 3$\times$ speedup in the PCG solve and up to 2.6$\times$ faster for the full simulation, thanks to more efficient precomputation and preconditioning.
	
	As seen from the similar average CG iterations across resolutions (the rows for \autoref{fig:octopus} in \autoref{tab:MAS_Preconditioner}), our preconditioner maintains the scalability of MAS, resulting in near-linear total timing with respect to resolution. In \autoref{fig:bunny2}, the 'Ours (16)' configuration shows only a 1.17$\times$ speedup in convergence, as the volumetric structure of the bunnies results in similar METIS and Morton code reorderings, limiting improvement. However, we still observe a 1.58$\times$ speedup in PCG time and a 1.47$\times$ overall speedup, due to the lower computational cost of our preconditioner compared to MAS. \modify{Notably, as the number of bunnies increases or their spatial distribution becomes more irregular, our preconditioner achieves greater speedup (up to 3.5$\times$ in PCG time). We also tested a heterogeneous scene with 9 bunnies, where the Young’s modulus varies from 20 KPa to 10 MPa, and observed consistent PCG solver speedup compared to the homogeneous counterpart. Additionally, in a sequence of simulations with an increasing number of contact pairs between two plates, our preconditioner also maintained consistent speedup. These additional experiments and analyses are provided in the supplemental document.}

	\begin{table*}
		\caption{\label{tab:MAS_Preconditioner} \textbf{Simulation statistics with different preconditioners.} Mesh DOFs: number of vertices (v), tetrahedra (t), and surface triangles (f). \texttt{PPC}, \texttt{PPR}, and \texttt{CGR} represent the time spent on preconditioner pre-computation, preconditioning, and CG iterations, respectively, with \texttt{PCG} = \texttt{PPC} + \texttt{PPR} + \texttt{CGR}. The remaining columns report the time for miscellaneous tasks (\texttt{misc}), total simulation time (\texttt{timeTot}), total CG iterations (\texttt{\#cg}), the average Newton iterations per frame (\modify{\texttt{avg. \#Newton}}), and the average CG iterations per Newton iteration (\texttt{avg. \#cg}). All times are in seconds.}
		\resizebox{0.85\linewidth{}}{!}{
			\begin{tabular}{r|r|r|ccccccccc}
				\toprule
				Example & Mesh DOFs: v, t, f & Preconditioner & \texttt{PPC} &\texttt{PPR} &\texttt{CGR} &\texttt{PCG} & \texttt{Misc} & \texttt{TimeTot} & \texttt{\#cg} & \modify{\texttt{avg. \#Newton}} & \texttt{avg. \#cg} \\ 
				\midrule
				\multirow{5}{*}{\autoref{fig:octopus}}&\multirow{5}{*}{$42K,171K, 45K$}&{GPU MAS} & 38 & 55  & 102 &195& 26 & 221 & 6.98e5&11.28 & 344\\&
				&{Block Diagonal} & \textbf{0.06} & 21  & 338 & 359 & 27 & 385 & 2.24e6&11.50&1082  \\&
				&{Ours (32)}& 32& \textbf{25}& \textbf{60}& 117 & 26  & 143 & \textbf{3.45e5} &11.22 & \textbf{171}  \\&
				
				&{Ours (16)}& \textbf{10} & 29  & 69& \textbf{108} & 26 &\underline{\textbf{134}} & 4.41e5 & 11.06 & 221 \\&
				&{\textbf{Speed Up}} & 3.80$\times$ & 1.90$\times$  & -& 1.81$\times$ & -& 1.65$\times$  & -& - & 1.51$\times$\\
				\midrule
				\multirow{5}{*}{\autoref{fig:octopus}} & \multirow{5}{*}{$114K,558K, 69K$} & {GPU MAS} & 105 & 162 & 257 & 524 & 69 & 593 & 9.51e5 & 12.78 & 412 \\&
				&{Block Diagonal}  & \textbf{0.16} & \textbf{43}  & 582 & 625 & 69 & 694 & 2.48e6&12.78&1079  \\&
				&{Ours (32)}   & 96& 79& \textbf{127}& 302 & 69  & 371 & \textbf{4.37e5} &12.83 & \textbf{189}  \\&
				&{Ours (16)} & 33 & 60  & 145& \textbf{238} & 69 &\underline{\textbf{307}}& 5.20e5 & 12.83 & 225 \\&
				&{\textbf{Speed Up}} & 3.18$\times$ & 2.7$\times$  & -& 2.20$\times$ & -& 1.93$\times$  & -& - & 1.84$\times$\\
				%\bottomrule
				\midrule
				\multirow{5}{*}{\autoref{fig:bunny2}} & \multirow{5}{*}{$203K,1122K, 50K$}&{GPU MAS} & 152 & 188  & 229 &569& 87 & 656 & 6.23e5&9.78 & 354  \\&
				&{Block Diagonal}  & \textbf{0.18} & \textbf{63}  & 918 & 981 & 88 & 1069 & 2.61e6&9.83&1474  \\&
				&{Ours (32)}   & 134& 137& \textbf{180}& 451 & 87  & 538 & \textbf{4.54e5} &9.78 & \textbf{258}  \\&
				&{Ours (16)} & 42 & 108  & 210& \textbf{360} & 87 &\underline{\textbf{447}}& 5.30e5 & 9.78 & 301 \\&
				&{\textbf{Speed Up}} & 3.61$\times$ & 1.74$\times$  & -& 1.58$\times$ & -& 1.47$\times$  & -& - & 1.17$\times$\\
				\midrule
				\multirow{5}{*}{\autoref{fig:sqball}} & \multirow{5}{*}{$123K,330K, 211K$}&{GPU MAS} & 371 & 836  & 1029 &2236& 217 & 2453 & 4.38e6&39.65 & 552  \\&
				&{Block Diagonal}  & \textbf{0.44} & \textbf{137}  & 1920 & 2057 & 216 & 2273 & 8.84e6&39.50&1119  \\&
				&{Ours (32)}   & 354& 196& \textbf{298}& 848 & 215  & 1063 & \textbf{9.98e5} &39.30 & \textbf{127}  \\&
				&{Ours (16)} & 76 & 215  & 439& \textbf{730} & 215 &\underline{\textbf{945}}& 1.81e6 & 39.25 & 231 \\&
				&{\textbf{Speed Up}} & 4.88$\times$ & 3.89$\times$  & -& 3.06$\times$ & -& 2.60$\times$  & -& - & 2.39$\times$\\
				\midrule
				\multirow{5}{*}{\autoref{fig:cardhouse}} & \multirow{5}{*}{$79K,298K, 110K$}&{GPU MAS} & 86 & 193  & 274 &553& 42 & 595 & 1.42e6&8.10 & 877  \\&
				&{Block Diagonal}  & \textbf{0.08} & \textbf{57}  & 798 & 855 & 43 & 898 & 5.09e6&8.20&3105  \\&
				&{Ours (32)}   & 47& 89& \textbf{124}& 260 & 42  & 302 & \textbf{6.84e5} &8.15 & \textbf{420}  \\&
				&{Ours (16)} & 12 & 68  & 136& \textbf{216} & 41 &\underline{\textbf{257}}& 7.53e5 & 8.10 & 465 \\&
				&{\textbf{Speed Up}} & 7.17$\times$ & 2.83$\times$  & -& 2.56$\times$ & -& 2.31$\times$  & -& - & 1.89$\times$\\
				\bottomrule
			\end{tabular}
		}
	\end{table*}

	\begin{figure}[htbp]
		\centering
		\begin{subfigure}[b]{0.23\textwidth}
			\centering
			\includegraphics[width=\columnwidth, trim=100 80 20 70, clip]{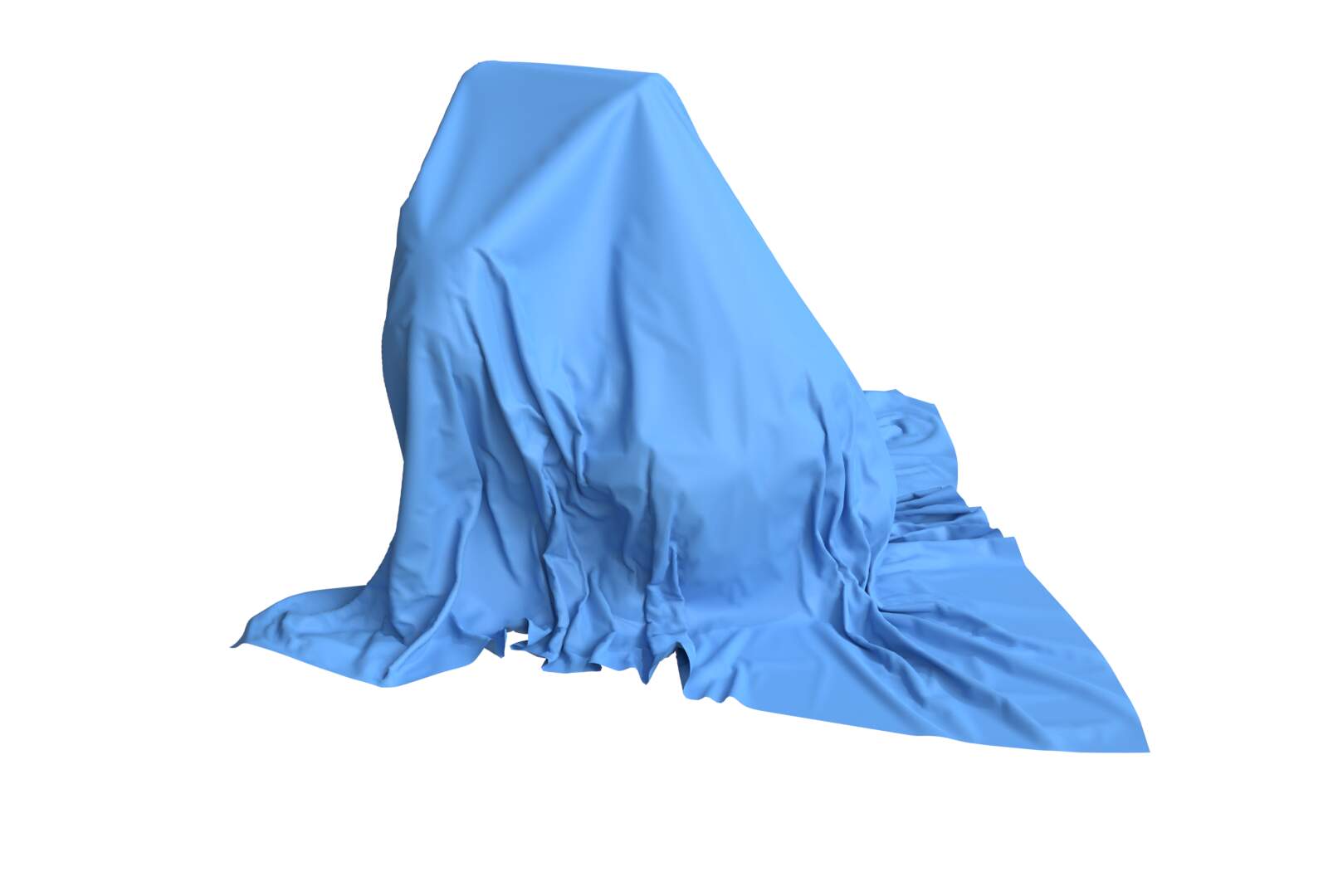}
			\subcaption{\label{fig:memLocking_5e6}$Y = 5MPa$, no SL}
		\end{subfigure}
		\begin{subfigure}[b]{0.23\textwidth}
			\centering
			\includegraphics[width=\columnwidth, trim=100 80 20 70, clip]{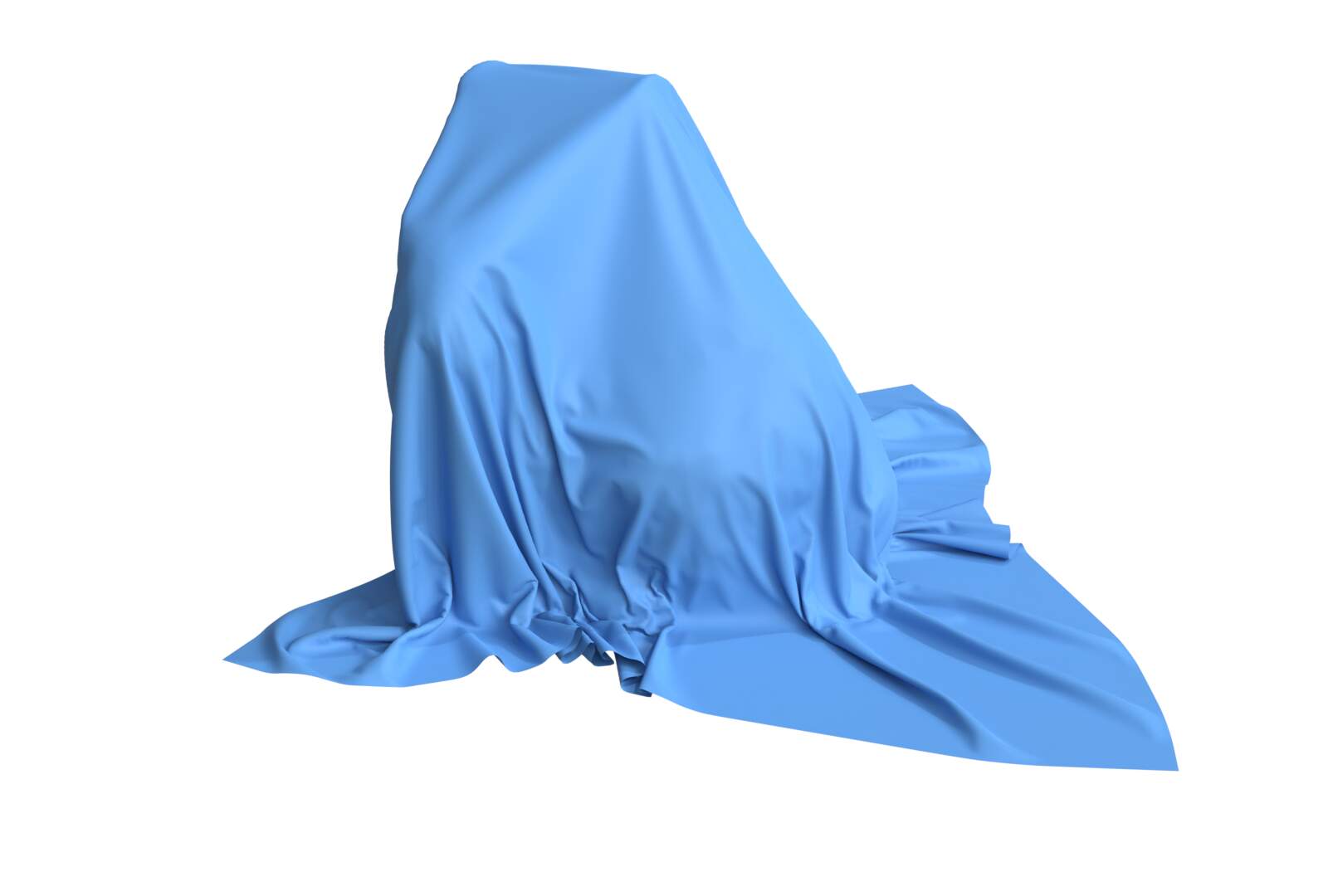}
			\subcaption{\label{fig:memLocking_5e5}$Y = 0.5MPa$, no SL}
		\end{subfigure}
		\begin{subfigure}[b]{0.23\textwidth}
			\centering
			\includegraphics[width=\columnwidth, trim=100 80 20 70, clip]{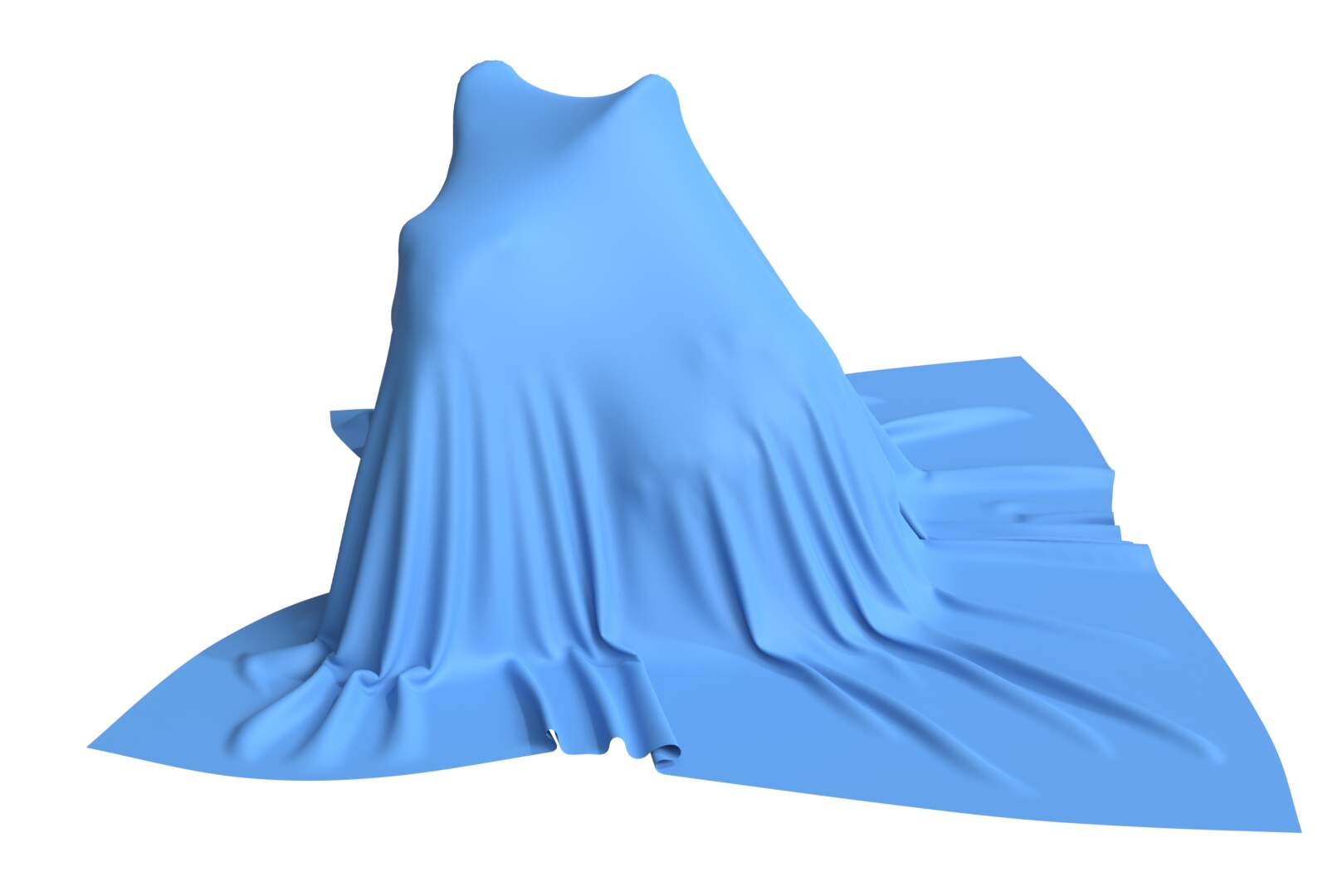}
			\subcaption{\label{fig:memLocking_5e4}$Y = 50KPa$, no SL}
		\end{subfigure}
		\begin{subfigure}[b]{0.23\textwidth}
			\centering
			\includegraphics[width=\columnwidth, trim=100 80 20 70, clip]{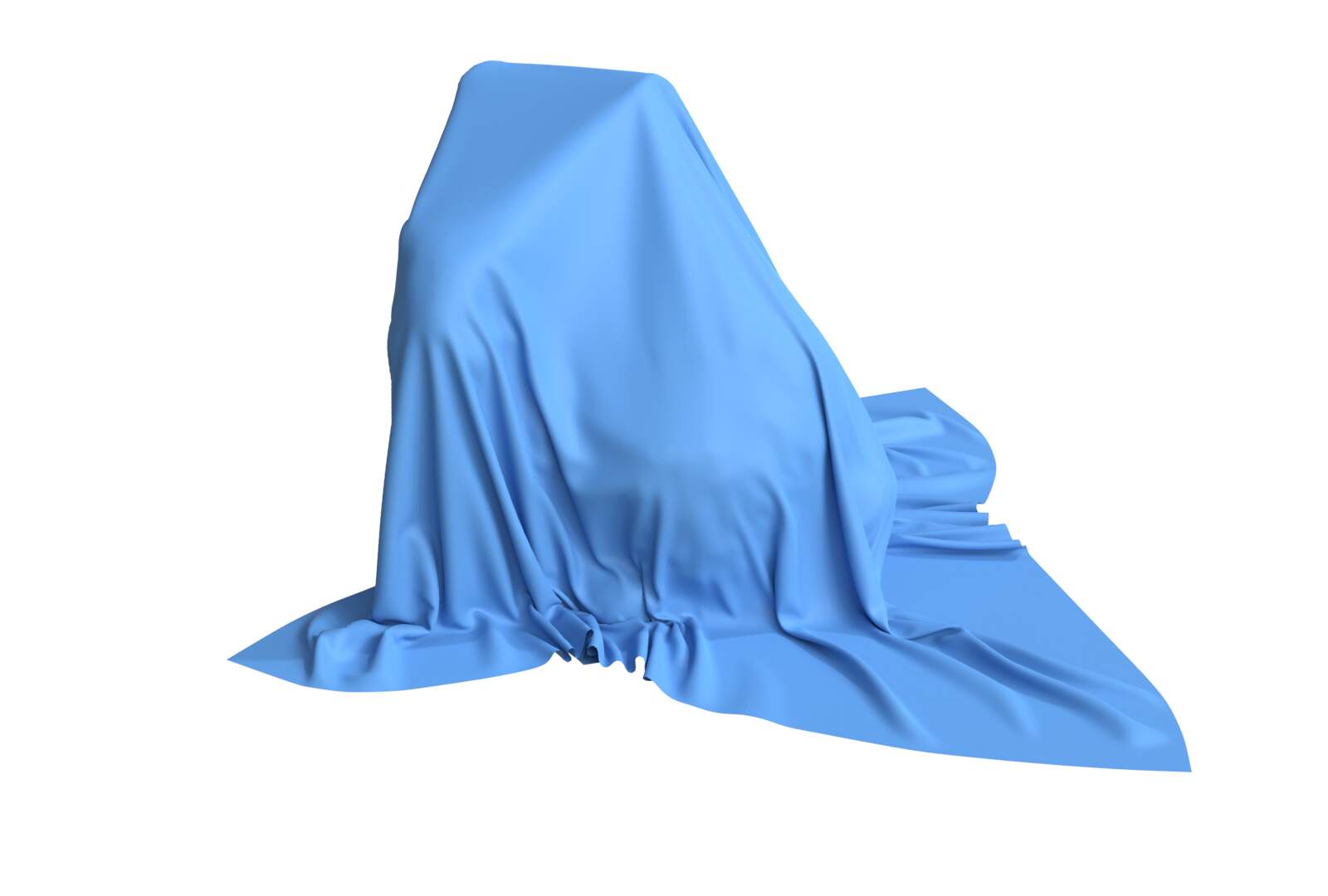}
			\subcaption{\label{fig:memLocking_sl}$Y = 50KPa$, with \citet{cipc}}
		\end{subfigure}
		\begin{subfigure}[b]{0.23\textwidth}
			\centering
			\includegraphics[width=\columnwidth, trim=100 80 20 70, clip]{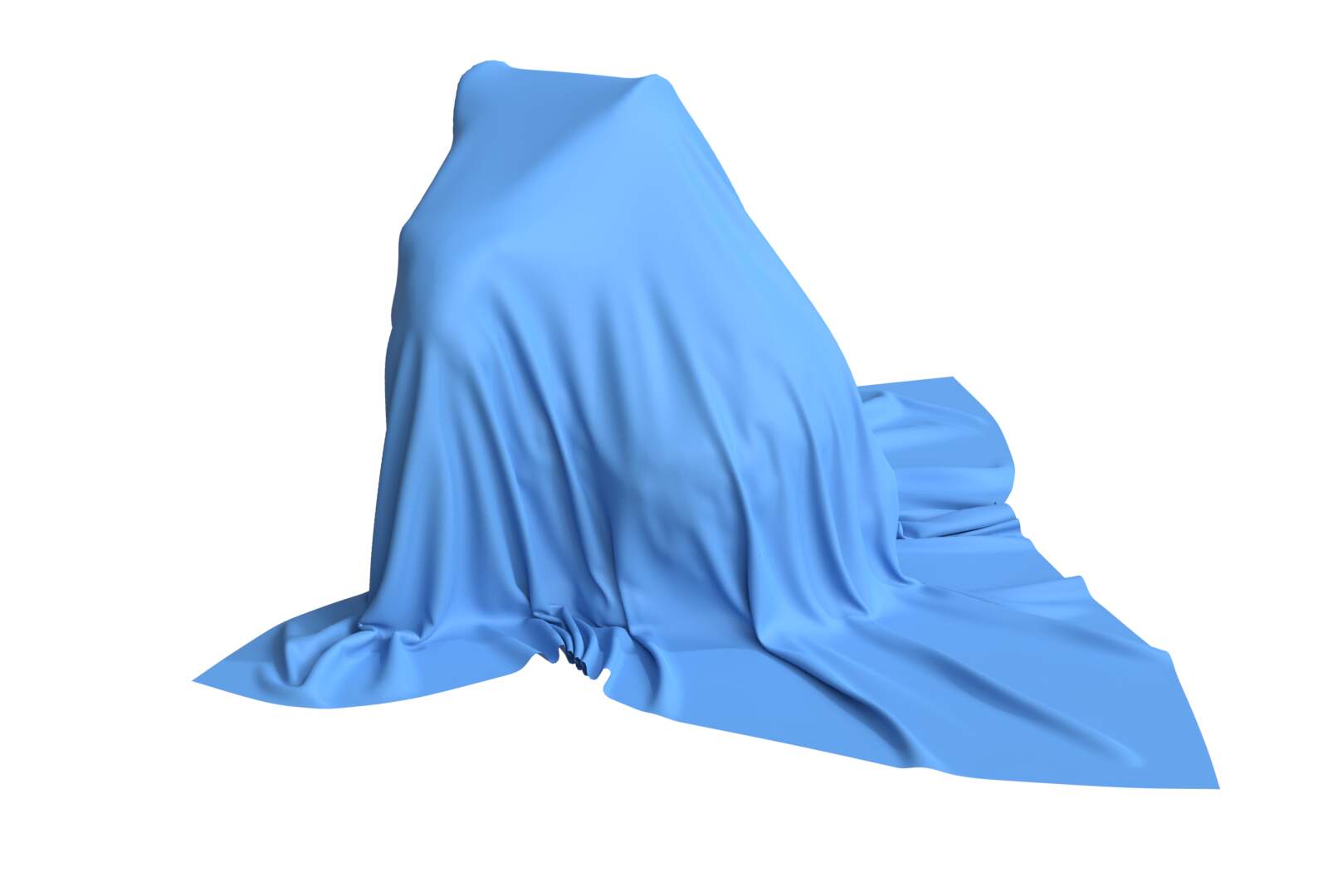}
			\subcaption{\label{fig:memLocking_new}$Y = 50KPa$, with our SL}
		\end{subfigure}
		\caption{\label{fig:memLocking}\textbf{Membrane locking and strain limiting.} The cloth consists of $100K$ vertices and $200K$ triangles, simulated with FEM, while the bunny is simulated using ABD. The simulation parameters are: $\Delta t = 0.01 \, \text{s}$, Poisson's ratio $\nu = 0.49$, density $\rho = 200 \, \text{kg}/\text{m}^3$, cloth thickness $10^{-3} \, \text{m}$, friction coefficient $\mu = 0.1$, Newton tolerance $\varepsilon_d = 10^{-2}l\Delta t$, membrane Young's modulus $Y = \{5 \times 10^6, 5 \times 10^5, 5 \times 10^4\} \, \text{Pa}$, and distance threshold $\hat{d} = 10^{-3}l$, where $l$ is the scene diagonal length. \citet{cipc} (d) and our method (e) simulate fine wrinkling details without membrane locking (a, b) or over-elongation (c), with our method being $1.24\times$ faster (\autoref{tab:memlocking}).}
	\end{figure}
	\paragraph{Strain limiting.} \autoref{fig:memLocking} presents a cloth simulation example using FBW~\cite{kim_baraf_witkin} with and without barrier-based \cite{cipc} or our cubic \modify{inexact} strain-limiting energy. The cloth consists of 100K nodes and is tested with various membrane stiffnesses. In cloth simulation, avoiding excessive stretching is crucial, as it can cause unrealistic, overly-elastic behavior. To prevent this, a large Young's modulus (e.g., $5MPa$), within the range of real cloth parameters \cite{penava2014determination}, is needed. However, such stiffness can lead to membrane locking in even medium-resolution meshes, resulting in sharp creases and artificial plastic appearances (\autoref{fig:memLocking_5e6}). Reducing the stiffness to $0.5MPa$ mitigates membrane locking but visible artifacts remain (\autoref{fig:memLocking_5e5}). Further reducing the stiffness to $50KPa$ smooths wrinkles but causes over-elongation (\autoref{fig:memLocking_5e4}).
	Our cubic \modify{inexact} strain-limiting energy provides sufficient stretching resistance as in real fabrics, avoiding excessive stretching and capturing smooth, realistic wrinkles with a small membrane stiffness of $50KPa$ for compression (\autoref{fig:memLocking_new}, and supplemental video).

	We also compare with the barrier-based strain-limiting method in CIPC using their default settings: a $1.1$ strain limit and $1$ activation threshold. We use a $1KPa$ stiffness for the barrier term, as smaller or larger values both lead to worse conditioned systems. With similar result quality (\autoref{fig:memLocking_sl}), our method is $1.24\times$ faster (\autoref{tab:memlocking}), with lower costs in local Hessian construction and line search (see \autoref{tab:Overall_comparison} for a more detailed breakdown). 
	While our method performs well for practical cloth manipulation scenarios, it does not guarantee strain limit satisfaction in extreme cases like the barrier-based method. 
    
\begin{figure}[htbp]{}
		
		\centering
		\begin{subfigure}[b]{0.47\textwidth}
			\includegraphics[width=\columnwidth, trim=0 0 0 0, clip]{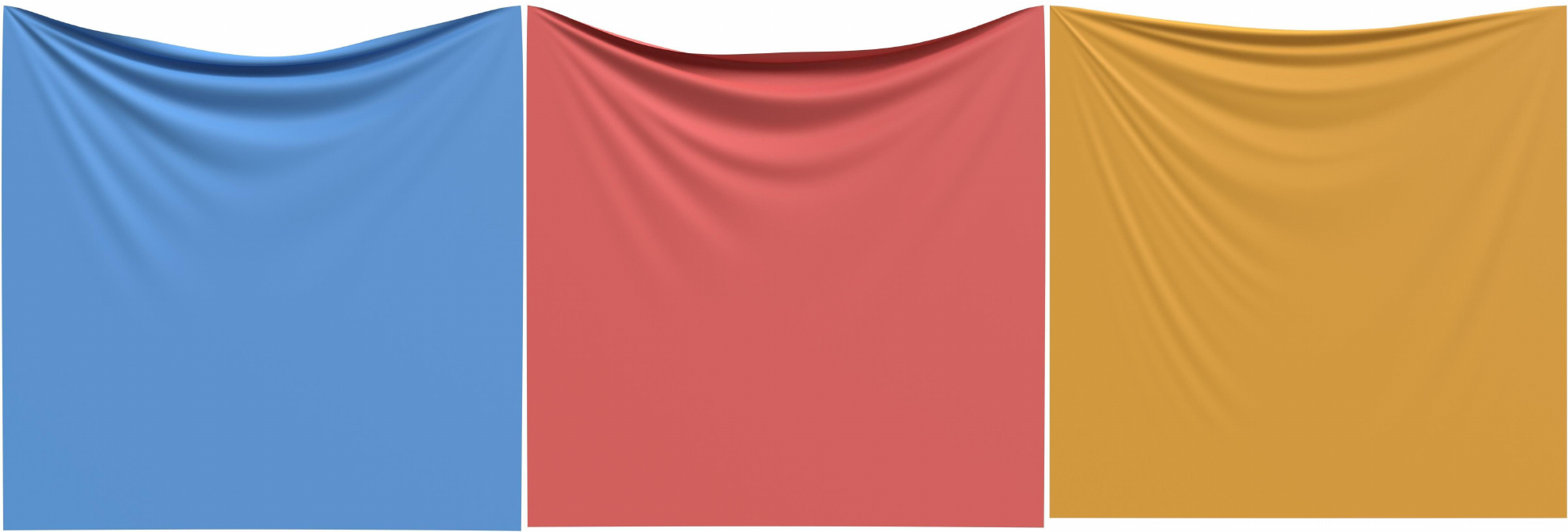}
            %\subcaption{\label{fig:cipcsl}SL}
		\end{subfigure}
				%\begin{subfigure}[b]{0.15\textwidth}
			%\includegraphics[width=\columnwidth, trim=150 150 150 100, clip]{figs/stretch_rate/0060_c2.jpg}
		%\end{subfigure}
        %\subcaption{\label{fig:cipcsl}Our SL}
        %\begin{subfigure}[b]{0.15\textwidth}
			%\includegraphics[width=\columnwidth, trim=150 150 150 100, clip]{figs/stretch_rate/0060.jpg}
            %\subcaption{\label{fig:cipcsl}Our SL}
		%\end{subfigure}
		\caption{ 
			% \textbf{Ablation study of multi-layer reduction for Hessian assembly.}
			% Strategy \textbf{A} constructs the global Hessian atomically and serves as the baseline. Strategy \textbf{B} uses global reduction of the affine body Hessian only for the overall Hessian reduction. Strategy \textbf{C} (ours) first reduces the contact Hessian before proceeding with the global reduction of the affine body Hessian.\todo{}
			\modify{\textbf{Quantitative evaluation of principal stretch.} We simulate a hanging cloth with pinned upper corners. The blue cloth uses the log barrier strain-limiting method \cite{cipc} with their default stiffness of 1 KPa and strain limit of 1.1. The red cloth applies our inexact strain-limiting method with a default stiffness of 5 MPa, resulting in a maximum principal stretch of 1.35 at the corners and an average of 1.01. The yellow cloth also uses our method but with an increased stiffness of 500 MPa, reducing the maximum stretch to 1.1 and the average to 1.001.
		}}
		\label{fig:stretchrate}
	\end{figure}

\begin{figure}[htbp]
		\centering
		\begin{subfigure}[b]{0.22\textwidth}
			\centering
			\includegraphics[width=\columnwidth, trim=400 50 400 650, clip]{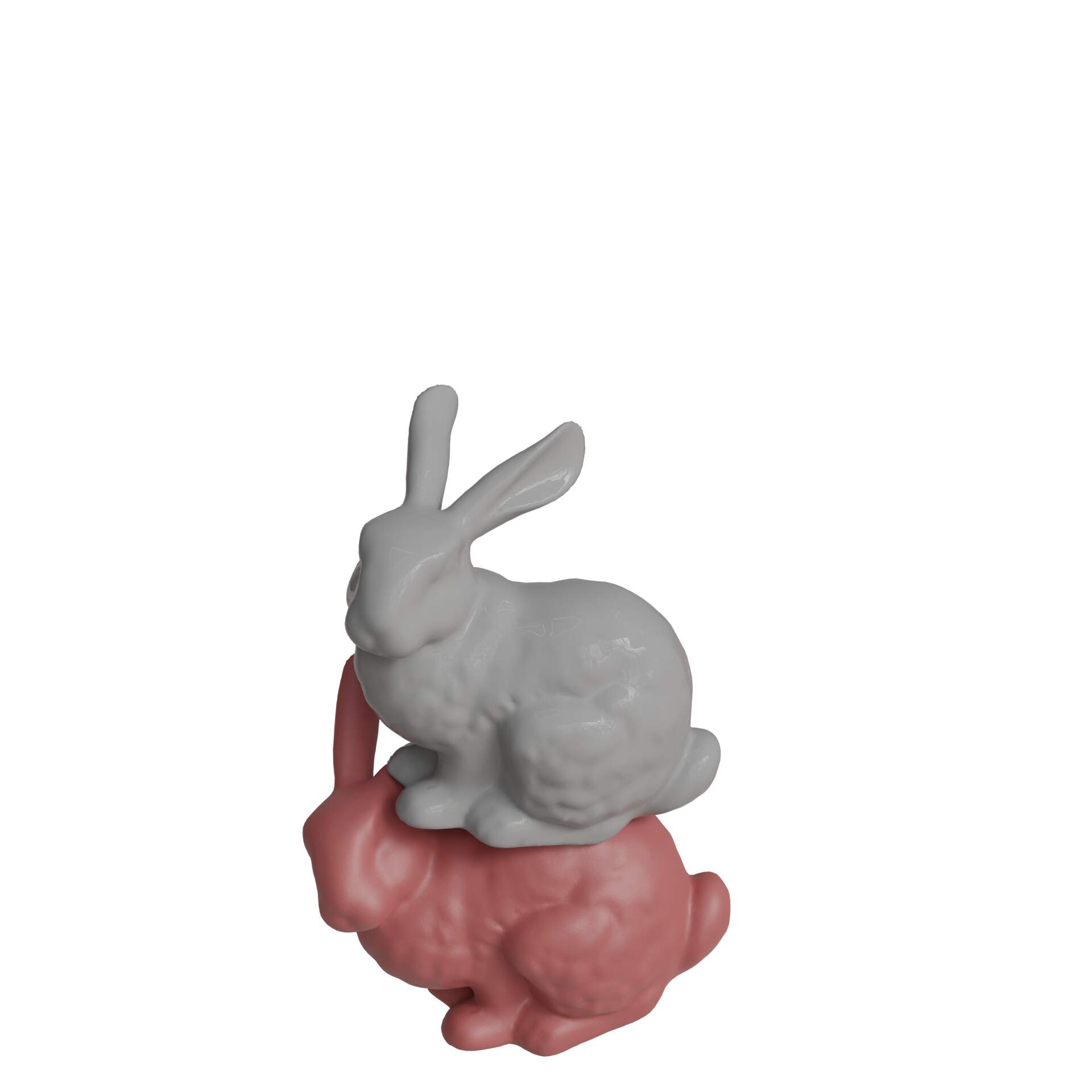}
		\end{subfigure}
		\centering
		\begin{subfigure}[b]{0.22\textwidth}
			\includegraphics[width=\columnwidth, trim=400 50 400 650, clip]{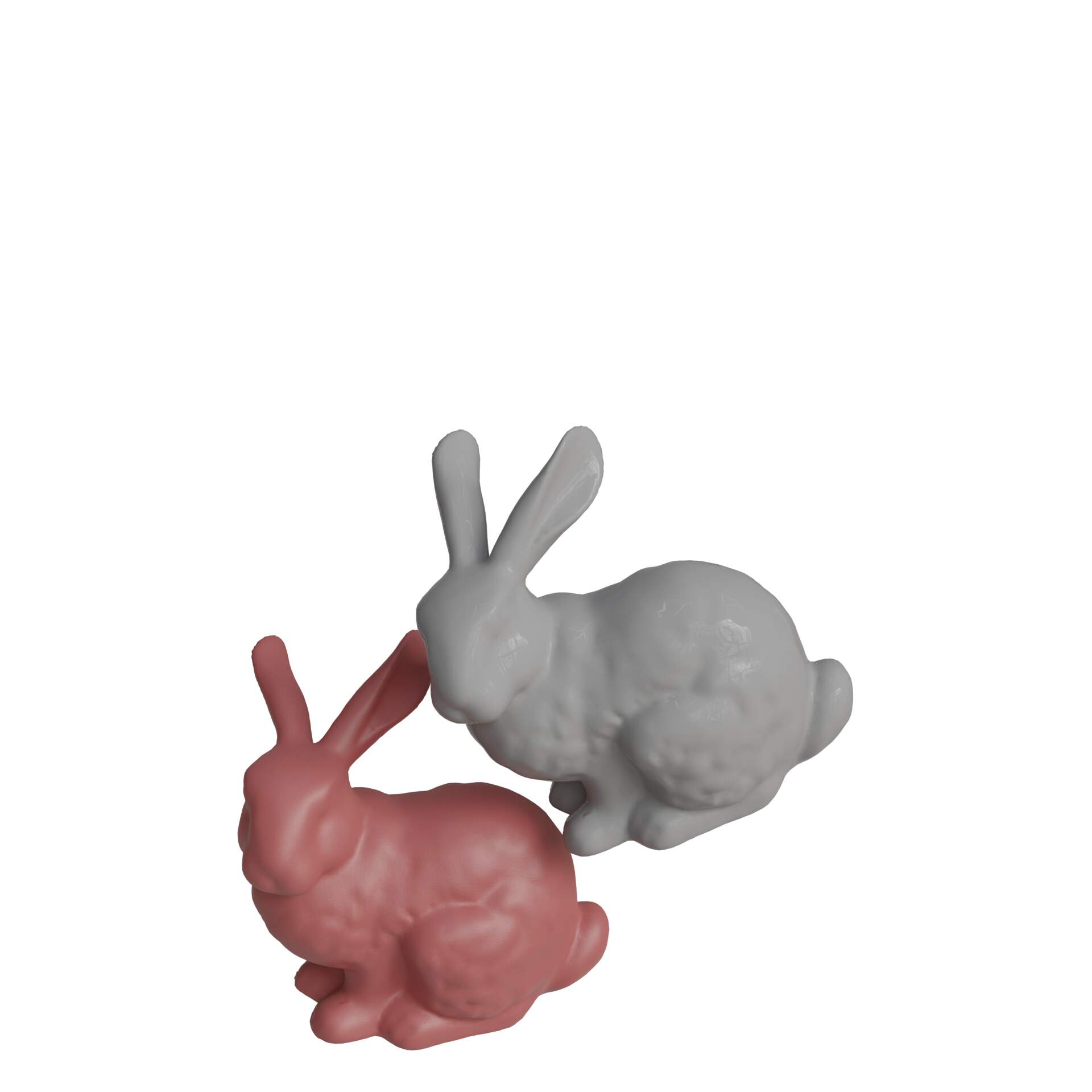}
		\end{subfigure}
		\caption{\label{fig:bunny2} \textbf{Stiff and soft bunnies.} The top bunny is nearly rigid (Young's modulus $10^8$ Pa), and the bottom red one is deformable (Young's modulus $10^6$ Pa). Each bunny consists of $101K$ vertices and $561K$ tetrahedra. The simulation uses $\hat{d} = 5\times10^{-4}l$, $\rho = 1000 \, \text{kg}/\text{m}^3$, $\Delta t = 5\times10^{-3} \, \text{s}$, and a Newton tolerance of $10^{-2}l\Delta t$. This test compares GIPC, ZeMa, and our method in hybrid scenarios. Our method shows a 4$\times$ speedup when simulating both bunnies using FEM and a 10$\times$ speedup when using ABD+FEM.}
	\end{figure}

    \modify{To quantitatively assess constraint satisfaction, we conducted an experiment by pinning the upper corners of a cloth and letting it hang under gravity (\autoref{fig:stretchrate}). The upper corner triangle experienced the most significant stretching. In this setup, the log barrier strain-limiting method strictly constrained the principal stretch to the predefined upper bound of 1.1, with a maximum of 1.098 and an average of 1.012.
    In contrast, our method with default stiffness (5 MPa) resulted in a maximum principal stretch of 1.35 while maintaining an average of 1.010. Although it does not strictly enforce strain limits at the upper corners, the overall simulation remains physically plausible (see the supplementary video). 
    For stress testing purposes, increasing our stiffness to 0.5 GPa brings the maximum principal stretch in line with the log barrier method, achieving a maximum of 1.100 and an average of 1.001.
    Furthermore, we also compare our inexact strain limiting with the log barrier method} in another stress test (see \autoref{fig:box_pipe}), which is discussed in detail later.

	% \mc{it'd be great to run the tablecloth pull stress test, as we cannot guarantee strain limit satisfaction, so this is a very challenging task}
	
	\begin{table}
		\caption{\label{tab:memlocking} \textbf{Simulation statistics for \autoref{fig:memLocking}.} Columns show total time for linear solve (PCG) and miscellaneous tasks (Misc), total simulation time (TimeTot), total PCG iterations (\#cg), total Newton iterations (\#Newton), and average PCG iterations per Newton iteration (\#cg per iter). All simulations use the same PCG relative tolerance of $10^{-4}$ with our connectivity-enhanced MAS preconditioner.
		}
		\resizebox{\columnwidth{}}{!}{
			\begin{tabular}{r|cccccc}
				\toprule
				Young's Modulus ($Pa$) & PCG & Misc & TimeTot & \#cg & \#Newton & avg. \#cg per iter \\ 
				\midrule
				{$5MPa$, no SL (\autoref{fig:memLocking_5e6})} 
				& 2.91e2 & 1.34e2  & 4.25e2 & 7.07e5 & 4.03e3 & 175  \\
				%&Diagonal& 1.74e2 & 1.52e2  & 3.26e2 & 5.61e5 & 4.18e3 & 134  \\
				\midrule
				{$0.5MPa$, no SL (\autoref{fig:memLocking_5e5})}  
				& 2.31e2 & 1.61e2  & 3.92e2 & 4.12e5 & 4.49e3 & 92  \\
				%&Diagonal& 2.51e2 & 1.59e2  & 4.1e2 & 4.60e5 & 4.69e3 & 98  \\
				
				%%
				\midrule
				{$50KPa$, no SL (\autoref{fig:memLocking_5e4})}  
				& 1.26e2 & 1.34e2  & 2.60e2 & 1.66e5 &3.63e3 & 46  \\
				
				\midrule
				{$50KPa$ + CIPC SL (\autoref{fig:memLocking_sl})}
				& 1.55e2 & 2.02e2  & 3.57e2 & 2.24e5 &3.99e3 & 56  \\
				
				\midrule
				{$50KPa$ + our SL (\autoref{fig:memLocking_new})}
				& 1.45e2 & 1.43e2  & 2.88e2 & 2.06e5 &3.76e3 & 55  \\
				%&Diagonal& 1.74e2 & 1.52e2  & 3.26e2 & 5.61e5 & 4.18e3 & 134  \\
				\bottomrule
			\end{tabular}
		}
	\end{table}
	
\begin{figure}[htbp]{}
		
		\centering
		\begin{subfigure}[b]{0.42\textwidth}
			\includegraphics[width=\columnwidth, trim=0 0 0 0, clip]{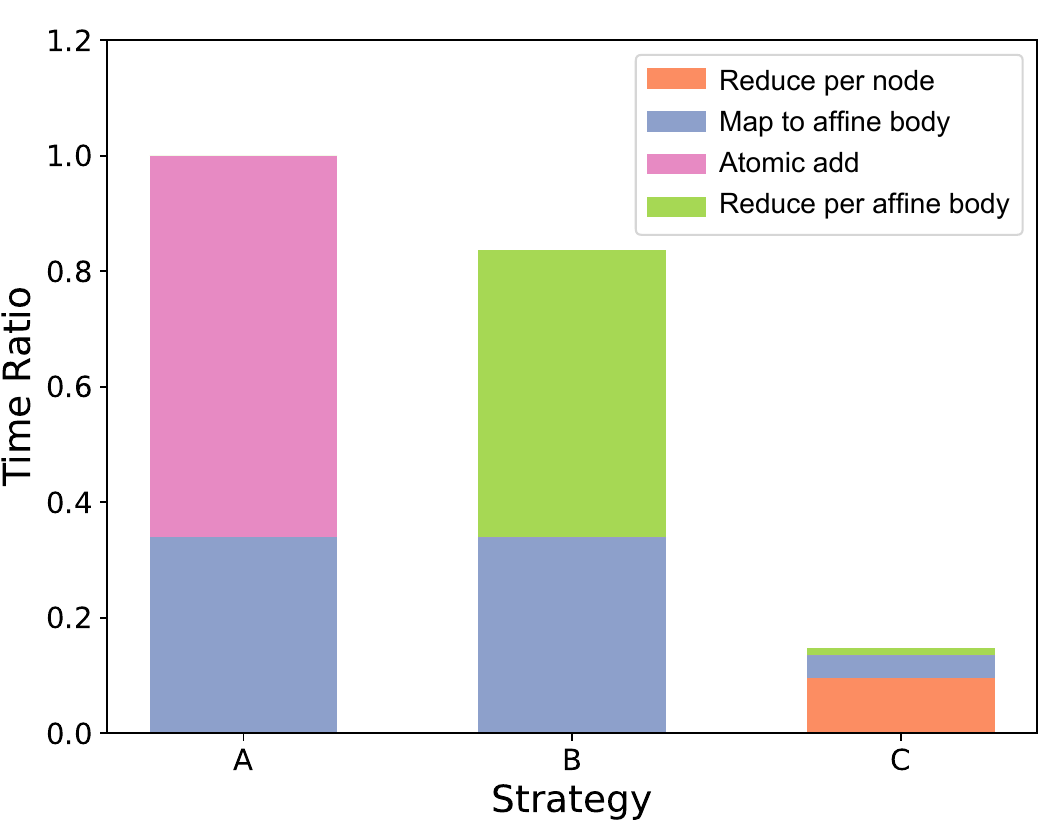}
		\end{subfigure}
		
		\caption{ 
			% \textbf{Ablation study of multi-layer reduction for Hessian assembly.}
			% Strategy \textbf{A} constructs the global Hessian atomically and serves as the baseline. Strategy \textbf{B} uses global reduction of the affine body Hessian only for the overall Hessian reduction. Strategy \textbf{C} (ours) first reduces the contact Hessian before proceeding with the global reduction of the affine body Hessian.\todo{}
			\textbf{Ablation study on \modify{two-level} reduction for Hessian assembly.} We compare our \modify{two-level} reduction strategy (C) against single-level reduction (B) and pure atomic operations (A) in the wrecking ball simulation (\autoref{fig:wrecking_ball}). Despite sorting the value array for optimal \emph{\textbf{atomicAdd}} performance, using atomic operations is slightly slower than single-level reduction, and ours achieves nearly $5\times$ speedup over Strategy B.
		}
		\label{fig:multi_level_reduction}
	\end{figure}

	\paragraph{Global Hessian assembly}
	As illustrated in \autoref{fig:multi_level_reduction}, we use the wrecking ball simulation (\autoref{fig:wrecking_ball}) to conduct an ablation study on our \modify{two-level} reduction strategy (Strategy C) for global Hessian matrix assembly, comparing it against single-level reduction (Strategy B) and pure atomic operations (Strategy A).
	In both Strategies A and B, the local Hessian matrices of all contact pairs are first mapped to the affine body DOFs by multiplying the Jacobian matrices. Strategy B then performs a reduction to accumulate these matrices into the global Hessian, while Strategy A uses atomic operations directly for accumulation. To ensure a fair comparison, we also sort the value array prior to atomic accumulation. This step optimizes the distribution of the value array, enabling aggregation optimization in the CUDA intrinsic function \emph{\textbf{atomicAdd}}, thereby achieving the best possible performance.
	Even with this optimization, pure atomic operations remain slightly slower than single-level reduction, as shown in the comparison of Strategies A and B in \autoref{fig:multi_level_reduction}. Our \modify{two-level} reduction strategy achieves an additional nearly $5\times$ speedup over single-level reduction.

\begin{figure}[htbp]
		\centering
		\begin{subfigure}[b]{0.42\textwidth}
			\centering
			\includegraphics[width=\columnwidth, trim=0 0 0 0, clip]{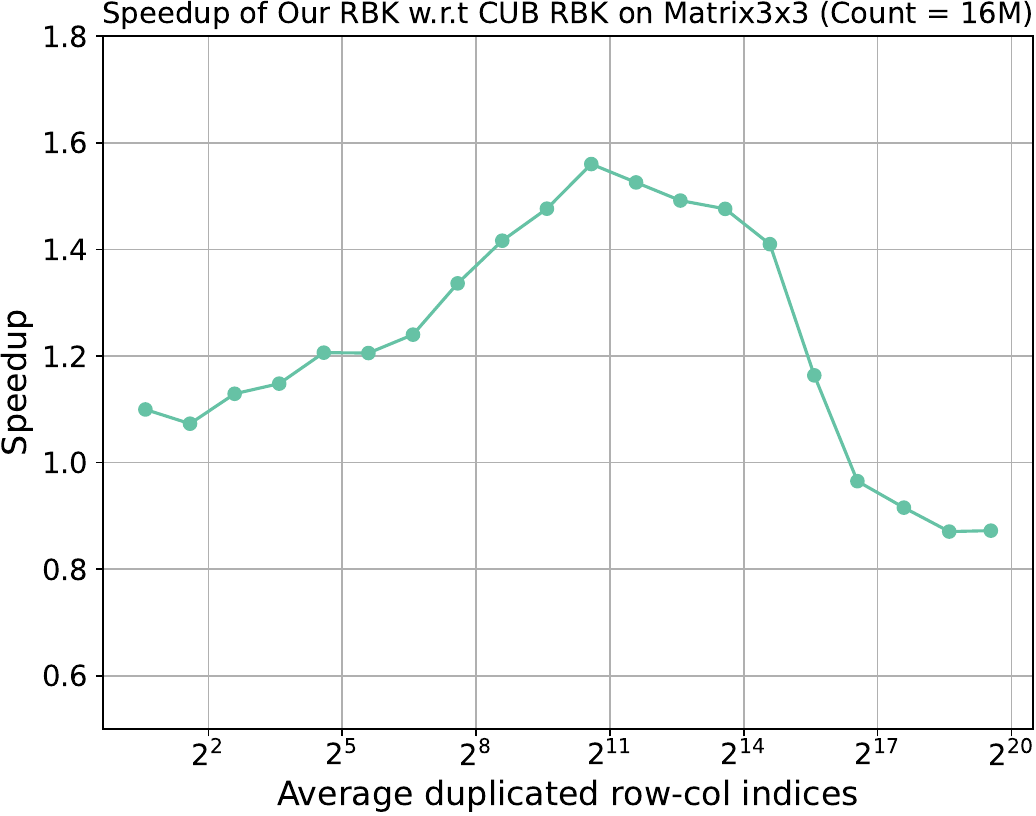}
		\end{subfigure}
		
		\caption{
			\textbf{Speedup of FastHashReduction vs. CUB ReduceByKey.} Speedup factor is shown relative to the average number of matrices sharing the same row and column indices. A peak speedup of $1.55\times$ occurs around $2^{11}$. For typical duplication levels in FEM simulations (up to $2^7$), FastHashReduction maintains a consistent speedup between $1.2\times$ and $1.4\times$.
		}
		\label{fig:rbk_speedup}
	\end{figure}

	\paragraph{Reduce by Key}
	We compare our \textit{FastHashReduction} method with the CUB intrinsic function \emph{\textbf{cub::DeviceReduce::ReduceByKey}} and plot our speedup relative to the number of values being accumulated (\autoref{fig:rbk_speedup}). To keep GPU memory usage constant, the reduction is performed on 16M $3\times 3$ matrices. The workload is controlled by adjusting the average number of matrices sharing the same row and column indices, ranging from $1$ to $2^{16}$. Our method shows superior performance across a wide range of accumulation workloads, achieving a maximum speedup of approximately $1.55\times$ at $2^{11}$. In FEM simulations, the average number of values with the same row and column indices typically remains below $2^{7}$, where our method achieves a consistent speedup of $1.2$ to $1.4\times$.
	Additionally, we observe that \emph{\textbf{cub::DeviceReduce::ReduceByKey}} fails to compile when reducing the $12 \times 12$ affine body Hessian matrix due to shared memory limitations. In contrast, our \textit{FastHashReduction} method remains effective by reusing warp registers, leveraging the element-independent nature of matrix summation.
	
\begin{figure}[htbp]
		\centering
		\begin{subfigure}[b]{0.44\textwidth}
			\centering
			\includegraphics[width=\columnwidth, trim=0 0 0 0, clip]{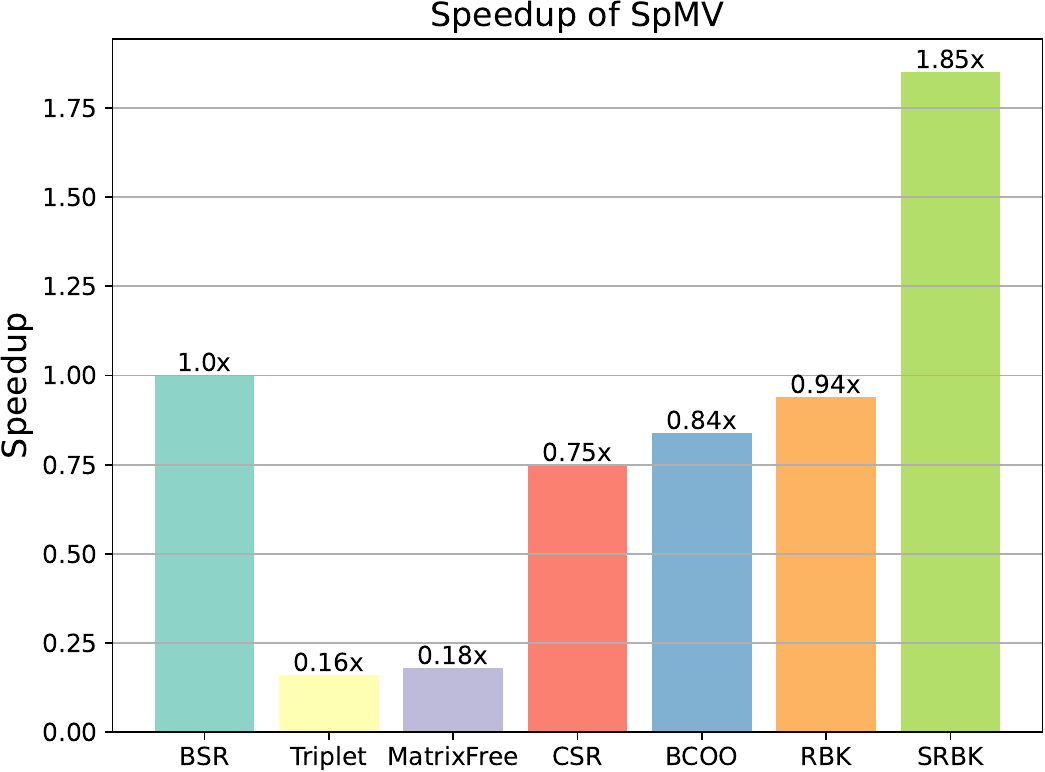}
		\end{subfigure}
		\caption{\label{fig:spmv_speedup} \modify{\textbf{Timing comparison of SpMV methods.} The BSR method from cuSparse serves as the baseline. Our method (SRBK) achieves a maximum speedup of 1.85$\times$ over the baseline.}}
	\end{figure}

	\paragraph{SpMV}
	We compare the performance of different SpMV methods using the stiff and soft bunnies example shown in \autoref{fig:bunny2}, simulating both bunnies with FEM. As shown in \autoref{fig:spmv_speedup}, the BSR SpMV (1$\times$) from cuSparse\footnote{\href{https://developer.nvidia.com/cusparse}{https://developer.nvidia.com/cusparse}} is treated as a baseline. Without matrix sorting, the Triplet SpMV (0.16$\times$) implemented by atomically adding the matrix-vector products and the MatrixFree SpMV (0.18$\times$) from GIPC both suffer from an \textit{atomic operation flood}, which undermines the parallelism. On the other hand, sorting-based methods have a significant speedup, benefiting from the reduction of atomic operations. The traditional CSR SpMV (0.75$\times$) from cuSparse has an almost 5$\times$ speedup compared to the unsorted methods. The Block Coordinates (BCOO) SpMV from MUDA\footnote{\href{https://github.com/MuGdxy/muda}{https://github.com/MuGdxy/muda}} and the BSR SpMV from cuSparse further improves the performance by applying blockwise matrix-vector multiplication. Our SRBK SpMV method ($1.85\times$) further mitigates the memory-bound issue of BSR SpMV and boosts the performance by almost $2\times$. \modify{Here, RBK refers to our Reduce-By-Key approach that loads the full Hessian matrix from global memory without leveraging the symmetric structure, leading to degraded performance.}
	
\begin{figure}[htbp]
		\centering
		\begin{subfigure}[b]{0.42\textwidth}
			\centering
			\includegraphics[width=\columnwidth, trim=0 0 0 0, clip]{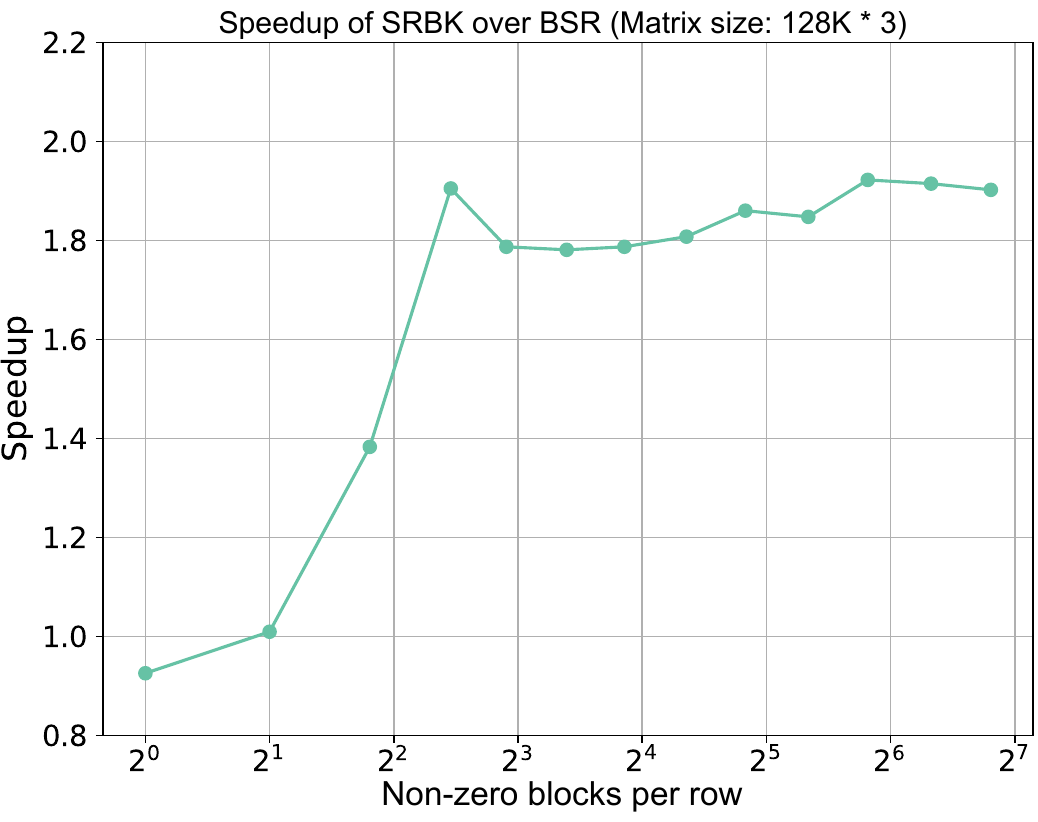}
		\end{subfigure}
		
		\caption{\textbf{Scalability of our SRBK SpMV.} Using the BSR from cuSparse as the baseline, we compare our SpMV's speedup over BSR on matrices with different numbers of non-zero blocks per row. The speedup increases from $0.93\times$ to $1.9\times$ as non-zero blocks per row approach \(2^3\). For typical FEM ranges (\(2^3\)–\(2^6\) blocks), speedup stabilizes between \(1.8\times\) and \(2\times\), demonstrating our strong scalability.}
		\label{fig:srbk_bsr_row_nonzero}
	\end{figure}

\begin{figure}[htbp]
		\centering
		\begin{subfigure}[b]{0.23\textwidth}
			\centering
			\includegraphics[width=\columnwidth, trim=50 80 400 0, clip]{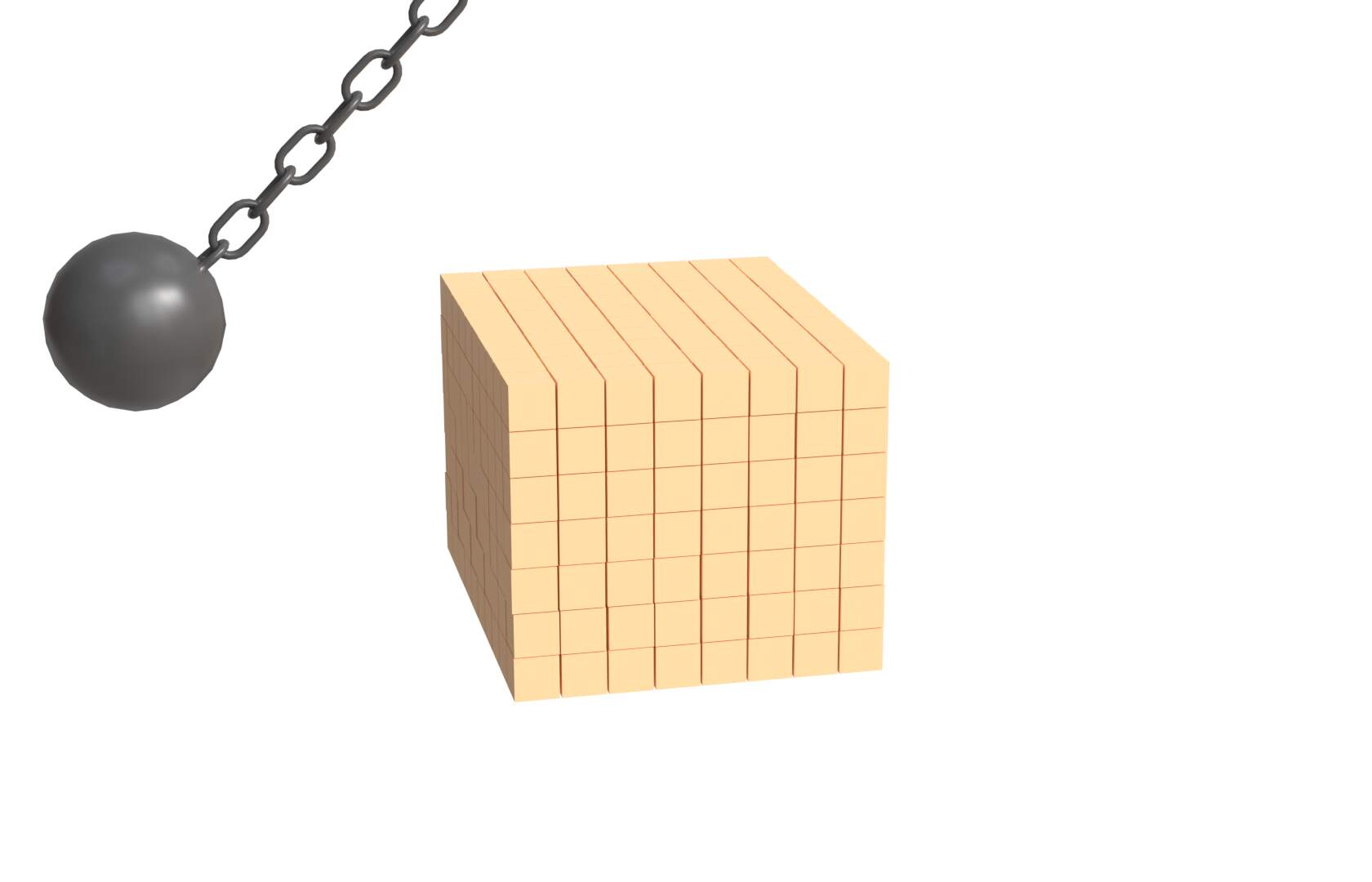}
			%\subcaption{\label{fig:memLocking_5e6}}
		\end{subfigure}
		\begin{subfigure}[b]{0.23\textwidth}
			\centering
			\includegraphics[width=\columnwidth, trim=50 80 400 0, clip]{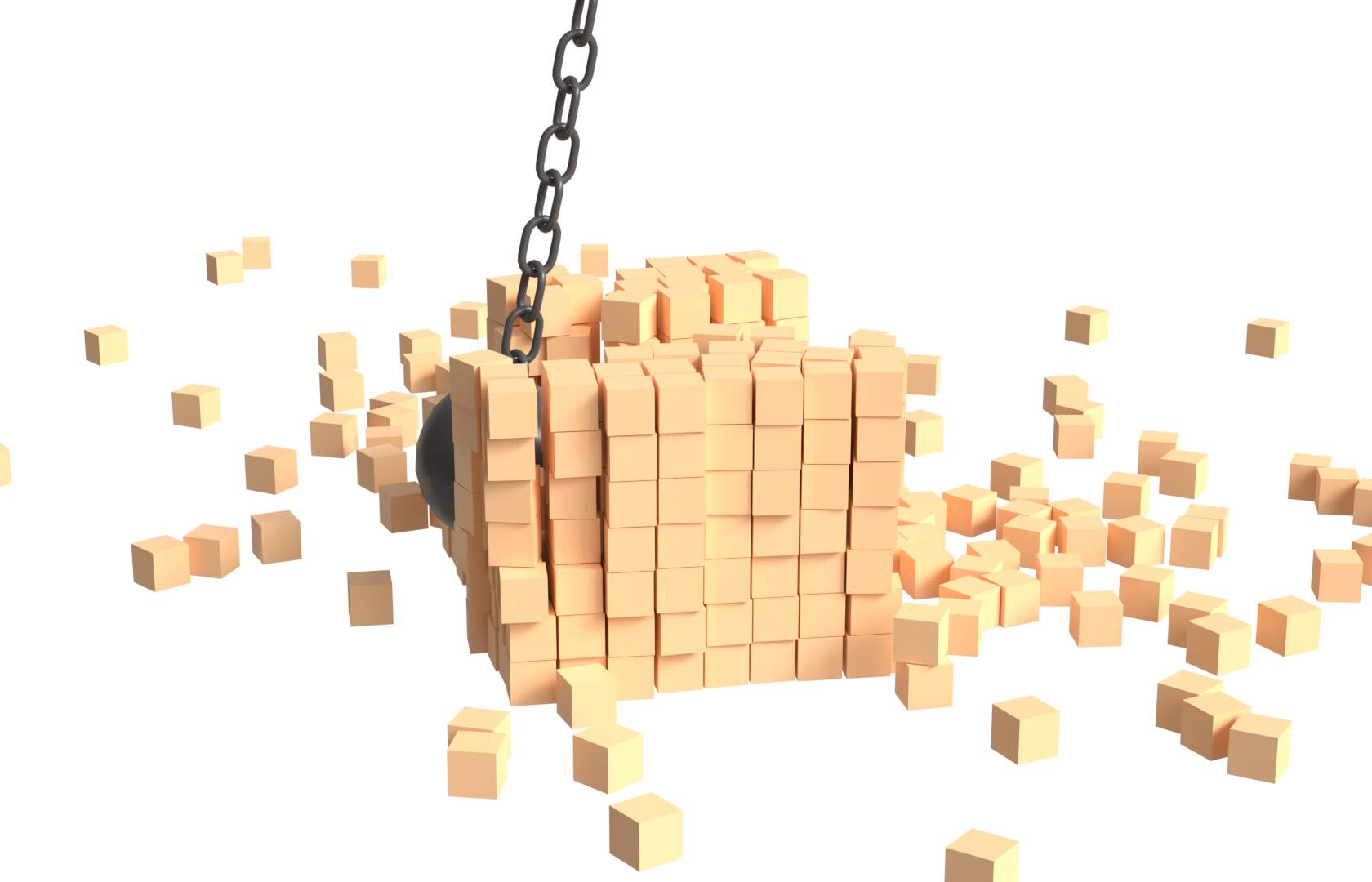}
			%\subcaption{\label{fig:memLocking_5e5}}
		\end{subfigure}
		\caption{\label{fig:wrecking_ball} \textbf{Wrecking Ball.} This example compares GPU ABD~\cite{abd} and our method using only stiff affine bodies. The simulation parameters are $\hat{d} = 10^{-3}l$, $\rho = 1000$ kg/m$^3$, $\Delta t = 0.01$ s, $\mu = 0.05$, static friction tolerance $10^{-3}l$, and Newton tolerance $10^{-2}l\Delta t$. Our method achieves a 2.6$\times$ speedup.}
	\end{figure}

\begin{figure}[htbp]
		\centering
		%\begin{subfigure}[b]{0.2\textwidth}
		%		\centering
		%		\includegraphics[width=\columnwidth, trim=500 50 500 600, clip]{figs/sqball/0041.jpg}
		% 	\end{subfigure}
		\begin{subfigure}[b]{0.225\textwidth}
			\centering
			\includegraphics[width=\columnwidth, trim=500 200 500 800, clip]{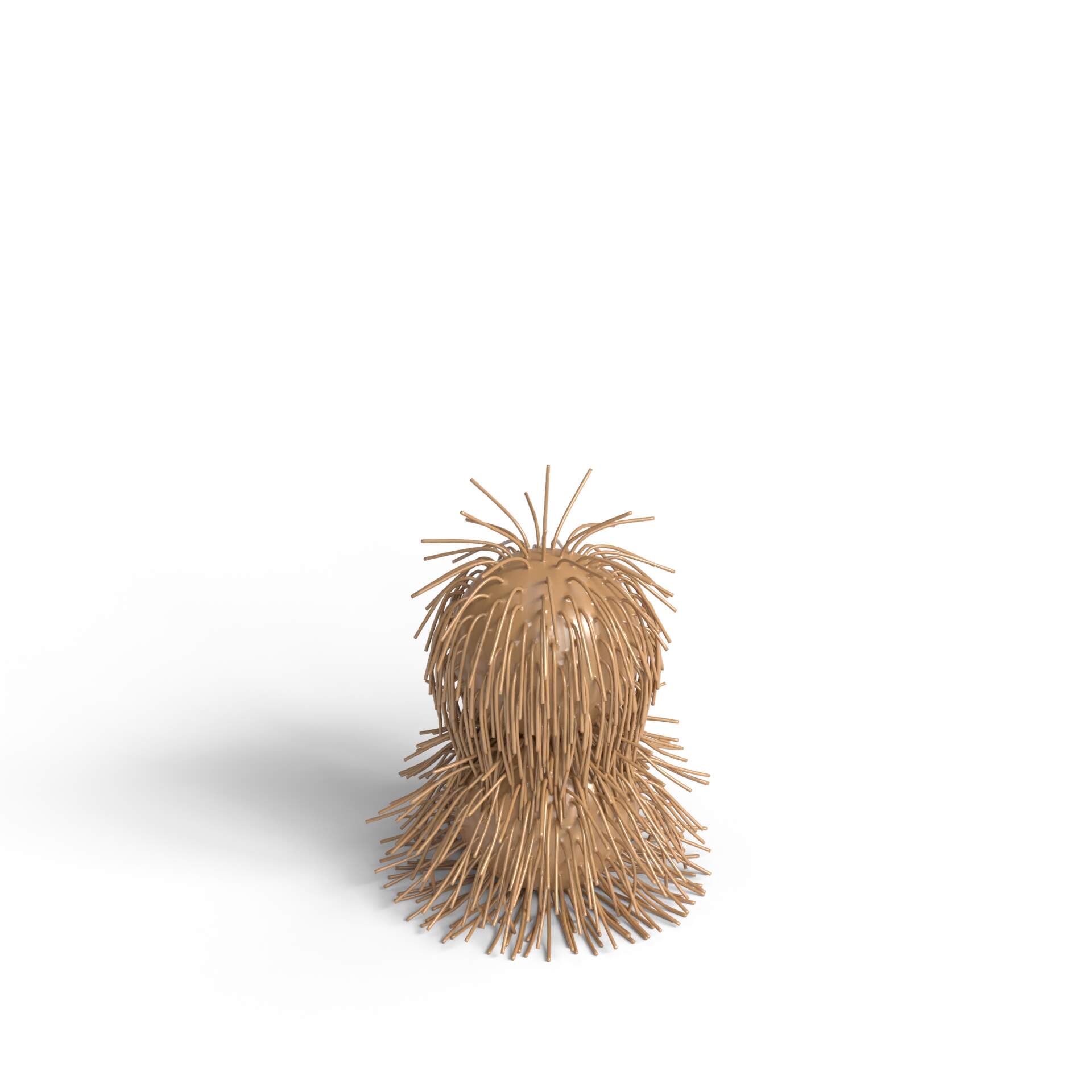}
		\end{subfigure}
		\begin{subfigure}[b]{0.225\textwidth}
			\centering
			\includegraphics[width=\columnwidth, trim=500 200 500 800, clip]{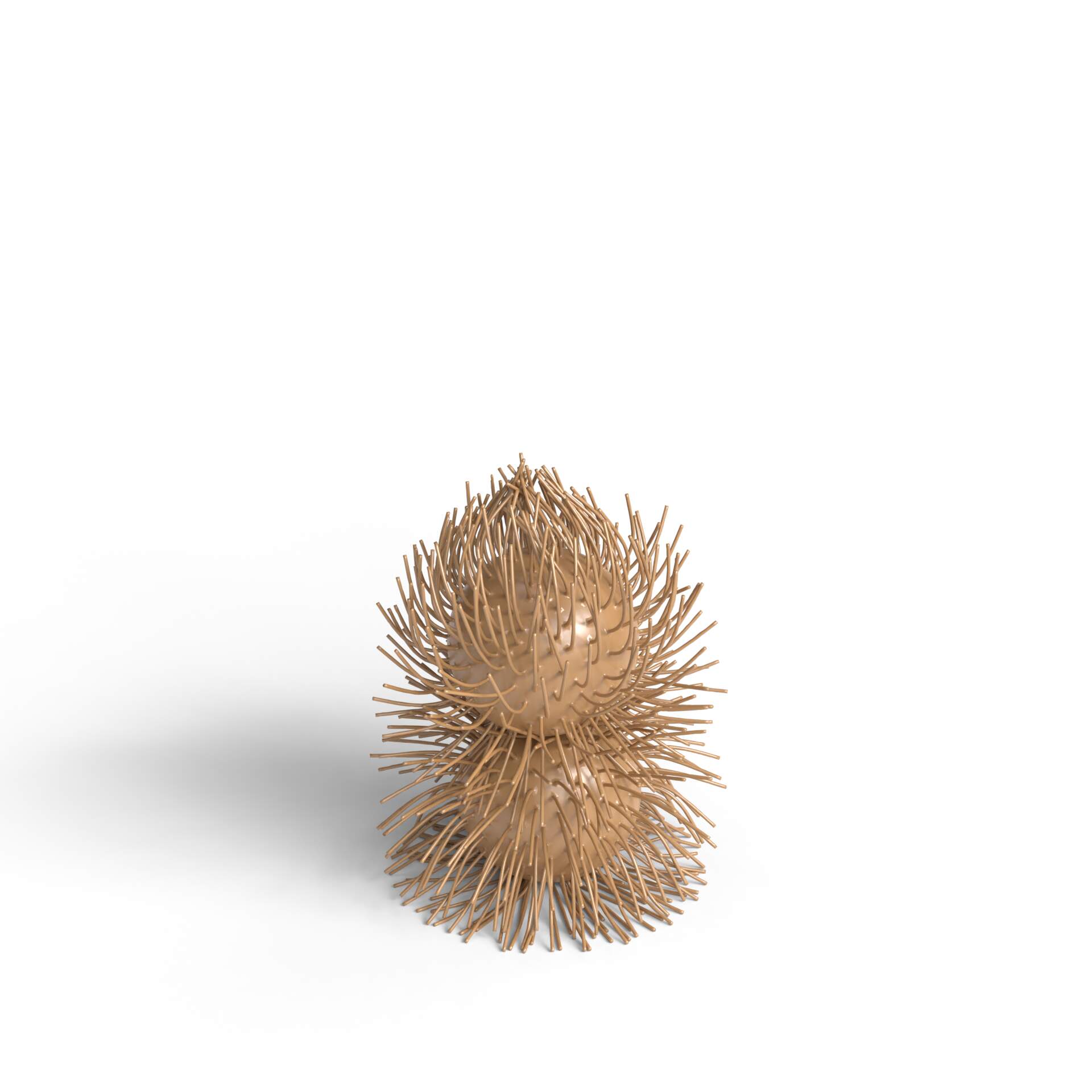}
		\end{subfigure}
		%\begin{subfigure}[b]{0.2\textwidth}
		%		\centering
		%		\includegraphics[width=\columnwidth, trim=500 50 500 800, clip]{figs/sqball/0164.jpg}
		% 	\end{subfigure}
		\caption{\textbf{Furry ball.}This example compares GIPC \cite{gipc} and our method. The simulation parameters are $\hat{d} = 4\times10^{-4}l$, $\rho = 1000$ kg/m$^3$,  $Y = 1 MPa$, $\nu = 0.49$, $\Delta t = 0.01$ s, and $\varepsilon_d = 10^{-2}l\Delta t$. Our method achieves a 4.42$\times$ speedup. } \label{fig:sqball}
	\end{figure}

\begin{figure}[htbp]
		\centering
		%\begin{subfigure}[b]{0.2\textwidth}
		%		\centering
		%		\includegraphics[width=\columnwidth, trim=500 50 500 600, clip]{figs/sqball/0041.jpg}
		% 	\end{subfigure}
		\begin{subfigure}[b]{0.225\textwidth}
			\centering
			\includegraphics[width=\columnwidth, trim=200 50 50 600, clip]{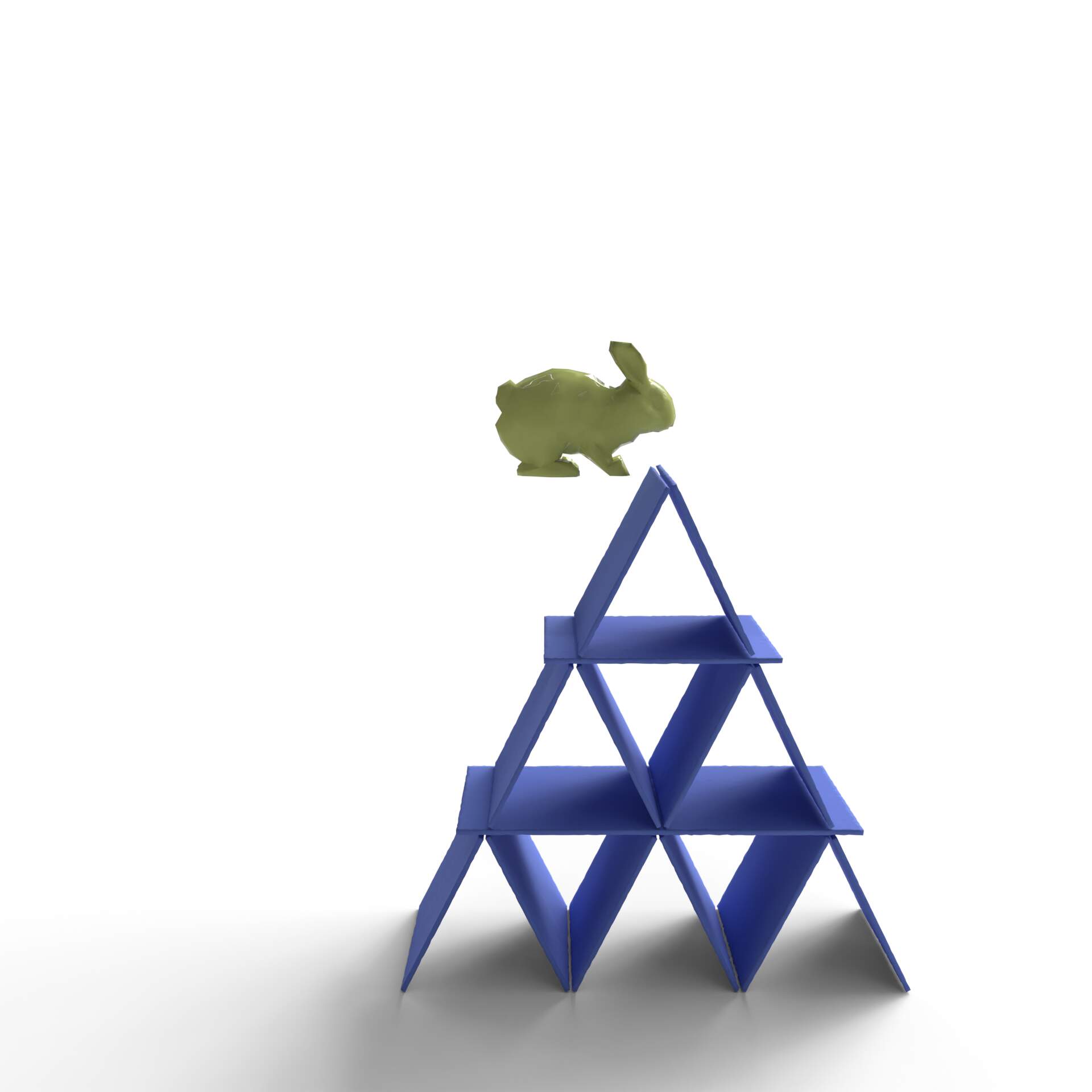}
		\end{subfigure}
		\begin{subfigure}[b]{0.225\textwidth}
			\centering
			\includegraphics[width=\columnwidth, trim=200 50 50 600, clip]{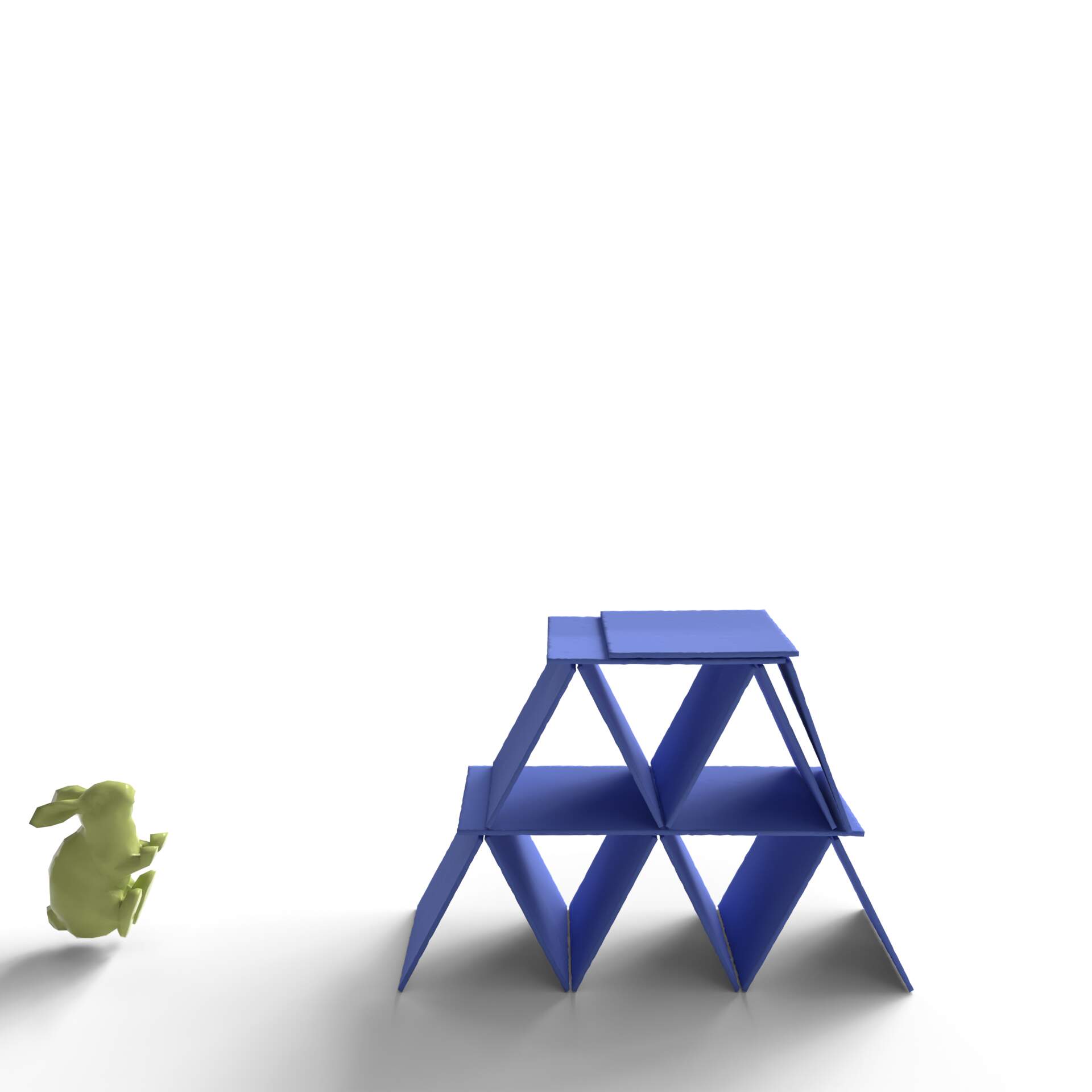}
		\end{subfigure}
		%\begin{subfigure}[b]{0.2\textwidth}
		%		\centering
		%		\includegraphics[width=\columnwidth, trim=500 50 500 800, clip]{figs/sqball/0164.jpg}
		% 	\end{subfigure}
		\caption{\textbf{Board House.} This example compares GIPC \cite{gipc} with our method. Simulation parameters are $\hat{d} = 1\times10^{-4}l$, $\rho = 1000 \, \text{kg/m}^3$, $Y = 0.1 \, \text{GPa}$, $\nu = 0.49$, $\Delta t = 0.01 \, \text{s}$, and $\varepsilon_d = 10^{-2}l\Delta t$. Our method achieves a 4.4$\times$ speedup using FEM and 6.24$\times$ using ABD.} \label{fig:cardhouse}
	\end{figure}

	To evaluate the scalability of our method, we compare its speedup factor to BSR on matrices with varying numbers of non-zero $3\times3$ blocks per row. As shown in \autoref{fig:srbk_bsr_row_nonzero}, the speedup factor increases from $0.93\times$ to $1.9\times$ as the number of non-zero blocks approaches \(2^3\) from 0. For average non-zero block counts between \(2^3\) and \(2^6\) -- the typical range for FEM global systems -- the speedup factor fluctuates between \(1.8\times\) and \(2\times\), which demonstrates \modify{improved} scalability.

	\paragraph{Comparison with GPU ABD} To evaluate the efficiency of our framework in simulating stiff affine bodies, we replicate the wrecking ball example (\autoref{fig:wrecking_ball}) from ABD~\cite{abd}. In our setup, the sphere and the boxes are assigned the same density, ensuring that the number of contacts remains large as less boxes will be punched out. This configuration makes the problem more challenging for the simulation framework.
	Our method utilizes PCG to solve the linear systems, using the $12 \times 12$ diagonal blocks as the preconditioner. In this scenario, ABD requires 37.2 ms per Newton iteration, whereas our method only takes 14.4 ms, resulting in a 2.6$\times$ speedup.
	% \todo{@kemeng: ours(67.5s),abd(174.5s)}
	% In this case, the time costs for the major processes in ABD are: 14.48 ms per Newton iteration for hessian and gradient construction, 17.01 ms per Newton iteration for the linear solver, 4.78 ms per Newton iteration for Continuous Collision Detection (CCD), and 0.9 ms per Newton iteration for the line search, resulting in a total of 37.17 ms. In contrast, the corresponding time costs in our simulation framework are 4 ms, 3.5 ms, 1.5 ms, 4 ms, and 13 ms in total, which exhibits a 3$\times$ speedup.
	% \mc{better show the timing in a table or pi chart. if not making the example harder, what will happen?}

	\begin{table*}
		\caption{\label{tab:Overall_comparison} \textbf{Simulation statistics for our examples.} 
			Columns show time for energy gradient and Hessian computation (\texttt{Hess}), linear system solves (\texttt{LSolve}), CCD (\texttt{CCD}), backtracking line search (\texttt{LineS}), miscellaneous tasks (\texttt{Misc}), and the entire simulation (\texttt{TimeTot}), followed by iteration counts, and finally, speedup relative to GIPC. All times are in seconds. Here, we compare our method with GIPC \cite{gipc}, ZeMa \cite{zema}, and the barrier-based SL \cite{cipc}. We also show how each of \textit{our innovation} contributes to the overall speedup.}
		\resizebox{0.95\textwidth{}}{!}{
			\begin{tabular}{r|r|r|cccccccccc}
				\toprule
				& \modify{\# verts, tets, faces} &Framework & \texttt{Hess} & \texttt{LSolver} &\texttt{CCD} &\texttt{LineS} & \texttt{Misc} & \texttt{TimeTot} & \texttt{\#cg} & \modify{\texttt{avg. \#Newton}} & \texttt{avg. \#cg}& \textbf{Speed Up} \\ 
				
				\midrule
				\multirow{3}{*}{\autoref{fig:octopus}}& 114K, 558K, 69K &{GIPC}  & 63  & 1134 &5& 13 & 6 & 1221&9.41e5&12.78 & 409&1.00$\times$\\& 114K, 558K, 69K
				&{Ours (SRBK)}& {45}  &{524}& {5} & {13}&{6} &{{593}} & {9.51e5} & {12.83} & {412}&2.06$\times$ \\& 114K, 558K, 69K
				&{Ours (CEMAS+SRBK)}& {45}  & {238}& {5} & {13}&6 &\underline{\textbf{307}} & {5.20e5} & {12.83} & \textbf{225}&3.98$\times$ \\
				
				\midrule
				\multirow{2}{*}{\autoref{fig:memLocking}}& 109K, 80K, 200K &{GIPC}  & 106  & 606 &20& 54 & 6 & 792&2.34e5&24.31 & 60&1.00$\times$\\& 109K, 80K, 200K
				&{Ours (SRBK)}& {75}  & {489}& {20} & {55}&{6} &{{645}} & {2.37e5} & {24.13} & {61}&1.23$\times$ \\& 109K, 80K, 200K
				&{Ours (CEMAS+SRBK)}& {75}  & {184}& {20} & {55}&6 &{{340}} & {2.61e5} & {24.69} & {66}&2.33$\times$ \\& 109K, 80K, 200K
				&{CIPC (CEMAS+SRBK)}& {104}  & {202}& {21} & {91}&6 &{{424}} & {2.89e5} & {26.25} & {69}& N/A \\& 100K, 0, 200K
				&{Ours (ABD+CEMAS+SRBK)}& {67}  & {145}& {19} & {51}&6 &\underline{\textbf{288}} & {2.06e5} & {23.50} & \textbf{55}&2.75$\times$ \\

				\midrule
				\multirow{4}{*}{\autoref{fig:bunny2}} & 203K, 1122K, 50K & {GIPC}  & 90  & 1481 &4& 14 & 7 & 1596&6.23e5&9.78 & 354& 1.00$\times$\\& 114K, 561K, 50K
				&{ZeMa}  & 42 & 54810 & 2.7 & 9 & 7 & 54871 & -&9.67&-&0.03$\times$  \\& 203K, 1122K, 50K
				&{Ours (SRBK)}& 63& 569& 3 & 14  & 7 & 656 &6.23e5 & 9.78& 354 & 2.43$\times$ \\& 203K, 1122K, 50K
				&{Ours (CEMAS+SRBK)}& 63& 360& 3 & 14  & 7 & 447 &5.30e5 & 9.78& 301 &3.57$\times$ \\& 114K, 561K, 50K
				&{Ours (ABD+CEMAS+SRBK)}& {37}  & {104}& {2.7} & {8.5}&7 &\underline{\textbf{159}} & {1.73e5} & {9.61} & \textbf{100} & 10.03$\times$\\

				\midrule
				\multirow{2}{*}{\autoref{fig:sqball}}& 123K, 330K, 211K &{GIPC}  & 137  & 3910 &35& 83 & 8 & 4173&4.31e6&39.35 & 548&1.00$\times$\\& 123K, 330K, 211K
				&{Ours (SRBK)}& {91}  & {2236}& {35} & {83}&{8} &{{2453}} & {4.38e6} & {39.65} & {552}&1.70$\times$ \\& 123K, 330K, 211K
				&{Ours (CEMAS+SRBK)}& {90}  & {730}& {35} & {82}&8 &\underline{\textbf{945}} & {1.81e6} & {39.25} & \textbf{231}&4.42$\times$ \\
				
				\midrule
				\multirow{2}{*}{\autoref{fig:cardhouse}}& 79K, 298K, 110K &{GIPC}  & 24  & 1080 &5& 13 & 8 & 1130&1.42e6&8.10 & 877&$1.00\times$\\& 79K, 298K, 110K
				&{Ours (SRBK)}& {16}  & {553}& {5} & {13}&{8} &{{595}} & {1.42e6} & {8.10} & {877}&1.90$\times$ \\& 79K, 298K, 110K
				&{Ours (CEMAS+SRBK)}& {16}  & {216}& {5} & {12}&8 &{{257}} & {7.53e5} & {8.10} & {465}&4.40$\times$ \\& 78K, 290K, 110K
				&{Ours (ABD+CEMAS+SRBK)}& {14}  & {148}& {4} & {8}&7 &\underline{\textbf{181}} & {4.50e5} & {7.05} & \textbf{319}&6.24$\times$ \\
				
				\midrule
				\multirow{2}{*}{\autoref{fig:mat-cloth-twist}}& 70K, 241K, 106K &{GIPC}  & 692  & 10686 &114& 637 & 14 & 12143&7.30e6&15.80 & 462&1.00$\times$\\& 70K, 241K, 106K
				&{Ours (SRBK)}& {460}  & {4511}& {113} & {594}&14 &{{5692}} & {7.58e6} & {15.80} & {480} &2.13$\times$\\&70K, 241K, 106K
				&{Ours (CEMAS+SRBK)}& {452}  & {1547}& {112} & {597}&14 &{{2722}} & {3.37e6} & {15.70} & {214} &4.46$\times$\\& 53K, 133K, 106K
				&{Ours (ABD+CEMAS+SRBK)}& {407}  & {696}& {103} & {535}&{12} &\underline{\textbf{1753}} & {1.48e6} & {14.60} & \textbf{101}&6.93$\times$ \\
				
				\midrule
				\multirow{2}{*}{\autoref{fig:box_pipe}}& 105K, 10K, 202K &{GIPC}  & 210  & 2473 &54& 211 & 9 & 2957&3.10e6&37.72 & 329&1.00$\times$\\& 105K, 10K, 202K
				&{Ours (SRBK)}& {142}  & {762}& {51} & {205}&{9} &{{1069}} & {3.13e6} & 37.08 & {338}&2.76$\times$ \\& 105K, 10K, 202K
				&{Ours (CEMAS+SRBK)}& {145}  & {451}& {52} & {207}&{9} &{{864}} & {8.21e5} & 38.00 & \textbf{86}&3.42$\times$ \\& 105K, 10K, 202K
				&{CIPC (CEMAS+SRBK)}& {219}  & {812}& {59} & {329}&{10} &{{1429}} & {1.75e6} & 38.76 & {180}& N/A \\& 105K, 0, 202K
				&{Ours (ABD+CEMAS+SRBK)}& {161}  & {468}& {47} & {178}&9 &\underline{\textbf{863}} & {7.78e5} & 34.04 & {91}&3.42$\times$ \\
				\bottomrule
			\end{tabular}
		}
	\end{table*}

	\paragraph{Comparison with GIPC and ZeMa on hybrid scenarios.}
	Here, we compare our method with GIPC \cite{gipc} and ZeMa \cite{zema} on the stiff and soft bunnies example (\autoref{fig:bunny2}). In this simulation, the Young's modulus is set to $10^6$ Pa for the bottom bunny and $10^8$ Pa for the top bunny. Our framework can treat the top bunny as a stiff affine body and the bottom bunny with FEM, while GIPC needs to simulate both using FEM. Results in \autoref{tab:Overall_comparison} indicate that our coupling framework achieves a 10$\times$ overall speedup compared to GIPC.
	To analyze the sources of this acceleration, we applied our contributions incrementally. ‘Ours (SRBK)' uses only our SRBK method, matching other components to GIPC, which already yields a $2.43\times$ speedup with more efficient Hessian assembly and SpMV. ‘Ours (CEMAS+SRBK)’ further adds connectivity-enhanced (CE) MAS preconditioner to ‘Ours (SRBK)', retaining other components, and shows an additional $1.47\times$ overall speedup due to faster PCG convergence enabled by CEMAS, resulting in a $3.57\times$ improvement compared to GIPC. Note that in both ‘Ours (SRBK)' and ‘Ours (CEMAS+SRBK)' cases, we simulate both bunnies using FEM.
	ZeMa performs the worst in this case, as direct solvers are not efficient for simulations with a large number of DOFs. Thus, for general IPC simulations, GIPC remains the best alternative. Therefore, in subsequent experiments, we compare our methods only against GIPC.

	We further compare our method with GIPC on the examples shown in \autoref{fig:octopus}, \autoref{fig:memLocking}, \autoref{fig:sqball}, \autoref{fig:cardhouse}, \autoref{fig:box_pipe}, and \autoref{fig:mat-cloth-twist}, with statistics reported in \autoref{tab:Overall_comparison}. Generally, ‘Ours (SRBK)' achieves a $1.23\times$–$2.76\times$ speedup over GIPC. Adding our CEMAS preconditioner further improves performance, providing an additional $1.24\times$–$2.6\times$ speedup as shown by ‘Ours (CEMAS+SRBK)', leading to an overall improvement of $2.33\times$–$4.46\times$ compared to GIPC. Notably, this is achieved without approximating stiff materials as affine bodies, which can further boost performance up to $2.75\times$–$10\times$ faster than GIPC.
	In \autoref{fig:memLocking}, comparing ‘Ours (SRBK)' and ‘Ours (CEMAS+SRBK)' reveals that while CEMAS does not improve PCG convergence due to similar ordering with Morton code sorting in GIPC, it significantly reduces PCG solver time due to lower construction and preconditioning costs. Here, PCG costs represent a smaller portion of total timing than in other examples due to fewer PCG iterations required, resulting in a modest $1.23\times$ speedup with SRBK alone.
	We will discuss \autoref{fig:mat-cloth-twist} and \autoref{fig:box_pipe} in detail in the next subsection on stress tests.

	\subsection{Stress Tests}\label{sec:stress_tests}
	
\begin{figure}
		\centering
		%\begin{subfigure}[b]{0.23\textwidth}
		%\centering
		\includegraphics[width=0.59\columnwidth, trim=0 300 0 300, clip]{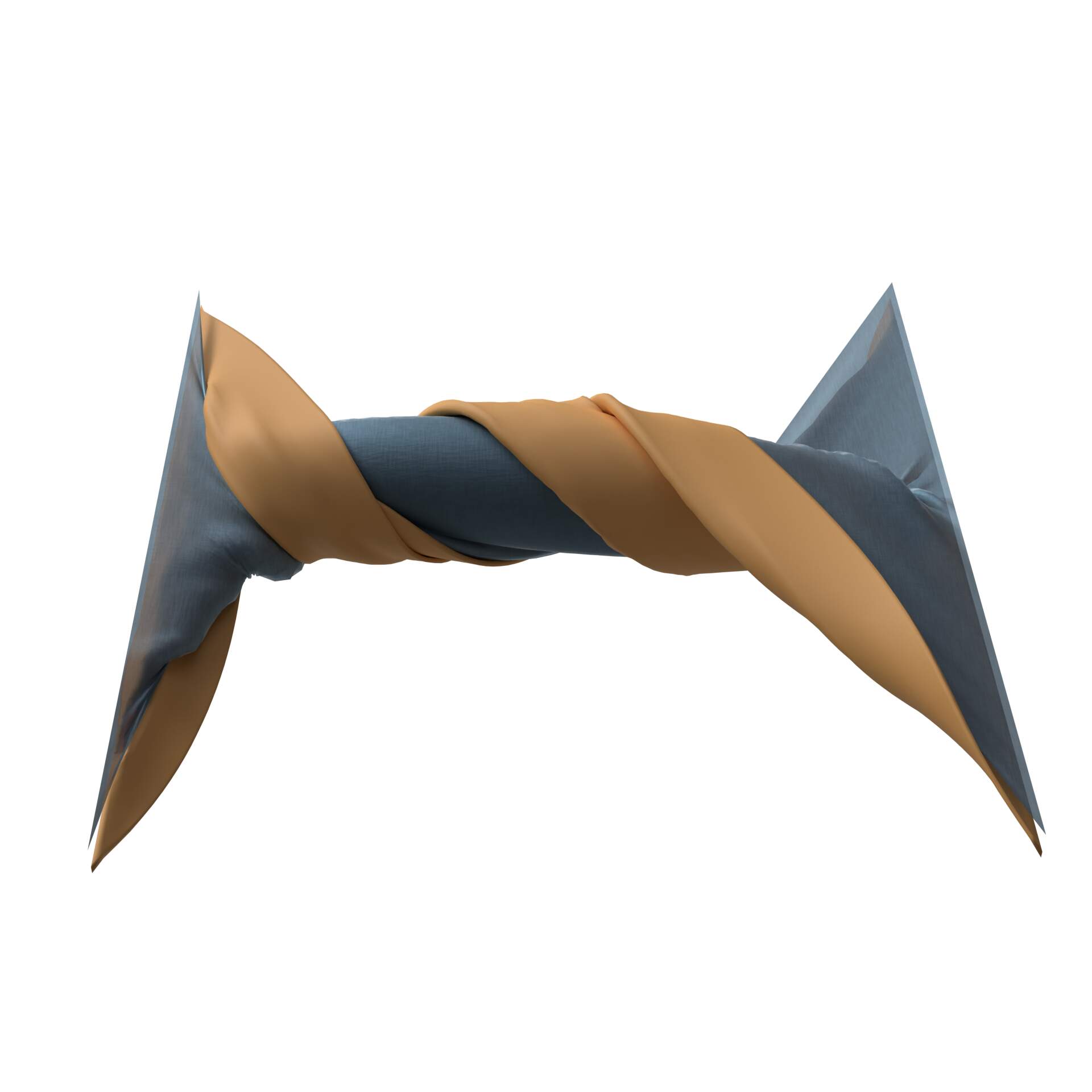}
		%\subcaption{\label{fig:memLocking_5e6}}
		%\end{subfigure}
		%       \begin{subfigure}[b]{0.35\textwidth}
		% \centering
		\includegraphics[width=0.39\columnwidth, trim=300 250 300 250, clip]{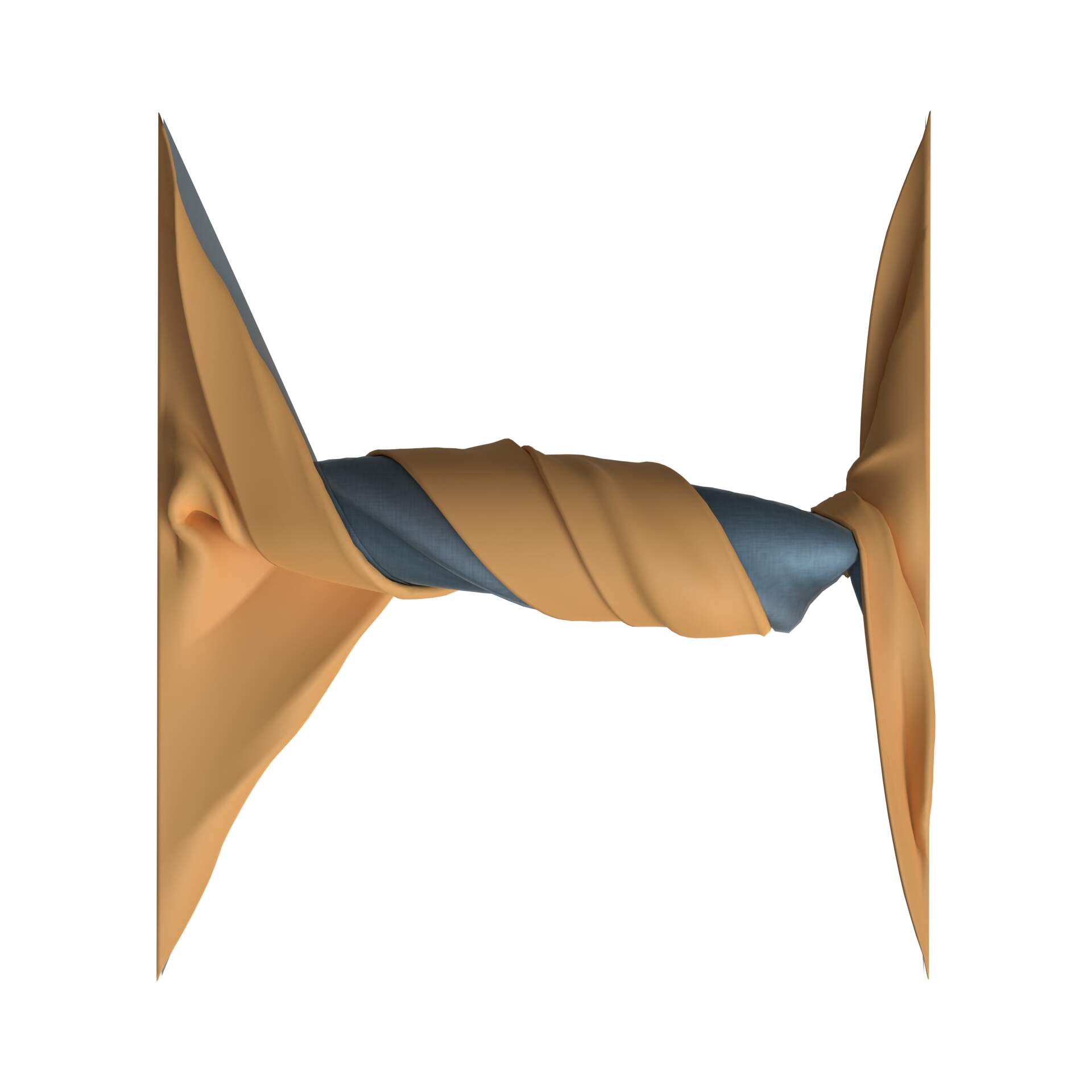}
		%\subcaption{\label{fig:memLocking_5e5}}
		% \end{subfigure}
		\caption{\textbf{Mat-cloth twist.} Stress test on extreme deformation with cloth, volumetric mat, and a stiff affine ellipsoid (inside). Our method simulates this stiff challenging scenario efficiently with 1.8s per time step ($\Delta t=0.01s$), achieving a $6.9\times$ speedup compared to GIPC.}
		\label{fig:mat-cloth-twist}
	\end{figure}

	\paragraph{Mat-cloth twist}
	To further validate the robustness and efficiency of our framework, we perform a simulation coupling volumetric solids, cloth, and stiff affine bodies, as shown in \autoref{fig:mat-cloth-twist}. The Young's modulus of the cloth and the mat are $50KPa$ and $0.5MPa$ respectively. They both have a Poisson's ratio of 0.49 and the friction coefficient of 0.4. We twist the cloth and mat by rotating their left and right sides for 2 rounds in 10 seconds, wrapping a stiff affine ellipsoid between them, creating extreme stretching, bending, and frictional contacts. The maximum number of contact pairs is 567K, averaging 351K per time step. The full simulation completed in 29 minutes for 1000 time steps (see the detailed time breakdown in \autoref{tab:Overall_comparison}). Our framework exhibit a 6.93$\times$ speedup compared to GIPC in this case.
	
\begin{figure}[htbp]
		\centering
		\begin{subfigure}[b]{0.15\textwidth}
			\centering
			\includegraphics[width=\columnwidth, trim=100 80 20 70, clip]{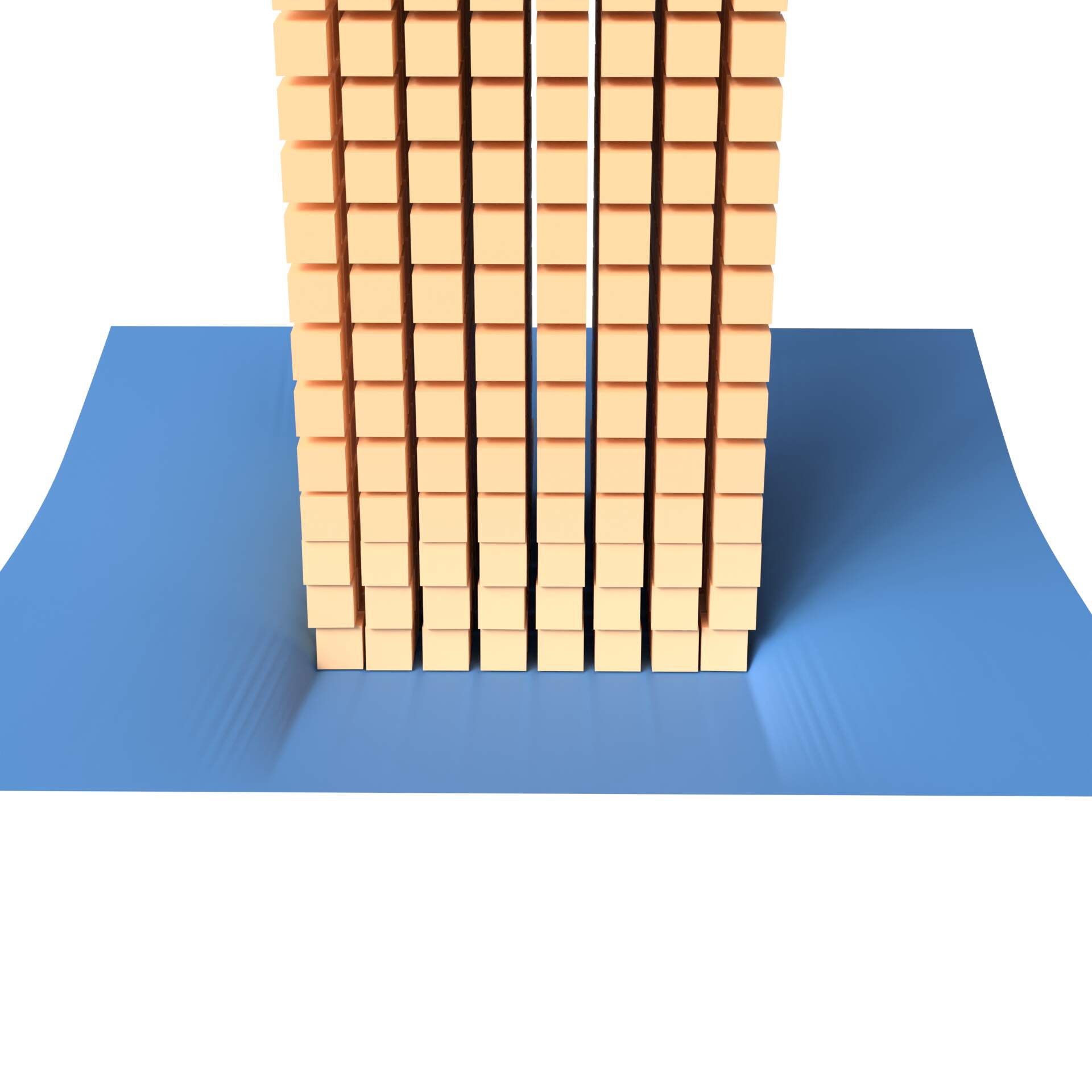}
		\end{subfigure}
		\begin{subfigure}[b]{0.15\textwidth}
			\centering
			\includegraphics[width=\columnwidth, trim=100 80 20 70, clip]{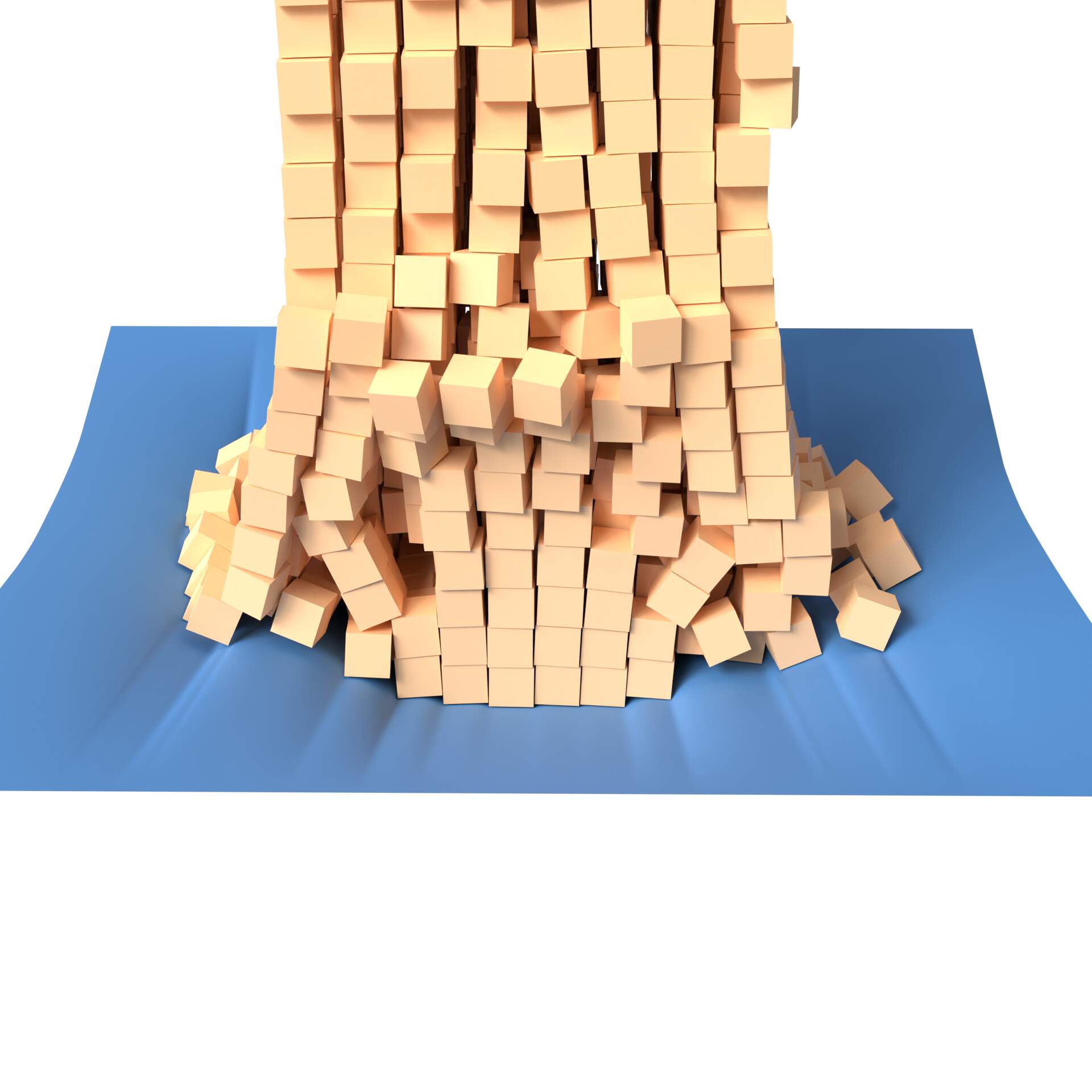}
		\end{subfigure}
		\begin{subfigure}[b]{0.15\textwidth}
			\centering
			\includegraphics[width=\columnwidth, trim=100 80 20 70, clip]{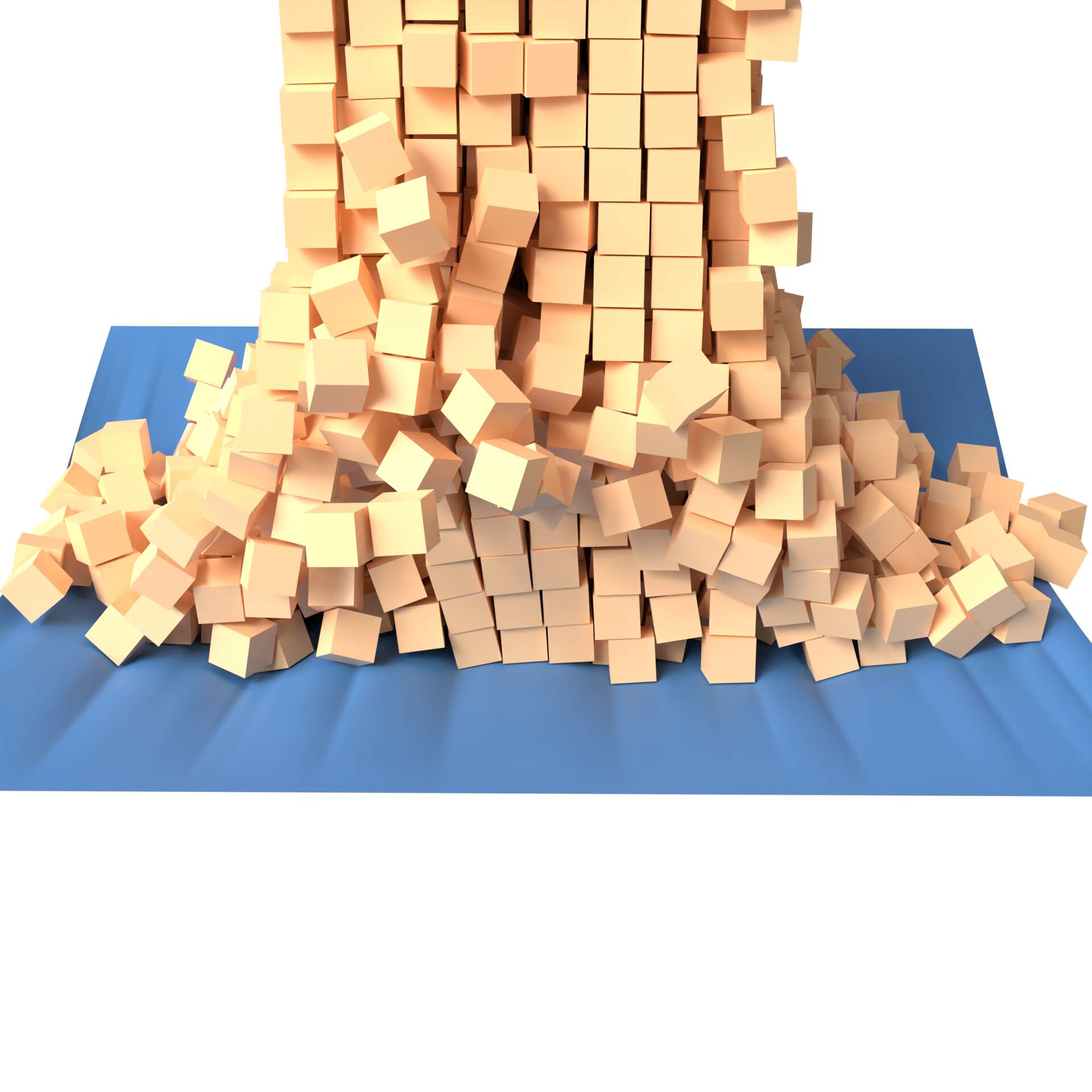}
		\end{subfigure}
		\begin{subfigure}[b]{0.15\textwidth}
			\centering
			\includegraphics[width=\columnwidth, trim=100 80 20 70, clip]{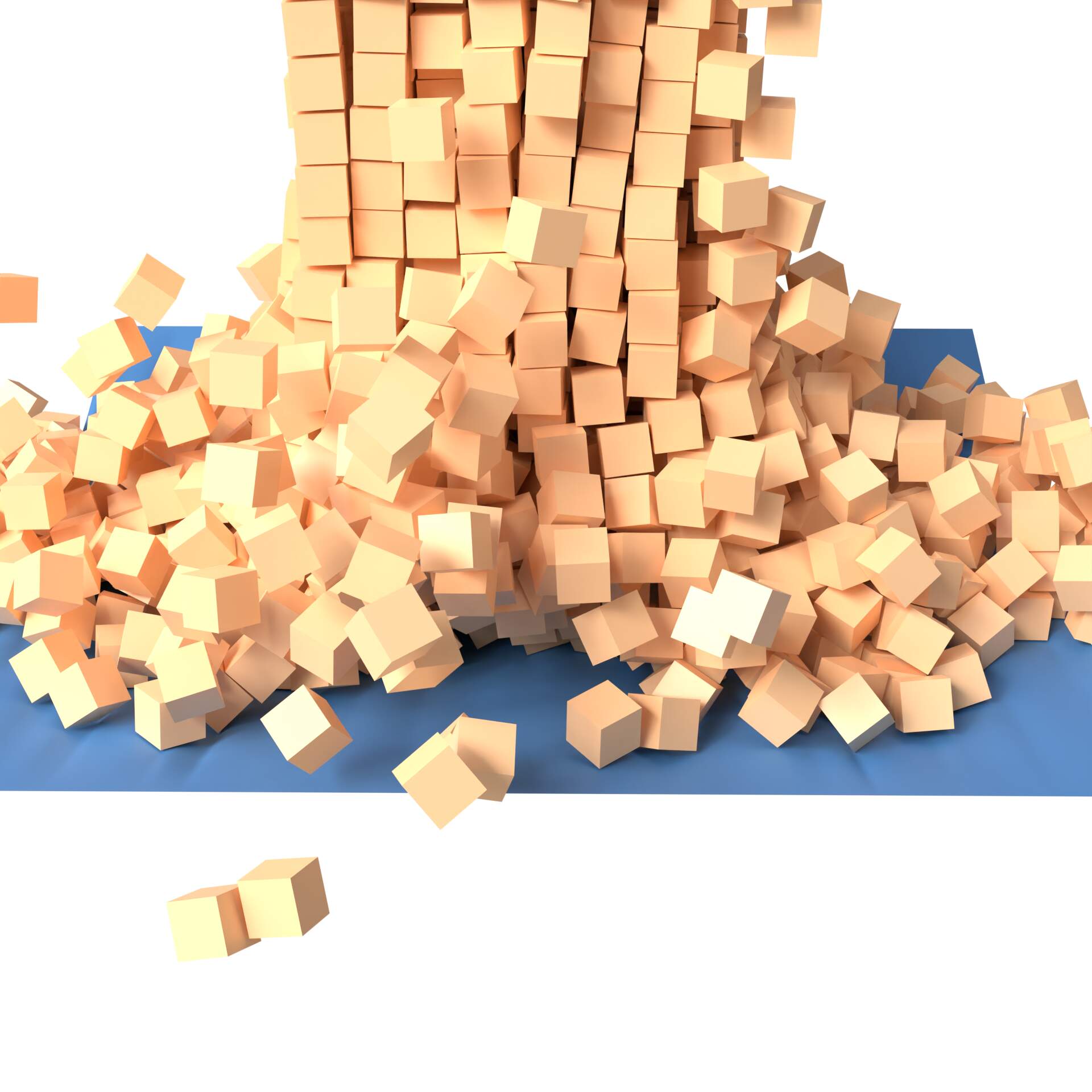}
		\end{subfigure}
		\begin{subfigure}[b]{0.15\textwidth}
			\centering
			\includegraphics[width=\columnwidth, trim=100 80 20 70, clip]{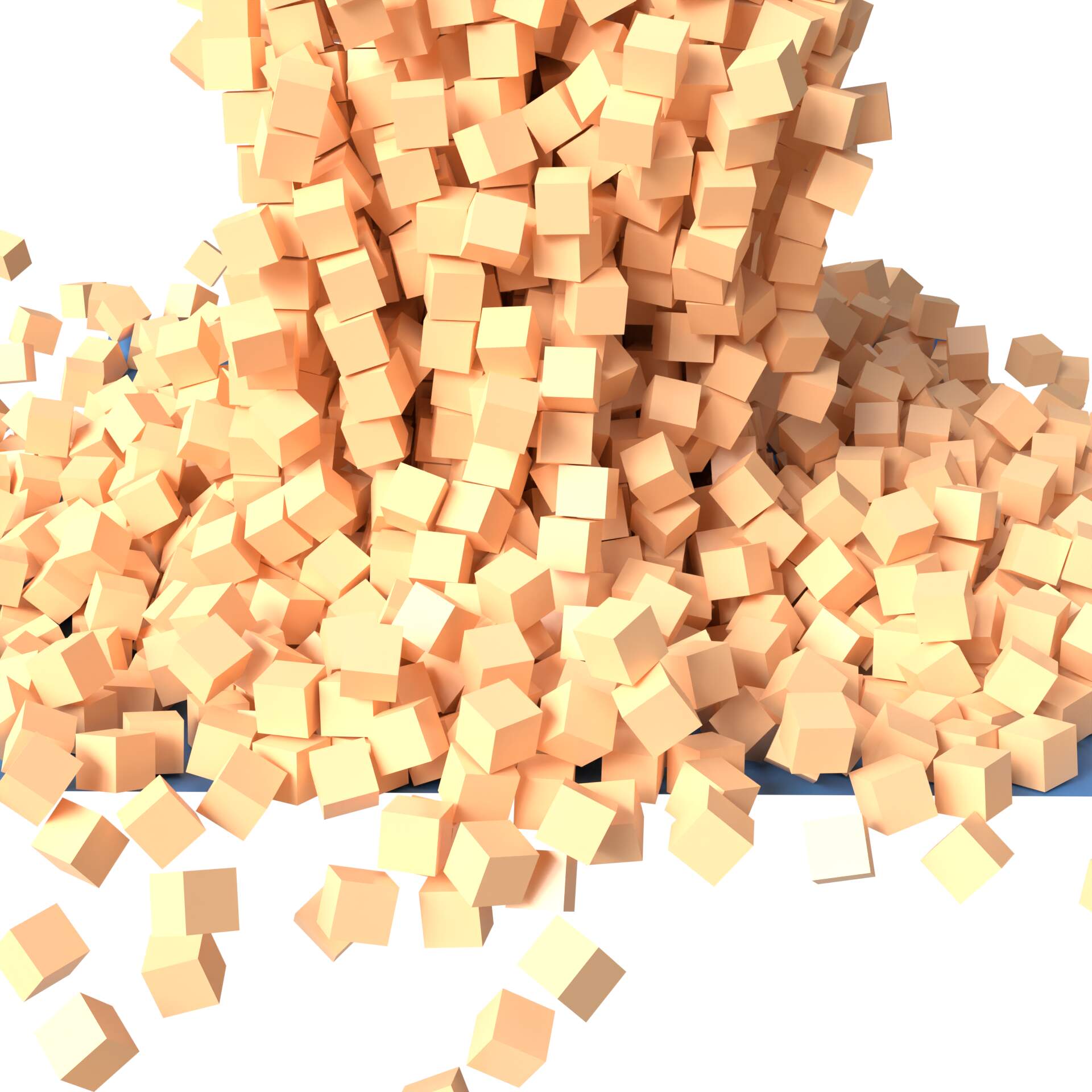}
		\end{subfigure}
		\begin{subfigure}[b]{0.15\textwidth}
			\centering
			\includegraphics[width=\columnwidth, trim=100 80 20 70, clip]{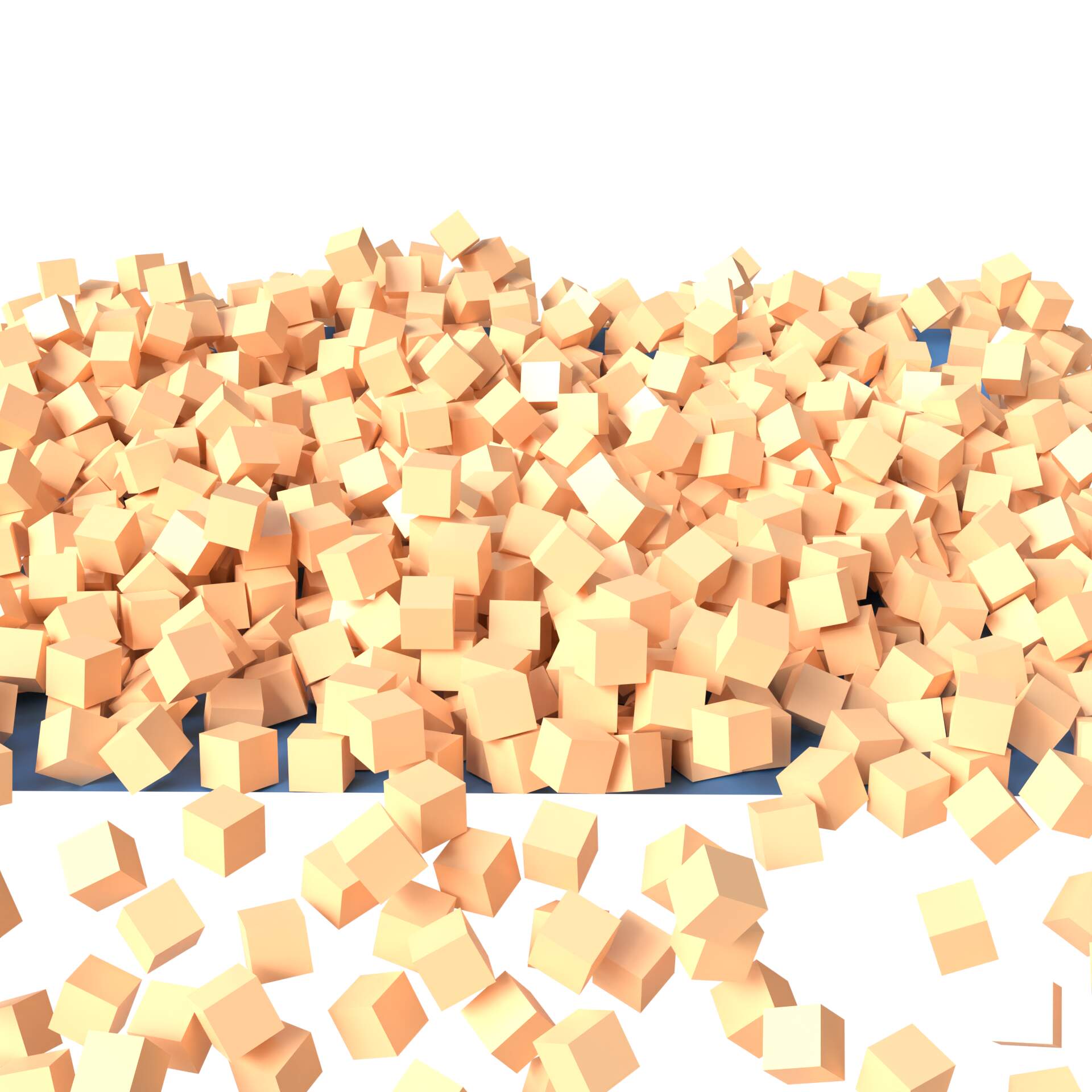}
		\end{subfigure}
		\caption{\label{fig:box_pipe}\textbf{Box pile.} 
			$8\times30\times8$ boxes dropped onto a high-resolution cloth with 100K nodes, showing our capability of efficiently simulating high-speed impacts, rapidly changing contacts, and fine wrinkling details in the same scene at 3.45s per time step.
		}
	\end{figure}

	\paragraph{Box Pile} 
	We then create a challenging example to stress test multibody contact and strain limiting (SL) by dropping an $8\times30\times8$ pile of $3.32 \, \text{m}^3$ boxes with $\rho=1000 \, \text{kg/m}^3$, Young's modulus of $0.1 \, \text{GPa}$, and Poisson's ratio of $0.49$ onto a $9 \, \text{m}^2$ thin shell with $100K$ vertices, $\rho = 200 \, \text{kg/m}^3$, and a thickness of $1 \, \text{mm}$, as shown in \autoref{fig:box_pipe}. The boxes weigh approximately $4 \, \text{t}$ in total; thus, with a $5 \, \text{MPa}$ stiffness for our SL energy, this is similar to dropping a sedan onto a bedsheet, which would result in fracture and is thus \textit{out of the scope} of our simulator. Therefore, \modify{\textit{for stress testing purpose}}, we raise our SL energy stiffness to $50 \, \text{GPa}$ to hold the heavy box pile on the thin shell. 
	Despite the high SL stiffness, our method robustly captures realistic wrinkling behavior. The fast-changing contact pairs, caused by the increasing falling speed of the boxes, significantly challenge the nonlinear solver, resulting in many Newton iterations per time step. Our method handles this challenging simulation efficiently, achieving $3.45 \, \text{s}$ per time step ($\Delta t = 0.01 \, \text{s}$). \modify{Since our method remains stable even with such high stiffness for the inexact strain-limiting energy, an interesting future work direction would be to dynamically adapt the stiffness to ensure constraint satisfaction, similar to \citet{cubicBarrier}.}
	
	We also compare our method with the SL method in CIPC, listed as ‘CIPC (SL+CEMAS+SRBK)' in \autoref{tab:Overall_comparison}. Due to the high-speed box collisions, CIPC’s SL requires a sufficiently large barrier stiffness to prevent the cloth from stretching near the SL limit; otherwise, \modify{strains that are extremely close to the limit will be needed for the logarithmic function to generate sufficiently large strain-limiting forces, where the numerical errors (vanishing displacement problem) often lead to optimization failure \cite{gipc}}. Here, we set the barrier SL stiffness to $0.1 \, \text{MPa}$. However, this higher stiffness results in more expensive linear solves due to worse system conditioning, and CIPC’s SL further incurs additional time from numerical eigendecomposition, singular value decomposition for local Hessian calculations, and backtracking line search to ensure SL feasibility, making it slower than our method. In this case, we achieve an approximately 30\% reduction in the line search time cost.
	
	In this example, we observe significantly faster PCG convergence with a diagonal preconditioner than with the MAS preconditioner in GIPC. Thus, we apply the diagonal preconditioner for both GIPC and ‘Ours (SRBK)'. This difference arises because Morton code-based reordering may spread nodes from the same box into different groups. Without connections between boxes, nodes from different boxes in the same group cannot be effectively merged, slowing down MAS performance. In contrast, our CEMAS minimizes inter-group connections, effectively grouping nodes from the same box, enabling faster PCG convergence as shown in ‘Ours (CEMAS+SRBK)'.
	Interestingly, approximating the boxes as stiff affine bodies does not significantly improve performance in this scenario, as each box’s small DOF count allows efficient computation with pure FEM.
	
	\begin{figure}[htbp]
		\centering
		\begin{subfigure}[b]{0.4\textwidth}
			\centering
			\includegraphics[width=\columnwidth, trim=300 0 200 0, clip]{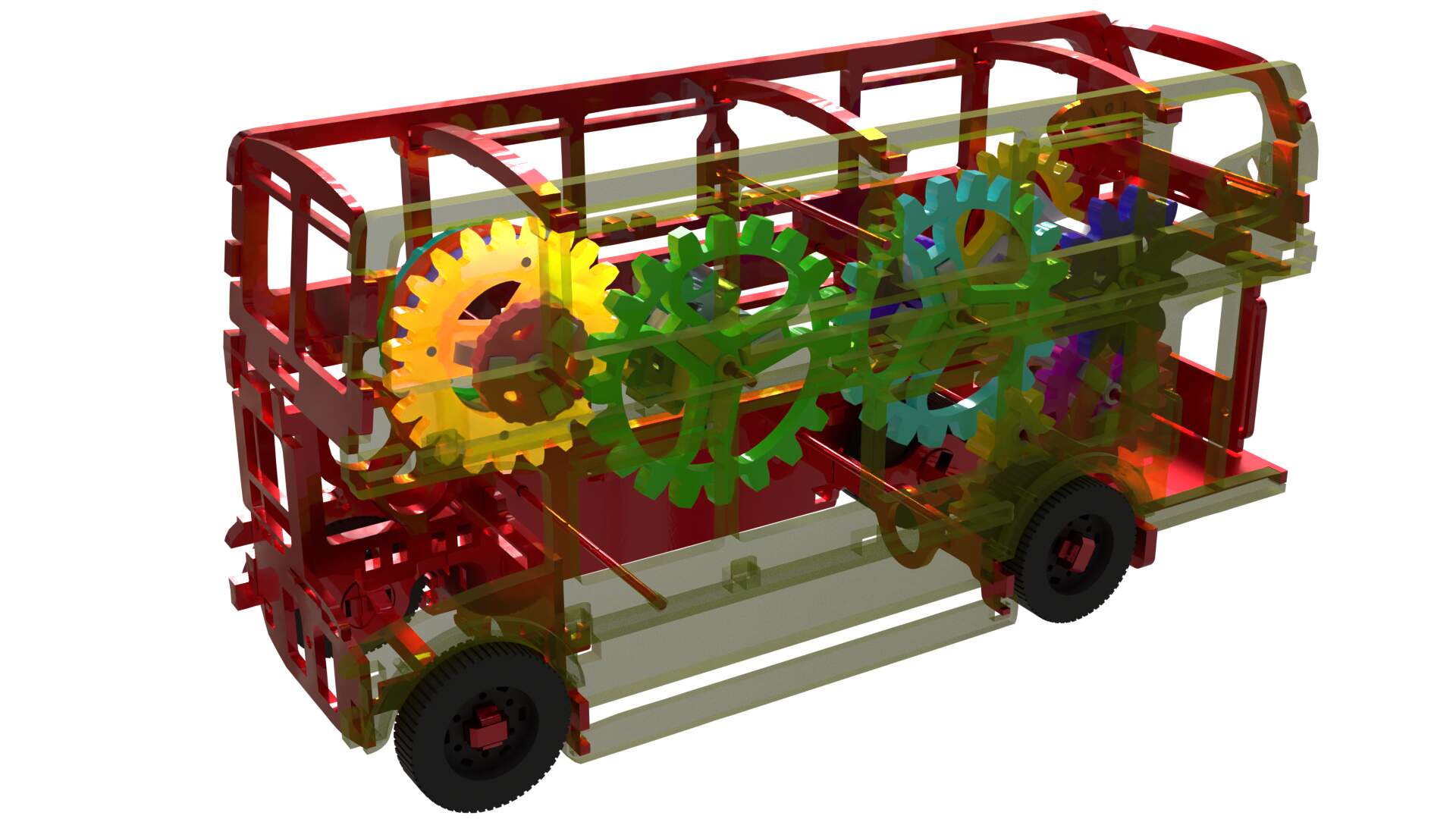}
		\end{subfigure}
		\caption{\label{fig:puzzle_london_bus} \textbf{Diagonal view of the London bus.} The yellow gear acts as the motor, driving the bus forward through rolling friction between the wheels and the ground. The wheels are simulated as rubber using FEM, while all other parts are modeled as stiff affine bodies. Detailed simulation setup is provided in \autoref{tab:bus_sim_info}.}
	\end{figure}
	\paragraph{3D puzzle bus simulation}
	%\todo{xinyu todo: car assembly and optimization settings }
	We digitalize a 3D puzzle, the London Bus from Wooden City$^\circledR$ (\autoref{fig:puzzle_london_bus}). 
	We extract the part contours in the illustrations of the puzzle instructions to create the 2D sketches in AutoCAD$^\circledR$. The 3D parts are created and assembled in SolidWorks $^\circledR$. 
	Our \textit{eye-norm} modeling accuracy of the 3D parts, as well as the rough assembling accuracy, is so far from the real-world vehicle manufacture standard that the gear teeth are not even well aligned, leading to a poor dynamic balance condition on the high-speed rotating parts. Thus, it is challenging to predict the behavior of the bus using the traditional analytic motion analysis (e.g., Mechanical Linkage Analysis), but our framework can still simulate it robustly with the yellow gear acting as the motor (see our supplemental document for details). In fact, our simulator robustly captures the realistic non-smooth motion caused by potential interference in the low-accuracy gear system and self-adjustment of motor speed (averaging $3m/s$) relative to resistance (see \autoref{fig:teaser} and our supplemental video).
	
	\begin{figure}[htbp]
		\centering
		\begin{subfigure}[b]{\linewidth}
			\centering
			\includegraphics[width=\linewidth, trim=0 0 0 0, clip]{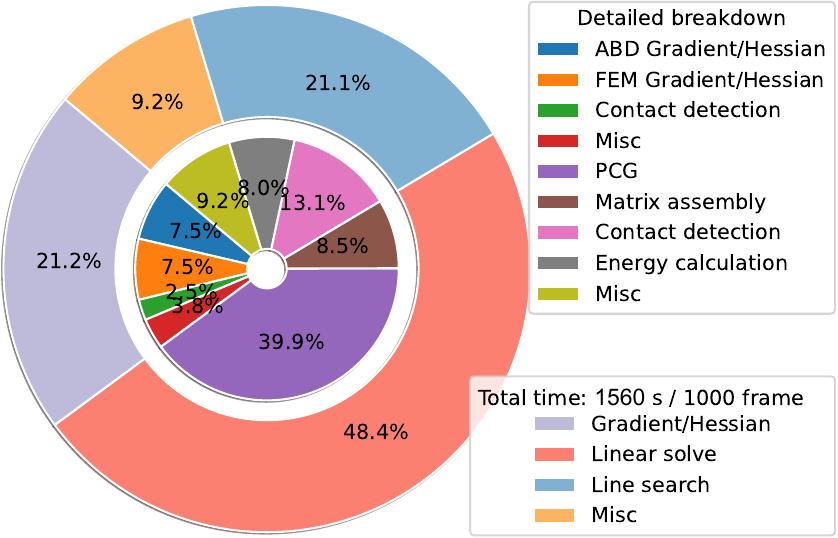}
		\end{subfigure}
		\caption{\label{fig:london_bus_pie} \textbf{Timing breakdown of the London bus simulation.} The outer ring shows the high-level breakdown into Gradient/Hessian (21.2\%), Linear solve (48.4\%), Line search (5.8\%), and Misc (24.6\%). The inner ring provides a more detailed breakdown.
		}
	\end{figure}
	
	A timing breakdown of the simulation is shown in \autoref{fig:london_bus_pie}. Since the bus is entirely driven by frictional contact forces between the FEM wheels and the ground, resulting in an average of 118K rapidly changing contact pairs, the scene complexity leads to a large number of Newton iterations. The median number of Newton iterations per time step is 29, with a maximum of 786. 
	Despite this challenging scenario, our simulator achieves an average of 1.56 seconds per frame. Detailed simulation setup is provided in \autoref{tab:bus_sim_info}.

	% \begin{table}
	% \caption{\label{tab:bus_sim_info} Simulation settings of the London Bus example. FEM Y/P refers to the Young's modulus and Poisson ratio; Gear fr. and Wheel fr. means the friction coefficient of gear-gear contact and wheel-ground contact.
	% }
	% \resizebox{\columnwidth{}}{!}{
	%     \begin{tabular}{cccccc}
	%         \toprule
	%         Mesh: v,t,f             & {$\Delta t$}  & FEM Y/P   & FEM density   & ABD stiffness & ABD density\\
	%         \midrule
	%         {$(191K, 675K, 132K)$}  & {$5\times10^{-3}s$}         & {$7MPa/0.49$}  & {$500kg/m^3$}           & {$100MPa$}           & {$600kg/m^3$}   \\
	%         \toprule
	%         FEM: v,t                & Gear fr.      & Wheel fr. & Relative {$\hat{d}$} & Bus Dimensions\\
	%         \midrule            
	%         {$(31K,120K)$}          & {$0.0$}           & {$0.3$}       & {$1\times10^{-3}$} & {$0.7m\times 1.2m\times 2.3m$}
	%         \bottomrule
	%     \end{tabular}
	% }
	% \end{table}
	\begin{table}
		\caption{\label{tab:bus_sim_info} \textbf{Simulation setup of the London Bus.} FEM Y/P refers to the Young's modulus and Poisson ratio; Gear fr. and Wheel fr. mean the friction coefficient of gear-gear contact and wheel-ground contact, respectively.}
		\resizebox{\columnwidth}{!}{
			\begin{tabular}{cccccc}
				\toprule
				Mesh: v, t, f & $\Delta t$ & FEM Y/P & FEM density & ABD stiffness & ABD density \\
				\midrule
				$(191K, 676K, 265K)$ & $5\times10^{-3}s$ & $7MPa/0.49$ & $500kg/m^3$ & $100MPa$ & $600kg/m^3$ \\
				\toprule
				FEM: v,t & Gear fr. & Wheel fr. & Relative $\hat{d}$ & Bus Dimensions \\
				\midrule
				$(31K, 121K)$ & $1\times10^{-3}$ & $0.99$ & $5\times10^{-4}$ & $0.7m\times 1.2m\times 2.3m$ \\
				\bottomrule
			\end{tabular}
		}
	\end{table}

	\section{Conclusion and Discussion}
	% This paper presents a novel, unified simulation framework that pushes the boundaries of simulator design. Our framework can uniformly handle a diverse range of simulations, including soft, highly stiff, and rigid bodies, as well as efficiently couple cloth with various volume solids. The key innovations in our framework are: a novel connectivity-enhanced MAS Preconditioner coupled with a highly optimized SpMV algorithm, creating a highly efficient linear solver that performs well for both soft and stiff simulations; and a novel locking-free membrane energy with eigensystems, proposed for stiff cloth simulation, which not only delivers better simulation quality but also achieves faster convergence rates in the PCG solver, further enhancing the efficiency of large, stiff simulations within our framework. By combining these technical advancements, our framework enables robust, accurate, and efficient simulations across a wide spectrum of material properties and coupling scenarios.
	
	In this paper, we presented a novel, fully GPU-optimized Incremental Potential Contact (IPC) framework capable of simulating materials of a wide range of stiffness with consistent high performance and scalability. Our framework introduces several significant advancements over previous methods, achieving up to 10$\times$ speedup compared to the state-of-the-art GIPC.
	Our contributions include:
	1) A novel connectivity-enhanced MAS preconditioner optimized for the GPU, which significantly improves convergence at a lower preconditioning cost;
	2) A cubic \modify{inexact} strain-limiting energy with an analytic eigensystem, designed to simulate stiff membranes such as cloth without membrane locking;
	3) An innovative hash-based \modify{two-level} reduction strategy, enabling fast matrix assembly for efficient {affine-deformable} coupling.
	Our framework's robust performance is demonstrated through extensive benchmark studies and rigorous performance analyses, showing superior results in various scenarios, including soft, stiff, and hybrid simulations. The framework efficiently handles high resolution, large deformations, and high-speed impacts, maintaining accuracy and robustness.
	We believe our advancements address critical challenges in IPC simulation, providing a scalable and high-performance solution for complex frictional contact behaviors. Our comprehensive approach not only improves the efficiency of IPC but also retains the method's reliability and accuracy, making it a valuable tool for future research and applications in a variety of fields.

	\paragraph{Limitations and future work}
    \modify{Our GPU optimizations leverage efficient mutual data access among threads within the same warp, making NVIDIA CUDA a natural demonstration platform due to its widespread adoption. These optimizations are generalizable to any platform with warp-like operations; however, without this capability, they would not be feasible.}
	Our preconditioner was designed under the assumption that the mesh connectivity would not change during the simulation. We precompute the partition when loading the meshes, which means our current design is not suitable for simulations with adaptive mesh refinement and/or fracture mechanics, where the mesh topology may change in any frame.
	Additionally, the main performance bottleneck of IPC comes from two main factors: the expensive computational cost of each Newton iteration and the large number of Newton iterations required in cases with challenging contact scenarios. Our preconditioner design and GPU optimizations have primarily focused on minimizing the cost of each Newton iteration. However, the slow convergence of the optimization in challenging cases can still result in slower overall performance compared to simulation methods that trade accuracy for efficiency.
	Therefore, it is very meaningful to keep investigating better preconditioners that can effectively handle simulations with mesh topology changes and to explore novel nonlinear solvers or discretization strategies that can significantly improve the convergence of the nonlinear solver.

        	\begin{acks}
This work was supported in part by the Junior Faculty Startup Fund of Carnegie Mellon University. It was also partially funded by the Research Grants Council of Hong Kong (Ref: 17210222), and by the Innovation and Technology Commission of the HKSAR Government under the ITSP-Platform grant (Ref: ITS/335/23FP) and the InnoHK initiative (TransGP project). Part of the research was conducted in the JC STEM Lab of Robotics for Soft Materials, funded by The Hong Kong Jockey Club Charities Trust. We thank the reviewers for their detailed and insightful feedback, and we are especially grateful to Xinyu Lu for his important contributions to the system implementation.
	\end{acks}

	\bibliographystyle{ACM-Reference-Format}
	\bibliography{ugipc}
        \newpage
        \clearpage
        

\section*{Supplementary material}

# Technical Supplement to "Advancing GPU IPC for Stiff Affine-Deformable Simulation"

**KEMENG HUANG**\*, Carnegie Mellon University, USA and The University of Hong Kong, TransGP, China
**XINYU LU**\*, TransGP, China
**HUANCHENG LIN**, Carnegie Mellon University, USA and The University of Hong Kong, TransGP, China
**TAKU KOMURA**, The University of Hong Kong, TransGP, China
**MINCHEN LI**, Carnegie Mellon University, USA

This document provides supplementary details about the technical background, derivations, and implementation of our methods.

CCS Concepts: • **Computing methodologies** → **Physical simulation**; *Parallel algorithms*.

Additional Key Words and Phrases: GPU Programming, Incremental Potential Contact, Elastodynamics, Finite Element Method, Affine Body Dynamics, Preconditioning, Cloth Simulation

## Contents

## 1 QUANTIZATION OF NEWTON ITERATION COST

In Newton's method for nonlinear optimization, the computational costs of assembling and solving the linear system are critical to overall performance. These costs can be quantified as

$$T^{\text{nt}} = T^{\text{ga}} + T^{\text{pa}} + N^{\text{pcg}} \cdot T^{\text{pcg}}, \quad (1)$$

where $T^{\text{nt}}$ is the time spent on the linear system per Newton iteration, $T^{\text{ga}}$ is the cost of assembling the linear system, $T^{\text{pa}}$ is the preconditioner construction time, $N^{\text{pcg}}$ is the number of iterations in the preconditioned conjugate gradient (PCG) solver, and $T^{\text{pcg}}$ is the time per PCG iteration. Non-bottleneck components, such as gradient/Hessian computation and collision detection, are excluded here.

The time per PCG iteration, $T^{\text{pcg}}$, can be further broken down as

$$T^{\text{pcg}} = T^{\text{spmv}} + T^{\text{ap}} + T^{\text{dot}} + T^{\text{axpby}}, \quad (2)$$

\*K. Huang and X. Lu contribute equally to this work

Authors' Contact Information: Kemeng Huang, kmhuang@connect.hku.hk, kmhuang819@gmail.com, Carnegie Mellon University, USA and The University of Hong Kong, TransGP, China; Xinyu Lu, lxy819469559@gmail.com, TransGP, China; Huancheng Lin, lamws@connect.hku.hk, Carnegie Mellon University, USA and The University of Hong Kong, TransGP, China; Taku Komura, taku@cs.hku.hk, The University of Hong Kong, TransGP, China; Minchen Li, minchernl@gmail.com, Carnegie Mellon University, USA.

where $T^{\text{spmv}}$, $T^{\text{ap}}$, $T^{\text{dot}}$, and $T^{\text{axpby}}$ represent the time costs of sparse matrix-vector multiplication, applying the preconditioner, dot products, and vector operations of the form $\mathbf{y} = \alpha \cdot \mathbf{x} + \beta \cdot \mathbf{y}$, respectively.

We can refine the quantization of $T^{\text{nt}}$ by introducing structure quality metrics: $Q^{\text{gs}}$ for the coefficient matrix of the linear system, $Q^{\text{ps}}$ for the preconditioner, and $Q^{\text{pi}}$ for the preconditioner's quality in approximating the inverse of the coefficient matrix. This gives:

$$\begin{aligned}T^{\text{nt}} =& T^{\text{ga}}(Q^{\text{gs}}) + T^{\text{pa}}(Q^{\text{pi}}, Q^{\text{ps}}) \\ &+ N^{\text{pcg}}(Q^{\text{pi}}) \cdot (T^{\text{spmv}}(Q^{\text{gs}}) + T^{\text{ap}}(Q^{\text{ps}}) + T^{\text{dot}} + T^{\text{axpby}}).\end{aligned} \quad (3)$$

Typically, $Q^{\text{gs}}$ is positively correlated with $T^{\text{ga}}$ but negatively correlated with $T^{\text{spmv}}$. For instance, compared to a system with a duplicated and unsorted format, a uniquely sorted and compressed format incurs a higher assembly cost but enables faster matrix-vector multiplication. Similar relations apply to $Q^{\text{ps}}$, $T^{\text{pa}}$, and $T^{\text{ap}}$. Generally, $Q^{\text{pi}}$ is positively correlated with $T^{\text{pa}}$ but negatively correlated with $N^{\text{pcg}}$—a higher-quality preconditioner requires more time to construct but leads to faster PCG convergence. This can result in significant differences in PCG iteration counts between preconditioners. $T^{\text{dot}}$ and $T^{\text{axpby}}$ are unaffected by $Q^{\text{gs}}$, $Q^{\text{ps}}$, or $Q^{\text{pi}}$, as they depend solely on the number of degrees of freedom (DoFs) in the system.

Thus, improving the performance of each Newton iteration requires enhancing $Q^{\text{gs}}$, $Q^{\text{ps}}$, and $Q^{\text{pi}}$ to reduce $N^{\text{pcg}}$, $T^{\text{spmv}}$, and $T^{\text{ap}}$ while ensuring that $T^{\text{ga}}$ and $T^{\text{pa}}$ do not increase significantly.

## 2 THE CONSTRUCTION OF CONNECTIVITY-ENHANCED MAS

The Multilevel Additive Schwarz (MAS) method is the state-of-the-art preconditioner for PCG solvers in elastodynamic contact simulations [Wu et al. 2022]. However, its Morton code based node reordering may result in slow performance in challenging cases. To further speedup PCG convergence, we proposed a connectivity enhanced MAS based on METIS [Karypis and Kumar 1998] in the main paper. In this section, we will start by introducing the background of MAS in subsection 2.1, and then detail our METIS-based node reordering approach in subsection 2.2, followed by an explanation of our GPU implementation for coarse space construction in subsection 2.3.

### 2.1 Multilevel Additive Schwarz (MAS) Background

For clarity and self-containment, we include the mathematical definition of MAS from Wu et al. [2022] in this subsection.

**Algorithm 1** METIS-based Node Reordering

**Input:**
    $Pos$   ▷ position of nodes
    $Ele$   ▷ node index in each element
    $DNum$   ▷ the number of partitions

**Output:**
    $sorted\_Pos$   ▷ updated order for each node
    $sorted\_Ele$   ▷ updated new node index in each element
    **map**   ▷ our mapping array
    **remap**   ▷ our remapping array

**Method:**
1: $P \leftarrow \text{Len}(Pos)$   ▷ get the number of nodes
  // find neighbor count and indices $j$ for each node:
2: $\{Neighbors[P]\{j\}, Counsts[P]\} \leftarrow$ **NeighborTraversal**$(Ele)$
3: $Offsets[P+1] \leftarrow \text{ExclusiveSum}(Counts)$
  // expand the neighbor information into an array for parallelization, which also aligns with the input format required by METIS:
4: **for** each node $i$ **do**
5:    $offset = Offsets[i]$
6:    **for** each neighbor $j$ of node $i$ **do**
7:      $Adjs[offset] = j$
8:      $AdjW[offset++] = Counts[j]$
9:    **end for**
10: **end for**
  // record the part index for each node after METIS partitioning:
11: $PIdx[P] \leftarrow$ **METIS_Partition**$(Adjs, AdjW, Offsets, DNum)$
  // sort mesh nodes according to the part index of nodes:
12: $idx[N] \leftarrow \{0, 1, 2, ..., P\}$
13: $\{sorted\_PIdx, sorted\_idx\} \leftarrow$ sort $\{PIdx, idx\}$ using $PIdx$
14: **for** $i = 0, 1, ..., P-1$ **do**
15:    $sorted\_Pos[i] = Pos[sorted\_index[i]]$
16: **end for**
  // update elements according to the new index of each node:
17: **for** each element $k$ **do**
18:    **for** each node index $t$ in element $k$ **do**
19:      $sorted\_Ele[k][t] = sorted\_index[sorted\_Ele[k][t]]$
20:    **end for**
21: **end for**
  // construct **map** and **remap** array:
22: $remap[DNum \times domain\_size] \leftarrow$ initialize all entries with -1
23: $s = 0$
24: **for** $i = 0, 1, ..., P-1$ **do**
25:    $remap[domain\_size \times sorted\_PIdx[i] + (s++)] = i$
26:    **if** i<node_number-2 **then**
27:      **if** $sorted\_PIdx[i] \neq sorted\_PIdx[i+1]$ **then**
28:        $s = 0$
29:      **end if**
30:    **end if**
31: **end for**
32: $s = 0$
33: **for** each entry $i$ in $remap$ array **do**
34:    **if** $remap[i] == s$ **then**
35:      $map[s++] = i;$
36:    **end if**
37: **end for**

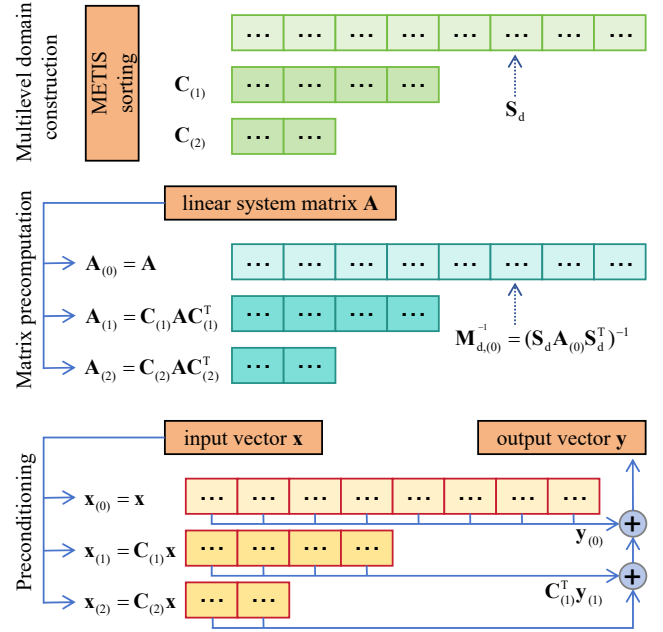

Fig. 1. *Illustration of MAS preconditioner.*

Let $\mathbf{A}$ be the system matrix, and let $\Omega_d$ represent a set of domains covering all nodes $\Omega$, such that $\Omega = \bigcup \Omega_d$. The selection matrix $\mathbf{S}_d \in \mathbb{R}^{3N_d \times 3N}$ extracts the $N_d$ nodes in domain $d$ from the $N$ nodes in $\Omega$. The additive Schwarz (AS) preconditioner is defined as:

$$\mathbf{M}_{(0)}^{-1} = \sum_d \mathbf{S}_d^T \mathbf{M}_{d,(0)}^{-1} \mathbf{S}_d, \qquad (4)$$

where $\mathbf{M}_{d,(0)} = \mathbf{S}_d \mathbf{A} \mathbf{S}_d^T \in \mathbb{R}^{3N_d \times 3N_d}$ is a symmetric positive definite matrix within domain $d$. We do not use overlapping domains here to improve parallel efficiency, making $\mathbf{M}_{(0)}^{-1}$ equivalent to a block diagonal preconditioner.

To further accelerate convergence in large linear systems, a multilevel approach can be adopted. In a two-level MAS, the MAS preconditioner is given by:

$$\mathbf{M}_{2,\text{MAS}}^{-1} = \mathbf{M}_{(0)}^{-1} + \mathbf{C}_{(1)}^T \mathbf{M}_{(1)}^{-1} \mathbf{C}_{(1)}, \qquad (5)$$

where $\mathbf{C}_{(1)} \in \mathbb{R}^{3N_1 \times 3N}$ is a coarsening matrix, and $\mathbf{M}_{(1)}$ is a single-level AS preconditioner for the coarsened system matrix $\mathbf{A}_{(1)} = \mathbf{C}_{(1)} \mathbf{A} \mathbf{C}_{(1)}^T$. The second term in Equation 5 is often referred to as the coarse space correction. By further coarsening $\mathbf{M}_{(0)}$, we can extend this to a general MAS preconditioner:

$$\mathbf{M}_{\text{MAS}}^{-1} = \mathbf{M}_{(0)}^{-1} + \sum_l^L \mathbf{C}_{(l)}^T \mathbf{M}_{(l)}^{-1} \mathbf{C}_{(l)}, \qquad (6)$$

where $\mathbf{C}_{(l)} \in \mathbb{R}^{3N_{(l)} \times 3N}$ is the coarsening matrix at level $l$. Unlike hierarchical multilevel methods, these coarseners directly map level 0 to each coarse level, allowing parallel preconditioning across all levels instead of sequentially.

A key challenge in MAS is the efficient construction of coarseners $C_{(l)}$. In GPU MAS [Wu et al. 2022], this process relies on Morton code sorting, which, while efficient, may present issues discussed in the main paper. We address these by incorporating mesh connectivity using METIS for sorting.

## 2.2 METIS-based Node Reordering

Based on the discussion in the main paper, considering mesh connectivity when reordering nodes is crucial for effective aggregation, which MAS [Wu et al. 2022] based on Morton code does not address. Instead, we partition the mesh nodes using METIS [Karypis and Kumar 1998] and map the partitions, often containing different numbers of nodes, to subdomains with padding, aiming for a dense connectivity within each subdomain to facilitate more effective aggregation at coarser levels.

Further details of our METIS-based sorting can be found in Algorithm 1. In Lines 1–10, *Adjs* is an expanded array of Neighbors, and *AdjW* is the corresponding weight array. We use the number of neighbors as weights to emphasize nodes with more connections. The *Offsets* array helps identify the neighbors for different nodes in *Adjs*. Using *Adjs*, *AdjW*, and *Offsets*, along with the required number of parts, we can perform the partition using the METIS library in Line 11. METIS then produces an array *PIdx*, which indicates the part index for each node, helping to facilitate sorting in Lines 12–21. As mentioned in the main paper, the METIS partitions may result in different number of nodes, which may results in isolation issues if directly mapped to MAS subdomains. To address this, we propose a map and remap method with padding in the main paper, with detailed implementation provided in Lines 22–39.

## 2.3 Coarse Space Construction on the GPU

This subsection details our GPU implementation for coarse space construction, focusing primarily on the level-1 (L1) coarse space, while higher levels are constructed similarly.

Our implementation leverages the CUDA library and efficiently utilizes CUDA warps. Testing with a maximum domain size of 32 (matching the warp size) ensures synchronized threads, as the warp is the fundamental execution unit. To address the artifact shown in Figure 6(b) of Wu et al. [2022], we adopt an aggregation strategy that considers node connectivity. Specifically, we evaluate each candidate supernode at L1, splitting it into multiple supernodes if nodes are not fully connected. To accomplish this, we assign a hash value to each node within a domain. Nodes sharing a hash value are connected, transforming the splitting process into a hashing operation, which we explain below.

*Initial Hash Encoding.* To identify all connected nodes, we first examine directly connected neighbors. In Algorithm 2, threads are launched based on the size of the remap array, ensuring no isolated nodes (as discussed in the main text; see Figure 6). First, we retrieve the actual node index using the remap array (line 4). This allows us to identify neighboring indices accurately (lines 11–14). Using the map array, we locate the global domain indices ('mapped_id') in the domain array for each neighbor (refer to Figure 7 in the main text). If a neighbor is in the same domain, we set the corresponding bit in a 32-bit hash integer to 1. This encoding uses a 32-bit integer,

**Algorithm 2** Initial hash encoding

**Input:**
  *Counsts*                     ▷ neighbor counts per node
  *Offsets*        ▷ same with the 'Offsets' in Algorithm 1
  *Adjs*              ▷ same with the 'Adjs' in Algorithm 1
**Output:**
  *con_hashs*           ▷ compressed domain connection
  *Counsts*                       ▷ updated 'Counsts'
  *Adjs*                             ▷ updated 'Adjs'
**Method:**
1: **for** $tid = 0, 1, ..., domain\_number \times domain\_size - 1$ **do**
2:    $domain\_id = tid/domain\_size$
3:    $lane\_id = tid\%domain\_size$   ▷ node index inside domain
4:    $node\_id = \mathbf{remap}[tid]$   ▷ get real node index
5:    **if** $node\_id < 0$ **then**
6:       **return**   ▷ filter the ghost node in subdomain
7:    **end if**
8:    $neighbor\_num = Counsts[node\_id]$
9:    $con\_hashs[node\_id] = 1 << lane\_id$   ▷ init hash value
10:    $remained\_neighbor = 0$
11:    $offset = Offsets[node\_id]$   ▷ get the first neighbor index in compacted neighbor array $adjs$
12:    $loop\_index = 0$
13:    **for** $loop\_index < neighbor\_num$ **do**
14:       $neighbor\_id = adjs[offset + loop\_index]$
        // get neighbor node index in the preconditioner domain:
15:       $mapped\_id = \mathbf{map}[neighbor\_id]$
16:       $neighbor\_domain\_id = mapped\_id/domain\_size$
17:       **if** $neighbor\_domain\_id == domain\_id$ **then**
        // update the hash value using the corresponding index information of neighbors within the same sub-domain:
18:          $neighbor\_lane\_id = neighbor\_id\%domain\_size$
19:          $con\_hashs[node\_id] \mathrel{|}= (1 << neighbor\_lane\_id)$
20:       **else**
        // update the remaining neighbor information:
21:          $adjs[offset + remained\_neighbor] = neighbor\_id$
22:          $remained\_neighbor = remained\_neighbor + 1$
23:       **end if**
24:       $loop\_index = loop\_index + 1$
25:    **end for**
26:    $Counsts[tid] = remained\_neighbor$
27: **end for**

sufficient for domain sizes up to 32. For instance, if a node in the first domain has index 1 and neighbors {2, 4, 5, 100} with mapped indices {2, 4, 5, 110}, only {2, 4, 5} are within the same domain, resulting in a bit representation of 00110110 for node 1's hash value. To prevent recalculation of visited neighbors at higher levels, we update the neighbor information as specified in lines 21–26. Algorithm 2 thus generates the initial hash value.

*Contact Handling.* The initial hash captures only mesh connectivity and does not account for contact or collision information. Here, we describe our GPU approach for incorporating contact. Our hash encoding simplifies this step: we update the hash value based on

### Algorithm 3 Contact handling

**Input:**
    $con\_hashs[N]$ ▷ compressed domain connection
    $coarse\_table[L][N]$ ▷ super node index in higher level
    $collision\_paris$ ▷ collision node index
    $l$ ▷ level index

**Output:**
    $con\_hashs$ ▷ updated 'con_hashs'

**Method:**
1: **for** $tid = 0, 1, ..., Len(collision\_pairs)-1$ **do**
2:     $col\_pair = collision\_pairs[tid]$ ▷ get collision pair
3:     $con\_Mask[col\_pair.size] \leftarrow$ Initialized with zero ▷ the temporary hash value for collision nodes
4:     **for** $i = 0, 1, ..., Len(col\_pair)-1$ **do**
5:         **if** $l == 0$ **then**
6:             $m\_pair[i] = map[col\_pair[i]]$ ▷ map the actual node index in the collision pair to the MAS domain index
7:         **else**
8:             $m\_pair[i] = coarse\_table[col\_pair[i] + (l-1) \times N]$ ▷ retrieve the corresponding super node index at higher levels
9:         **end if**
10:     **end for**
11:     **for** $i = 0, 1, ..., Len(col\_pair)-1$ **do**
12:         **for** $j = i + 1; j < col\_pair.size; j = j + 1$ **do**
            // fetch all node pairs within the collision stencil to construct their collision connectivity information:
13:             $nid\_0 = m\_pair[i]; nid\_1 = m\_pair[j]$
14:             **If** $nid\_0 == nid\_1$ **then** $continue$
15:             **if** $nid\_0/domain\_size == nid\_1/domian\_size$ **then**
                // update the temporary hash value with collision nodes within the same MAS subdomain:
16:                 $con\_Mask[i] \mathrel{|}= (1 << (nid\_0 \% domian\_size))$
17:                 $con\_Mask[j] \mathrel{|}= (1 << (nid\_1 \% domian\_size))$
18:             **end if**
19:         **end for**
20:     **end for**
        // merge the temporary collision hash value with the domain connection hash value:
21:     **for** $i = 0, 1, ..., Len(col\_pair)-1$ **do**
22:         **atomicOr**$(\&con\_hashs[i], con\_Mask[i])$
23:         **if** $l == 0$ **then**
24:             **atomicOr**$(\&con\_hashs[col\_pair[i]], con\_Mask[i])$
25:         **else**
26:             **atomicOr**$(\&con\_hashs[m\_pair[i]], con\_Mask[i])$
27:         **end if**
28:     **end for**
29: **end for**

contact pair information. In Algorithm 3, threads are launched for each contact pair, starting with the initial hash value from Algorithm 2. The variable *collision_pair* stores the indices of colliding nodes, which are mapped to domain or supernode indices at higher levels (lines 4–10). The *coarse_table* maps L0 node indices to corresponding supernode indices at higher levels, detailed in Algorithm 5. Each node in the contact pair is treated as fully connected; if both nodes belong to the same domain at the current level, the corresponding bit in the local hash is set to 1 (lines 12–22). Finally, we atomically merge the local hash into each node's hash to incorporate contact information. Note that colliding nodes may reside in different domains at lower levels but are usually aggregated at higher levels.

### Algorithm 4 Connection expansion

**Input:**
    $con\_hashs[N]$ ▷ compressed domain connection
    $remap$ ▷ 'remap' array

**Output:**
    $con\_hashs[N]$ ▷ updated 'conn_hashs'
    $supNode\_num[DNum]$ ▷ super node count per domain

**Method:**
1: **for** $tid = 0, 1, ..., domain\_number \times domain\_size - 1$ **do**
2:     $domain\_id = tid/domain\_size$
3:     $btd\_id = tid \% blockSize$ ▷ local thread id inside the CUDA block
4:     $bDomain\_id = (btd\_id)/domain\_size$ ▷ get the local subdomain index within CUDA thread block
5:     $lane\_id = tid \% domian\_size$
6:     $node\_id = $ **remap**$[tid]$
7:     $Masks[blockSize] \leftarrow$ **shared memory** ▷ for loading the hash value for sharing within each subdomain
8:     $counts[blockSize/domain\_size] \leftarrow$ **shared memory** ▷ for counting the number of super nodes per subdomain
9:     **If** $node\_id < 0$ **then** $continue$
10:     **If** $lane\_id == 0$ **then** $counts[bDomain\_id] = 0$
11:     $conMask = con\_hashs[node\_id]$
12:     $Masks[btd\_id] = conMask$
13:     $visited = (1 << lane\_id)$
14:     **while** $conMask \neq (1 << domain\_size) - 1$ **do**
15:         $todo = (visited \oplus conMask)$ ▷ 32-bit integer marking the connected nodes that have been traversed
16:         **If** $todo == 0$ **then** $break$ ▷ all connected nodes have been visited
17:         $next = \_\_\mathbf{ffs}(todo) - 1$ ▷ fetch the remaining connected node with the minimum index
18:         $visited \mathrel{|}= (1 << next)$ ▷ mark the visited node
19:         $conMask \mathrel{|}= Masks[next + bDomain\_id \times domain\_size]$
        ▷ merge the neighbor's connected nodes that are not included in the current hash value
20:     **end while**
21:     $con\_hashs[node\_id] = conMask$
22:     $prefix = \_\_\mathbf{popc}(conMask \text{ And } ((1 << lane\_id) - 1))$
        // count the super nodes in each MAS subdomain:
23:     **if** $prefix == 0$ **then**
24:         **atomicAdd**$(\&counts[bDomian\_id], 1)$
25:     **end if**
26:     **if** $lane\_id == 0$ **then**
27:         $supNode\_num[domain\_id] = counts[bDomain\_id]$
28:     **end if**
29: **end for**

*Expanded Hash Value.* After encoding connectivity and collision information, we merge hash values for nodes indirectly connected within a domain, minimizing the number of connection segments. In Algorithm 4, each node accesses neighboring hash values in shared memory (line 15) for merging. Since the hash is based on domain indices, we reverse map each bit set to 1 to access the neighbor's hash values (lines 16–26; __ffs finds the position of the least significant set bit). Consequently, connected nodes share the same hash value, where bit positions indicate their local domain indices. After merging, the number of unique hash values in the domain determines the supernode count at the next level (lines 27–33; __popc counts set bits in a 32-bit integer).

*L1 Construction.* At this point, all connected nodes within a domain share the same hash value, and the number of supernodes per domain is known. We construct the final L1 coarse space, consisting of two mapping arrays: *coarseTable* and *goNext*. *coarseTable* maps node indices to supernode indices, while *goNext* maps each node to its matrix or vector index. Although *coarseTable* provides supernode indices, *goNext* ensures consistent domain size, even when domains lack sufficient nodes, by setting empty entries to zero. Additionally, since all domains are stored in a single array, *goNext* aligns node and array indices. In Algorithm 5, we identify the thread assigned to the node with the lowest domain index for each hash value. These threads match the supernode count, setting a shared 32-bit integer per domain and marking bit positions based on domain index. Counting bits set to 1 before each bit position determines the supernode index, with the global supernode offset allowing correct settings in *coarseTable* and *goNext*.

This completes our GPU approach for L1 coarse space construction. Higher levels can be constructed similarly, and our source code will be made available upon acceptance.

## 3 LONDON BUS SIMULATION DETAILS

We partition the bus into several components:

(1) the bus frame, a relatively static structure that supports other parts;
(2) the gear system transferring the motion and power from the motor to the wheels;
(3) the back wheels contacting the ground and driving the bus through rolling friction; and
(4) the front wheels contacting the ground but acting as resistance to the bus.

We pack the bus frame and gear axles together into one affine body [Lan et al. 2022] and keep every gear as a separate affine body. The constraints between the gear axles and the gears are achieved directly by the contact force, which is purely physical. The wheels are simulated using FEM, with a 7$MPa$ Young's modulus and a 0.49 Poisson's ratio, a common rubber material in reality. Other configurations are listed in the main paper. The time step size limitation (5$ms$) comes from the fastest rotating speed of the gears.

To drive the bus, we designed an ideal motor that applies any extra kinetic energy we want to a single affine body. A generalized motor energy $P$ for a single affine body is defined as

$$P = \frac{1}{2}\|\dot{\mathbf{q}}^P\|^2_{\mathbf{M}^P}, \quad (7)$$

**Algorithm 5** L1 construction

**Input:**
  $supNode\_num$ ▷ super node number is each L0 subdomain
  $con\_hashs$ ▷ the output 'con_hashs' in Algorithm 4
  $remap$
**Output:**
  $goNext$ ▷ map to the global super node index
  $coarseTable$ ▷ map to the local super node index in higher levels
  $sNodes$ ▷ the number of super nodes in L1
**Method:**
1: $Offsets \leftarrow \text{ExclusiveSum}(supNode\_num)$ ▷ start from 0
2: **for** $tid = 0, 1, ..., domain\_number \times domain\_size - 1$ **do**
3:   $dom\_id = tid/domain\_size$
4:   $btd\_id = tid\%blockSize$
5:   $bDomain\_id = (btd\_id)/domain\_size$ ▷ CUDA blockSize is set with $domain\_size^2$
6:   $lane\_id = tid\%domian\_size$
7:   $electMask[domain\_size] \leftarrow$ **shared memory space** ▷ used to encode each subdomain with supernode information
8:   $PrefixId[domain\_size^2] \leftarrow$ **shared memory space** ▷ used to record the supernode index within the subdomain
9:   **If** $lane\_id == 0$ **then** $electMask[bDomain\_id] = 0$
10:  **if** $is\_the\_last\_thread(tid)$ **then**
11:    $sNodes = Offsets[dom\_id] + supNode\_num[dom\_id]$ ▷ get the number of super nodes in L1
12:  **end if**
13:  $node\_id = \textbf{remap}[tid]$
14:  **If** $node\_id < 0$ **then** continue
15:  $conMask = con\_hashs[node\_id]$
16:  $prefix = \_\_\textbf{popc}(conMask \text{ And } ((1 << lane\_id) - 1))$ ▷ locate the thread of the leaf node with the minimum node index for each supernode
    // only the located thread can set the bit value for the $electMast$ of each subdomain, so the order of the bits with a value of 1 can be used to determine the order of supernodes within each subdomain:
17:  **if** $prefix == 0$ **then**
18:    $\textbf{atomicOr}(\&electMask[bDomian\_id], 1 << lane\_id)$
19:  **end if**
20:  $PrefixId[btd\_id] = \_\_\textbf{popc}(electMask[bDomian\_id] \text{ And } ((1 << lane\_id) - 1))$
21:  $PrefixId[btd\_id] = PrefixId[tid\%block\_size] + Offsets[domain\_id]$
22:  $elect\_lane = \_\_\textbf{ffs}(conMask) - 1$
23:  $lane\_prefix = Prefix[elect\_lane + domian\_size \times bDomain\_size]$
24:  $coarseTable[0][node\_id] = lane\_prefix$
25:  $goNext[node\_id] = lane\_prefix + \lceil N/domain\_size \rceil \times domain\_size$
26: **end for**

where $\dot{\mathbf{q}}^P$ is the effective velocity of the motor driver, $\mathbf{M}^P$ is the power mass matrix, which we expand to $K^P \cdot \mathbf{M}$, where $K^P$ is a parameter indicating how strong the motor is relative to the body mass, and $\mathbf{M}$ is the mass matrix of the affine body.

The time-discretized version of this energy has a form of soft position constraints weighted by the power mass matrix:

$$\hat{P}(\mathbf{q}) = \frac{1}{2\Delta t^2} \|\mathbf{q}^P - \mathbf{q}\|_{\mathbf{M}^P}^2, \tag{8}$$

where $\mathbf{q}^P$ is the target state the motor is trying to drive the system towards and $\Delta t$ is the time step size. To explain it in more details, at the beginning of time step $t+1$, we have the state $\mathbf{q} = \mathbf{q}^t$, and we can add the motor energy to the system by simply adding $\hat{P}(\mathbf{q}^t)$, where $(\mathbf{q}^P - \mathbf{q}^t)/\Delta t \approx \dot{\mathbf{q}}^P$. Then, by solving $\mathbf{q}^{t+1}$, the added motor energy decreases to $\hat{P}(\mathbf{q}^{t+1})$ (since $\mathbf{q}^{t+1}$ will get closer to $\mathbf{q}^P$ than $\mathbf{q}^t$), and the lost energy transfers to other forms in the system, e.g. kinetic energy, heat, etc.

To compute $\mathbf{q}^P$, we first calculate the movement in material space as

$$\bar{\mathbf{x}}_i^P = \bar{\mathbf{x}}_i + \Delta \bar{\mathbf{x}}_i^P, \tag{9}$$

where $\bar{\mathbf{x}}_i$ is the initial position of the $i$-th node, $\Delta \bar{\mathbf{x}}_i^P$ is the movement in the material space, and $\bar{\mathbf{x}}_i^P$ is the final position of the $i$-th node in the material space. Due to the linearity of the affine transformation, we can then calculate the target world-space position as

$$\mathbf{x}^P = \mathbf{J}(\bar{\mathbf{x}}_i^P)\mathbf{q}^t, \tag{10}$$

by changing the basis of the affine deformation modes in $\mathbf{J}$ while keeping using $\mathbf{q}^t$ (note that $\mathbf{x}^t = \mathbf{J}(\bar{\mathbf{x}}_i)\mathbf{q}^t$).

Next, based on the relation

$$\mathbf{x}^P = \mathbf{J}(\bar{\mathbf{x}}_i)\mathbf{q}^P, \tag{11}$$

we can now solve for $\mathbf{q}^P$ by choosing 4 mesh nodes that are not coplanar to form $\mathbf{J}(\bar{\mathbf{x}}_i)$. As illustrated in Figure 2, we pick $\bar{\mathbf{x}}_1 - \bar{\mathbf{x}}_0$ as the rotation axis of the motor, and $\bar{\mathbf{x}}_2 - \bar{\mathbf{x}}_0$ and $\bar{\mathbf{x}}_3 - \bar{\mathbf{x}}_0$ to form two orthogonal vectors to the rotation axis.

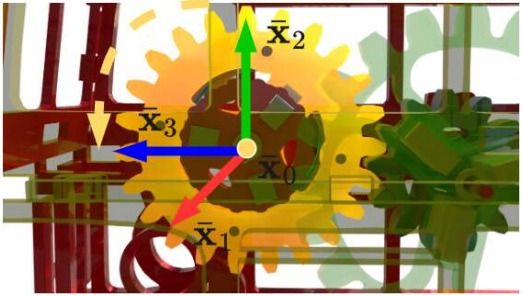

**Fig. 2. Illustration of the local frame for calculating $\mathbf{q}^P$.** Here, $\bar{\mathbf{x}}_1 - \bar{\mathbf{x}}_0$ is picked as the rotation axis of the motor, and $\bar{\mathbf{x}}_2 - \bar{\mathbf{x}}_0$ and $\bar{\mathbf{x}}_3 - \bar{\mathbf{x}}_0$ are two orthogonal vectors to the rotation axis.


	
\end{document}